\documentclass[12pt,preprint]{aastex}

\newcommand\kms{km~s$^{-1}$}
\newcommand\msun{\ifmmode{M_{\odot}}\else $M_{\odot}$\fi}
\newcommand\rsun{\ifmmode{R_{\odot}}\else $R_{\odot}$\fi}

\begin{document}

\title{A Comparison of the Chemical Evolutionary Histories
of the Galactic Thin Disk and Thick Disk Stellar Populations}

\author{Mary-Margaret Brewer\altaffilmark{1} \& Bruce W.\ Carney}
\affil{Department of Physics \& Astronomy, University of North Carolina,
Chapel Hill, NC 27599-3255; email: mmbrewer@physics.unc.edu; bruce@physics.unc.edu}
\altaffiltext{1}{Now at William Jewell College; 
email: brewerm@william.jewell.edu}

\clearpage

\begin{abstract} 
We have studied 23 long-lived G dwarfs that belong to
the thin disk and thick disk stellar populations. The stellar
data and analyses are identical, reducing the chances for
systematic errors in the comparisons of the chemical abundance
patterns in the two populations. Abundances have been derived
for 24 elements: O, Na, Mg, Al, Si, Ca, Ti, Sc, V, Cr, Mn, Fe,
Co, Ni, Cu, Zn, Sr, Y, Zr, Ba, La, Ce, Nd, and Eu.
We find that the behavior
of [$\alpha$/Fe] and [Eu/Fe] vs.\ [Fe/H] are quite different for
the two populations. As has long been known, the thin disk 
O, Mg, Si, Ca, and Ti ratios are enhanced relative to iron
at the lowest metallicities, and decline toward
solar values as [Fe/H] rises above $-1.0$. For the
thick disk, the decline in [$\alpha$/Fe] and
[Eu/Fe] does not begin at [Fe/H] = $-1.0$,
but at $-0.4$. Other
elements share this same behavior, including Sc, Co, and Zn,
suggesting that at least in the chemical enrichment history of
the thick disk, these elements were manufactured in similar-mass
stars. The heavy $s$-process elements Ba, La, Ce, and Nd
are over-abundant in the thin disk stars relative to the
thick disk stars.
On the other hand, the constancy of the [Ba/Y] ratio
suggests that only one $s$-process site was manufacturing these
elements, or, possibly, that the $r$-process was responsible for
the bulk of the nucleosynthesis of these elements. We combine
our results with other studies (Edvardsson et al., Prochaska et al.,
Bensby et al., and Reddy et al.), who had already found very
similar trends, in order to further explore
the origin of the thick disk. The signs for
an independent (parent galaxy) evolution of the thick disk are
clear, in terms of the different metallicities at which
the [$\alpha$/Fe] ratios begin to decline,
as well as ``step function" behavior of some elements,
including
[Eu/Y], [Ba/Fe], and, possibly, [Cu/Fe], at [Fe/H] $\approx -0.2$.
\end{abstract}


\section{INTRODUCTION}

On the basis of star counts, Gilmore \& Reid (1983) argued
that the Galaxy contains a second disk population, with a
larger vertical scale distribution and hotter
kinematics than the ``classical old disk". 
This ``thick disk", as it is now commonly
called, is of considerable relevance to the understanding of
the evolution of our Galaxy and, by extension, other disk
galaxies. Does the development of all disk galaxies involve
such a thick disk and, if so, why? If the thick disk
is such a ``pre-destined" evolutionary stage, can we determine
its relationship with the other Galactic stellar populations,
including the halo, the thin disk, and the bulge?

Or does the thick disk
of the Milky Way represent an entanglement or a merger with
another, presumably much smaller, galaxy? If so, are such
encounters stochastic and infrequent among disk galaxies or
are such encounters expected as part of some sort of
hierarchical assembly of large galaxies out of cannibalization
of smaller ones? 

The salient properties of the thick disk, and clues to answering
these questions, include the relative masses of the thick and
thin disks, the vertical scale height, the mean metallicity,
the metallicity gradient, the kinematics, the age (and age spread, if one
is detectable), and the chemical evolution. The last property
is the subject of this paper, but, as we discuss, it is not
independent of the others.

The mid-plane density normalization of the thick disk relative to
the thin disk has been the subject of serious study since the
thick disk's discovery, and the subject is well reviewed by
Chen et al.\ (2001). They employed the Sloan Digital Sky Survey
to provide a comprehensive new attack on the problem, finding
preferred vertical scale heights of between 580 and 750~pc for
the thick disk, compared to 330~pc for the thin disk. Since
density normalization is directly related to scale
heights, this range in values implies mid-plane density
normalizations of 13\% down to 6.5\%. It is worth recalling
that heating over a long period of time of Galactic thin disk
stars by spiral arms or a central bar, giant molecular
clouds, or accretion and enlargement of the Galactic
disk are unable to explain such a difference
in the scale heights (Jenkins 1992; Asiain, Figueras, \& Torra 1999;
Fuchs et al.\ 2001).

The local mean metallicity,
$<$[Fe/H]$>$, of the thick disk is between $-$0.5 and $-$0.7 (Gilmore \&
Wise 1985, Carney et al.\ 1989, Gilmore et al.\ 1995; Chiba \& Beers 2000).
The mean metallicity and spread are very similar to
what is seen among the disk globular clusters identified by Zinn (1985;
see also Armandroff 1989), and
there is some evidence of a metal-weak
tail extending down to [Fe/H]=$-$1.6 (Norris et al.\ 1985, Martin \&
Morrison 1998). 
Though the average metallicity
of the thick disk is only one third that of the 
thin disk, and much higher than that of the halo, the metallicity ranges 
certainly overlap, making it difficult to separate the two components based solely
on metallicity.

The thick disk may be most readily distinguished from the thin disk in
kinematics.  The asymmetric drift of the thick disk has been measured
between 20 \kms\ (Chiba \& Beers 2000) and 50 \kms\ (Carney et al.\ 1989).  
Soubiran et al.\ (2003) and Bensby et al.\ (2003) agreed well of the
values for the asymmetric drift, $<$V$>$, and the 
velocity ellipsoid [$\sigma$(U):$\sigma$(V):$\sigma$(W)] of
the Galactic thin disk, thick disk, and halo stellar populations,
with $<$V$>$ $\approx$ $-15$, $-50$, and $-220$ \kms, and
velocity ellipsoids of, approximately, 
(35:20:16), (67:38:35), and (160:90:90) \kms,
respectively. These differences may be exploited to assign probabilities
of membership in the thin disk, thick disk, and halo populations,
and throughout this paper we follow the formulations employed by
Venn et al.\ (2004).

A key property, and one overlooked occasionally, is that the
thick disk appears to be about as old as the Galactic halo. 
Bensby et al.\ (2004b) have attacked the question of
the relative ages of the thin disk and thick disk field stars, using field
stars whose ages may be estimated individually, and whose
population membership may be assigned probabilistically
on the basis of kinematics. They find no evidence for
age differences for thick disk stars with [Fe/H] $\leq -0.4$,
but for the minority of stars more metal-rich than this
limit, an age spread of several Gyrs may exist. As they
noted, this is also consistent with the appearance of
contributions by Type~Ia supernovae at that metallicity
level, and which we address also in this paper.

The Milky Way is not the only galaxy which possesses a thick disk
population. Indeed, Dalcanton \& Bernstein (2002) have found
them in all of the 47 disk systems they surveyed. However, it is not
clear if the envelopes they detected are thick disks or 
a galactic halo, because the limiting surface brightnesses they reached
are comparable to where Zibetti, White, \& Brinkmann (2004)
have detected the appearance of faint spheroidal components.
Nonetheless, the work of
Dalcanton \& Bernstein (2002)
breaks new ground since earlier work did not reveal
thick disks in all edge-on disk galaxies
(Morrison, Harding, \& Boroson 1994; Fry et al.\ 1999). The
absence of thick disks in other galaxies would support the idea
that thick disks arise from ``accidents", probably encounters
with other galaxies, rather than being a natural part of the
formation of a thin disk. However, as Dalcanton \& Bernstein (2002)
argued, the near-universality of thick disks does not necessarily
mean that they arise independently of encounters. While it is true
that the red colors
of extra-galactic thick disks indicate great ages, consistent
with the ``monolithic" formation idea, the thick disks do not
appear to show any correlations with the sizes of the thin disks,
nor with the presence or absence of galactic bulges. Dalcanton \&
Bernstein (2002) suggested, instead, that thick disks could be expected from
the hierarchical assembly of galaxies, in which case no such
correlations would be expected. A particularly nice piece of work
showing the dissimilarity between some thin and thick disk populations
is that of Yoachim \& Dalcanton (2005).
We note that Dalcanton \& Bernstein (2002) commented on 
the apparent similarity of the Galactic
thick disk and thin disk stellar populations' chemical
abundance patterns rules out independent origins, a conclusion
that now appears to have been premature, and which is the subject
of our study.

\section{THE ORIGIN OF THE THICK DISK}

Several groups (Gilmore 1984; Wyse \& Gilmore 1988; Sandage 1990; Burkert,
Truran, \& Hensler 1992; Pardi, Ferrini, \& Matteucci 1995) have advocated
that the thick disk was a precursor
to the ``old thin disk", and thus a thick disk is a natural early step in
the chemodynamical evolution of galactic disks. Burkert et al.\ (1992)
argued for a rapid (300-400~Myr) thick disk formation phase. 
Such a rapid process would result in
a thick disk with a very small age range and little or no metallicity
gradient, as has been observed. However, such
models predict a smooth transition to
the thin disk predicting continuous age, [Fe/H], [X/Fe], and kinematic
relationships.

Rather than a ``pre-ordained" evolutionary scenario, the thick
disk may have resulted from an encounter with a small galaxy,
with a mass roughly like that of the Magellanic Clouds.
Simulations by Quinn et al.\ (1993) and Sellwood, Nelson,
\& Tremaine (1998) have shown that a thick disk can form
during a merger event by the heating of a pre-existing cold, thin disk.
Because the old disks are wrecked in the merging process, any thin disk
we see today would have re-formed after the merger event, 
and the thick disk
would be older than the bulk of the oldest stars in the 
reincarnated thin disk. 

A thick disk may
also form by the direct accretion of stars from the merging galaxy.  
Such a view was offered by
Statler (1988), whose model involved the accretion of a ``mature",
mostly stellar, galaxy. Statler showed 
that accretion can give rise to a boxy
distribution similar to a thick disk, though this requires special
initial conditions for the cannibalized galaxy's orbit. 
In merger models there need not be continuity between the
kinematics or chemical abundances of the thick and thin disks.

Distinguishing between evolutionary models and merger models requires age,
kinematic, and abundance data for the halo, thick disk, and thin disk,
especially at the transitions between the populations. 
Carney, Latham, \& Laird (1989) argued that the thick
disk is too distinct in its chemodynamical behavior, manifested, for
example, by [Fe/H] vs.\ $|$W$|$ velocity, to share a common origin with the thin
disk. Instead, they argued that the thick disk is likely to have arisen from a merger
event, but a major one that involved already-assembled galaxies. The
star formation in the victim may have been largely completed long
before the actual merger.

\section{TESTS OF THE MODELS}

Mixing of stars from an accreted galaxy into the Milky Way,
and scattering of stellar orbits via encounters with spiral
arms, the central bar, or giant molecular clouds, will blur
the kinematic signatures of a merger event. 
The orbital angular momentum, represented in the solar
neighborhood by the Galactic V velocity, may be the
best conserved kinematic property. In addition,
stars retain the history of star formation in their abundance patterns,
what Freeman \& Bland-Hawthorn (2002) have called ``chemical tagging".

An ensemble of gas and stars should initially yield stars whose abundance
patterns reflect only the ejecta from SNe~II, which is thought to be the main
production site of the $\alpha$-elements (O, Mg, Si, Ca, Ti) and $r$-process
elements. SNe~II arise from relatively massive stars, and hence the timescale
for the appearance of their ejecta should be $\approx\ 10^{7}$ years.
Low-mass (M $<$ 4 M$_\odot$) AGB stars are thought to be a major
production site for $s$-process elements (Busso et al.\ 1999), and 
such stars begin
releasing these elements into the ISM in about $10^8$ yrs, significantly
later than $r$-process elements from SNe~II.  The $r$- to
$s$-process ratio should therefore be higher at the earliest epochs in a
population's history and the metallicity where this ratio begins to
decrease can give a star formation rate time-scale. SNe~Ia begin to appear even
later, perhaps within about $10^{9}$ years 
(see Matteuchi \& Recchi 2001), and if
star formation is slow enough for their ejecta to
be incorporated into gas still undergoing star 
formation, the signature should
appear at the mean metallicity the interstellar medium and newly-forming
stars had reached. This is the basic explanation for the enhanced
abundances of $\alpha$- elements seen in stars in the 
solar neighborhood at the lowest
metallicities, and the steady decline in [$\alpha$/Fe] abundances from
[Fe/H] $\approx$ $-1.0$ and higher. In other words, the solar neighborhood 
interstellar medium had
reached a mean metallicity of [Fe/H] $\approx$ $-1$ before the SNe~Ia
began to significantly enrich the interstellar medium. Had 
star formation been more rapid, the metallicity would be higher
before the SNe~Ia contribution begin to appear.
Thus one
possible test of the inter-relationship between thin disk and thick disk
populations is whether they share a common [$\alpha$/Fe] vs.\ [Fe/H]
relation. If they do, the common history model is strengthened. If they do
not, then one must either conclude that the thick disk evolved
independently of the thin disk, or that the chemical enrichment models of
the joint thick disk-thin disk evolution are much more complicated.  

\section{PRIOR WORK}

One of the pioneering papers in this field is that by Edvardsson et al.\ (1993),
who studied the abundances of 13 elements in 189 relatively bright,
nearby dwarf stars. The stars were selected so that the sample covered
a wide range in [Fe/H], and with an admixture of some halo population
stars, judged by both metallicity and kinematics. The stars were also
selected so that ages could be estimated for individual stars, and
that those ages covered a very wide range, from less than 1 Gyr to
roughly the age of the halo.
Photometric metallicities enabled
them to select stars on the basis of a range of ages. The kinematical
data (proper motions and radial velocities) led to the caculation of
space velocities and Galactic orbits, a truly comprehensive approach
to the problem of the Galaxy's chemodynamical evolution. At the risk
of doing injustice to such a major piece of work, we summarize their
three primary results briefly as the following. 1. As had been
described earlier (see Wheeler, Sneden, \& Truran 1989), [$\alpha$/Fe]
is elevated by about 0.3 dex with respect to the Sun for stars with
[Fe/H] $\leq\ -1.0$, but that it declines 
as a function of [Fe/H] for higher metallicities, reaching a solar
value at [Fe/H] = 0.0. 2. This trend appears to depend on the mean
Galactocentric distance of the stars involved. The trend is different
for stars with smaller mean Galactcocentric distances, which
could be explained by a higher star formation rate in the denser
inner regions of the Galactic disk. We point out that thick disk
stars in the solar neighborhood, with larger values of the asymmetric
drift, may also have smaller mean Galactocentric distances, and hence
the interpretation of these results could also be explained by a separate chemical
evolutionary history for the thick disk rather than provide
insight into the chemical evolution of the Galactic disk itself.
3. The Galactic disk does not appear to reveal an age-metallicity
relation, as simple models of chemical evolution predict. The
explanation offered most readily is that this might be explained
by a combination of vigorous mixing of supernovae ejecta within the disk and
infall of metal-deficient gas into the solar neighborhood. An alternative
not considered by Edvardsson et al.\ (1993), but recently raised
by Bensby, Feltzing, \& Lundstr\"{o}m (2004b), is that the lack of
a correlation between age and metallicity could have resulted from
a sample that inter-mixed two populations with differing mean
metallicities and differing age-metallicity relations. 

One good example of the challenges of target selection is the
work of Chen et al.\ (2000). They studied 90 disk dwarfs with
accurate parallaxes and kinematical data, and for which they
could estimate ages for individual stars. They found that
stars that kinematically resemble the thick disk follow
a very similar trend in [$\alpha$/Fe] vs.\ [Fe/H] as do
the stars of the thin disk. This supports the
evolutionary model. However, the temperature regime
for their target stars essentially rules out any long-lived
stars in their sample, and if age is indeed a defining
characteristic of the thick disk, then their adopted sample 
of thick disk stars may have simply included
stars from the higher velocity tails of the thin disk
velocity distributions. Thus the apparent agreement between
the chemical evolution of the thin disk and thick disk
may be spurious.

Gratton et al.\ (2003) added kinematics to the chemical
abundance analyses of about 200 stars from Carretta, Gratton,
and Sneden (2000). Their careful analyses, especially in the
difficult matter of oxygen abundances, confirmed the onset
of the decline in [O/Fe] at [Fe/H] $\approx\ -1.0$ for halo stars.
However, their program stars were drawn from a wide variety of
sources, so selection effects are hard to disentangle, and the
ages of individual stars were not considered.

Fuhrmann (1998) initiated a large program of chemical abundances
analyses of nearby F and G stars, mostly dwarfs but extending
to subgiants near the main sequence turn-off. He employed 
kinematics (in a manner not well described) to distinguish between
thin disk, thick disk, and halo stars, and was among the first
to notice that the thick disk stars do not appear to follow
the same [Mg/Fe] vs.\ [Fe/H] trend as do thin disk stars. His work
has been continued by Mashonkina \& Gehren (2000, 2001, 2003) in
studies of [Ba/Fe] and [Eu/Fe] in the first two studies,
and of more metal-poor stars in the last study. They confirmed
differences in the [X/Fe] vs.\ [Fe/H] trends of thick disk
vs.\ thin disk stars. Fuhrmann (2004) also undertook a comprehensive
re-evaluation of the relative ages of the thin and thick disks
based on a sample of roughly 150 field stars, concluding that
the thick disk represents the first stage of the formation of the
Galactic disk, and that a hiatus of perhaps 3 Gyrs ensued prior to
the chemical evolution of the thin disk. A recent comprehensive
literature study by Soubiran \& Girard (2005) suggests a similar
age difference, and confirms again the separate chemical enrichment
histories of the two populations.

A much smaller sample, only ten stars, was studied by Prochaska
et al.\ (2000). However, these stars were selected on the basis
of both kinematics and ``ages", in the form of life expectancies
that equal or exceed the age of the Galaxy, thereby minimizing
the dilution of the sample by younger, hence thin disk, stars,
which we have noted afflicted the study by Chen et al.\ (2000).
Venn et al.\ (2004) confirmed the high probability of thick
disk membership for all ten stars. A potential systematic
effect may exist, however, since Prochaska et al.\ (2000)
did not analyze a similar sample of thin disk stars. Possible
differences in adopted $gf$ values or other analytical techniques
raise the
possibility of systematic effects creating differences
where none exist or, alternatively, blurring or even
erasing them when they do exist.
Prochaska found that the thick disk
behaves differently than the thin disk, supporting the
accretion model,
especially in the abundances of the $\alpha$
elements and the $r$-process element europium,
a result obtained as well by Mashonkina et al.\ (2003).

Reddy et al.\ (2003) analyzed a much larger sample, 181 F and G
dwarfs, studying abundances of 27 elements that span the
critical classes of
nucleosynthesis: CNO, $\alpha$, iron-peak, $s$-process,
and $r$-process. The stars were selected employing kinematics and ages
as the primary criteria, and the large majority of their
stars belong to the thin disk. While they also noted different
chemical evolutionary behaviors between the thin disk and the thick
disk, perhaps their most important result was clear evidence for the
expected thin disk
age-metallicity relation with the remarkably small scatter
in the relation. 
This suggests that the study of Edvardsson et al.\ (1993)
may indeed have been compromised by mixing of the thin disk
and thick disk in uncertain proportions.

In our opinion, the most important ensemble of studies regarding
the chemical evolution of the thin disk and thick disk populations
are those of Feltzing, Bensby, \& Lundstr\"{o}m (2003), 
Bensby, Feltzing, \& Lundstr\"{o}m (2003, 2004a, 2004b),
and Bensby et al.\ (2005). These
authors studied a sample comparable in size to those of
Edvarsson et al.\ (1993), but were careful to distinguish
thin disk and thick disk stars on the basis of their kinematics,
using adopted mid-plane density ratios, asymmetric drifts,
and velocity ellipsoids. Further, their sample covered a critical
range of metallicities, including thick disk stars at the
higher metallicities than have been traditionally, but probably incorrectly,
assumed to be ``reserved" to the thin disk, [Fe/H] $>$ $-0.5$.
Bensby et al.\ (2003, 2004a, 2005), found, as had others, that the
thick disk and thin disk populations begin to diverge in
[$\alpha$/Fe] vs.\ [Fe/H] for key elements at [Fe/H] $\approx\ -1.0$,
with the thin disk [$\alpha$/Fe] declining while the thick disk
[$\alpha$/Fe] values remain constant. Further,
they also found that the thick disk population also begins
to show a decline in [$\alpha$/Fe] toward solar values beginning
at [Fe/H] $\approx\ -0.4$. Thus the metal-rich tail of the thick
disk appears to have formed long enough after the onset of
star formation for SNe~Ia to contribute their ejecta to
the remaining gas, or, perhaps, that the remaining gas was mixed
into a larger reservoir in which the [$\alpha$/Fe] ratio was
already reduced. We have noted above that Bensby et al.\ (2004b)
found that even among the stars in their sample most likely
to belong to the thick disk (their TD/D $>$ 10 sample), the
most metal-rich stars appear to be several Gyrs younger than
the more metal-poor stars. 

Finally, Mishenina et al.\ (2004) obtained abundances
for  magnesium and silicon 
($\alpha$ elements), as well as
iron and nickel, in 174 F, G, and K dwarfs, and confirmed the
break in the [$\alpha$/Fe] vs.\ [Fe/H] of the thick disk at
[Fe/H] $\approx\ -0.5$, and that the thin disk and thick disk
are chemically distinct. Unfortunately, none of their program
stars had estimated ages, and some of the stars appear to have
relatively short life expectancies.

\section{OBSERVATIONS}

\subsection{Target Selection}

We have selected our targets for abundance determinations following
an unpublished study by Nauomov (1999) of the Galaxy's thin disk and thick
disk stellar populations.  His work comprised two parts. In the first, he
undertook an exhaustive
review of all available {\em uvby} photometry and parallax data and he
selected dwarfs generally lying within 100~pc of the Sun. Temperatures
were determined using the color-temperature calibrations of
Alonso, Arribas, \& Martinez-Roger (1996). Nauomov employed 
the Olsen (1993) catalogue in an unbiased manner, selecting all
stars with with $b-y$ $\ge$ 0.4, corresponding
to stars cooler than $\approx$5700~K, and having 
life expectancies on order of the thick disk's age.  
This criterion 
was chosen based on the estimated turn-off temperature of 47~Tuc,
and resulted in a sample of 269 stars.

In the second part,
Nauomov obtained objective prism spectra near the Galactic mid-plane
using narrow-band interference filters to enable classification
of stars as faint as $V \approx 11$ mag. Spectral indices enabled
him to distinguish dwarfs from giants, and to determine effective
temperatures. Again, there were no selection effects that
might bias the sample in favor of thin disk or thick disk
stars on the bases of either kinematics or chemistry.
Short-exposure direct imaging CCD observations taken
with the same telescope produced $V$ magnitudes and $V-I_{C}$ color
indices. Nauomov compared the temperature estimates from the
reddening-independent line indices and the observed $V-I_{C}$ colors
to estimate the reddening for each star.
This second sample was likewise trimmed so that only long-lived
stars remained.

The resultant sample, comprising roughly 1200 stars near the
Sun or at least near the Galactic plane, was then studied using the
Center for Astrophysics radial velocity speedometers. From
these spectra, Nauomov was able to determine radial velocities with
individual measurement precisions of better than 1 \kms. Further,
with confidence that all the program stars are dwarfs, and with
knowledge of the temperatures, Nauomov could determine mean
metallicities following the procedures described by Carney et al.\ (1987, 1994)
and Laird et al.\ (1988). Figure~\ref{fig:lsrv} shows the results
for the stars selected from the Olsen (1993) catalog, as well as
the stars selected from the objective prism observations with $\ell \approx 90$
and $b \approx 0$, so that the radial velocity is a direct measure
of the V velocity. The Figure merits careful consideration. As expected,
one sees the thin disk with $<$V$>$ $\approx$ 0~\kms, and dispersion
of order 20 \kms. The thin disk range in [Fe/H] is approximately
from $-1$ to slightly above solar. Note that there are very few stars
in the Figure with V velocities as high as +50 \kms, 
but a considerable number
with V $\approx -50$ \kms. That excess, we believe, represents
the thick disk. 

In Figure~\ref{fig:vhistfem03m10} we show the V velocity histogram for
stars with $-1.0 \leq$\ [Fe/H\ $\leq\ -0.3$. The thick disk
signature is distinct, with a peak, as expected, near $-50$ \kms, and it
is clear that the thick disk is a minor component near mid-plane,
consistent with the results of Chen et al.\ (2001).

From Figure~\ref{fig:lsrv} we note also that the thin disk
and the thick disk stars do not appear to overlap in the V vs.\ [Fe/H] plane
at any location. The thick disk stars appears to maintain a characteristic
signature V velocity at {\em all} values of [Fe/H]. This 
supports an independent origin, presumably later
involving accetion, for the origin of the thick disk, and that it is
not an antecedent to the thin disk. For the purposes of this paper,
Figure~\ref{fig:lsrv} also enables us to select our program stars. 
We chose to limit our subsequent studies of stars to those
with $uvby$ photometry.

We re-calculated U, V, and W velocities for the 269 stars in Naumov's (1999)
sample based on parallaxes and proper motions from the Hipparcos Catalogue
(Perryman et al.\ 1997) and his and new radial velocities from 
Latham et al.\ (2002 and private communication).  
Since the most distinctive kinematic signature of the thick disk is its
asymmetric drift, we used V velocities as a primary selection criterion.  
We employed a ``Toomre diagram" that plots the V velocity against
the combined U and W velocities, as shown in 
Figure~\ref{fig:toomre}. The distance from the origin in
such a diagram is a measure of a star's energy of motion, itself also
an approximately conserved quanitity. Stars were also chosen on
the basis of scientific goals, in the sense that we wished to have
comparable numbers of thin disk and thick disk stars, that they span
similar ranges of metallicity, and that all stars have similar effective
temperatures to minimize any potential systematic effects in our analyses.

We assigned nine of our program
stars to the thin disk and fourteen stars to the thick disk. 
Open circles in Figure~\ref{fig:toomre} depict stars we classifed
as belonging to the thin disk, while filled circles represent the
thick disk candidates. Subsequent
to this, well after the observing and analyses had been completed, we
attempted a more quantitative population membership assignment.
We followed the prescriptions of Venn et al.\ (2004) to assign
such membership probabilities, which are given in Table~\ref{tab:stars}.
All nine stars assigned by our selection criteria to the thin disk
appear to be members, with probabilities exceeding 75\% in all cases.
For the thick disk assignments, membership assignments are less clear
for several stars, including BD+65~3 (only 51\%) and BD+13~1655 (only 57\%).
We retain these stars as thick disk candidates throughout this work, however.

We present the basic data for our program stars in Tables~\ref{tab:radvels}, 
\ref{tab:photometry}, and \ref{tab:stars}. Table~\ref{tab:radvels} summarizes
the radial velocity data obtained using the CfA speedometers. The
summary includes the number of observations, N, the span (in days), the
mean radial velocity and its associated error of the mean, the ``external"
error, E, derived from the individual velocities, the ``internal" error, I,
estimated from the ratio of the height of the peak in the power spectrum
to the mean noise level, and the ratio E/I. Values above 1.5 generally
indicate velocity variability. A better indicator is $P$($\chi^{2}$),
the probability that the resultant value of $\chi^{2}$ could have arisen
from the internal errors. A sample of constant-velocity stars should show
a uniform distribution of $P$($\chi^{2}$). As the Table reveals, five of
our program stars 
have very small $P$($\chi^{2}$) values, and
are single-lined spectroscopic binaries, a more-or-less
typical fraction seen among field stars, independent of metallicity
(see Latham et al.\ 2002). Table~\ref{tab:photometry} summarizes the available
photometry for our program stars, including $JHK$ photometry from the
2MASS\footnote{The
Two Micron All-Sky Survey is a joint project of the
University of Massachusetts and the Infrared Processing and
Analysis Center/California Institute of Technology, funded by
the National Aeronautics and Space Administration and the
National Science Foundation.} survey.
In Table~\ref{tab:stars} we summarize the metallicity
and temperature estimates obtained from the photometry, and our
calculated Galactic UVW velocities.

Because the stars in this study are more metal rich
than 47~Tuc, we compared our stars' temperatures
and metallicities with the model isochrones of
Kim et al.\ (2002) to ensure that these
stars are unevolved and have life expectancies of at 
least 10 Gyr. Temperature estimates from both Str\"{o}mgren photometry 
and $V-K$ colors using 2MASS data were employed to calculate
the stellar effective temperatures, again relying on the color-temperature
relations of Alonso et al.\ (1996).

\subsection{Observing Runs \& Data Reductions}

We obtained high resolution echelle spectra for each star. The majority of
the data were taken with the echelle spectrograph on the 4m Mayall
telescope at Kitt Peak
\footnote{Observations reported here were made with the
Mayall 4-meter telescope of the Kitt Peak National Observatory,
which is operated by the National Optical Astronomy Observatory
under contract with the National Science Foundation.}.
To obtain the necessary large wavelength range,
each star was observed with two different set-ups, one for the red end of
the spectrum and one for the blue end.  Both set-ups included the 31.6
grooves mm$^{-1}$ echelle grating, a 2048 x 2048 CCD chip with 24 $\mu$m
pixels, and the T2KB long camera.  The red set-up used the 226-1 cross
disperser and the long red camera.  The blue set-up used the
226-2 cross disperser in the second order, the long blue camera,
and was binned by 2 in the spatial direction.  The stellar spectra were
taken with a 1\arcsec\ slit and a 
2.9\arcsec\ decker, while the flats were taken with
a 4\arcsec\ decker.  Both setups gave continuous coverage throughout the
wavelength range with a resolving power of
R = $\lambda/\Delta\lambda$ = 32,000.

Higher resolution (R = 60,000) spectra were also acquired by 
Inese Ivans using the 2.7m Harlan J. Smith telescope at 
McDonald Observatory using the ``2d-coud\'e" echelle.  The spectrograph 
was set in the cs23-e2 configuration and employed a Tektronix 2048 x 2048 chip 
with 24 $\mu$m pixels and a 1\arcsec\ slit.  The wavelength coverage is 
continuous at the blue end, with increasing inter-order gaps starting 
at $\approx$~5700~\AA.
Table~\ref{tab:observations} gives the dates and wavelength 
coverage for each observing run.  
Table~\ref{tab:spectra} gives the observing dates, 
total exposure time in minutes, 
and S/N per resolution element for each stellar spectrum. These
are cited at 4130 \AA\ and 6645 \AA, which are near key lines of the
$r$-process element europium. These are not at the centers of their
orders' blaze, and so are representative of a typical S/N level, and
perhaps an underestimate in the blue region.
Each spectrum has continuous wavelength 
coverage from 3900-7400~\AA. 

Standard echelle spectra reduction was undertaken using tools 
running inside the IRAF\footnote{IRAF
(Image Reduction and Analysis Facility) is distributed by the National
Optical Astronomy Observatories, which are operated by the Association
of Universities for Research in Astronomy, Inc., under contract with
the National Science Foundation.} environment (Tody 1986). 
Bad pixels were interpolated over and the overscan region was trimmed. 
A set of projector flats was taken each night, averaged, and 
applied to the spectra.  A fit to the overscan region was used 
for bias subtraction.  Scattered light was removed by subtraction 
of a low-order surface fit to the inter-order light.  The spectra 
were extracted, continuum flattened, and normalized.  
Thorium-argon lamp comparison spectra were used for wavelength calibration of the data.
Each run also included spectra of hot, rapidly rotating stars
to aid in the removal of telluric lines. Each such spectrum 
used in the analysis of any program star was taken
at a comparable airmass, and always had S/N
levels of over 500 per resolution element. We took care to 
measure equivalent widths 
before and after the division to be certain that this
calibration step did not itself introduce errors into our results.

\section{ANALYSES} 

\subsection{Spectroscopic Atmospheric Parameters}

The Fe~I and Fe~II lines were used to set the atmospheric parameters:
effective temperate ($T_{\rm eff}$), surface gravity (log~g), and
microturbulent velocity ($\xi$). Our list of
iron lines comes from Prochaska et al.\ (2000).
In general, their 
spectra were of higher resolution than ours, and several of the lines 
they employed were blended in our spectra and had to be discarded.
Lines were used only if they were measurable
in at least two of our program stars.   Equivalent widths for each 
line were measured in the SPLOT package of IRAF,
using a Gaussian profile for weak lines ($\le$ 40~m\AA) or 
a Voigt profile for stronger lines.   Lines stronger than 
150~m\AA~were not employed in our analyses, since they generally form
at such shallow depths that we have greater concerns about
the validity of the LTE model atmosphere $T-\tau$ relations.

ATLAS9 (Kurucz 1993) was used to
produce plane-parallel, atomic and molecular line-blanketed,
flux-conserving stellar models.
Local thermodynamic equilibrium (LTE) was assumed throughout the stellar
atmosphere. 
WIDTH9 was used, along with the Kurucz model atmospheres, to
derive an elemental abundance from each line. At the
temperatures of our program stars, the equivalent widths of weak Fe~I
lines are relatively unaffected by changes in log~g and $\xi$, 
so they were used to
set $T_{\rm eff}$. Using only Fe~I lines weaker 
than about 40~m\AA, the slope
of the derived Fe~I abundance versus excitation
potential ($\chi$) was determined. If the slope deviated
from zero by more than 50\% of the uncertainty of the slope,
a new model atmosphere was computed with a different temperature,
and the process was repeated until the slope was, effectively, zero.
The typical uncertainty in the slope for 
our stars translates to $\sigma$($T_{\rm eff}$) $\approx$ 60~K.
With the temperature set,
$\xi$ was found using all Fe~I lines and repeating the analyses
until the slope of abundance
versus W$_\lambda$/$\lambda$ was zero to well within
the uncertainty, with 
$\sigma$($\xi$) $\approx\ 0.1$ \kms.

The surface gravity was then found by requiring that the 
abundances from Fe~I lines and
Fe~II lines agree. Again, if the results disagreed, a new
model atmosphere was computed, and the iterations continued
until the two species 
gave an iron abundance agreement within 0.01~dex.  Based on the uncertainties
in the Fe~II abundances, we estimate the typical internal
uncertainty in log~g (excluding systematic effects such
as the $gf$ values) to be 0.05~dex. A case has been made for
systematic effects due to the failure of the LTE approximation
(Th\'{e}venin \& Idiart 1999), but the effects are likely to be
very small for metal-rich dwarfs (Korn, Shi, \& Gehren 2003).
With these values of $T_{\rm eff}$, $\xi$, and log~g, a final model 
atmosphere was
produced with ATLAS9, and this model atmosphere for
each star was then used in all subsequent
abundance analyses.  The final atmospheric parameters 
for each star are listed in the
Table~\ref{tab:parameters}, which gives the 
$T_{\rm eff}$, metallicity, $\xi$, and 
log~g of the model atmosphere and the derived Fe~I and Fe~II 
abundances, the number of absorption lines used for each, and 
the standard deviation of the abundances from individual 
absorption lines from the mean abundance. 

Figures~\ref{fig:tby} and \ref{fig:vkt} compare the effective 
temperatures derived spectroscopically to those from the photometry. 
On average, the spectroscopic temperatures are 48~K higher 
($\sigma$ = 47~K) than those 
from estimated using Str\"{o}mgren photometry and 71~K
($\sigma$ = 57~K) higher than those obtained using $V-K$. We have used
the relations derived by Schuster \& Nissen (1989) to estimate
[Fe/H] photometrically, and
Figure~\ref{fig:fecomp} compares the the results with our
spectroscopic values. Agreement is remarkably good, with  
no significant offset with and a remarkably small
scatter of $\sigma$ = 0.11~dex.

\subsection{Elemental Abundances}

The abundances for 23 other elements were found using the adopted
Kurucz model atmospheres and measured equivalent widths.
As in the case of the iron lines, the equivalent widths were
measured using SPLOT in IRAF employing a Gaussian or Voigt profile, 
depending on the strength of the line.
Atomic parameters were taken primarily from the sources cited by
Prochaska et al.\ (2000), with $gf$ values for several of the 
neutron-capture elements taken from Cowan et al.\ (2002). 
The wavelengths, excitation potentials, and $gf$ values of the 
lines are given in Table~\ref{tab:ew}\footnote{The full table
is available electronically. Table~\ref{tab:ew} contains
only an illustration of its contents.} along with the 
measured equivalent widths for each star.
We adopted the photospheric solar abundances of Grevesse \& Sauval (1998) with the 
exception of iron, for which the meteoritic value of log $\epsilon$ = 7.51 
(Anders \& Grevesse 1989) was chosen in order to match that adopted by
Edvardsson et al.\ (1993).
This value was also 
derived directly from our own photospheric abundance
analysis of a solar spectrum using the same lines and $gf$
values employed for our program stars (see Section 6.3 and
Table~\ref{tab:solar}. It is only 0.01 dex larger
than that obtained by Grevesse \& Sauval (1998).

\subsubsection{Hyperfine and Isotopic Splitting}

Isotopes with an odd number of protons or neutrons exhibit hyperfine
splitting of their spectral lines.  Hyperfine
interactions split the lines
into multiple components with typical separations of a few m\AA.  The
splitting of the lines inhibits saturation of
strong lines and must be taken into account to accurately
derive the abundances from these lines, since, in
general, abundances are overestimated if hyperfine splitting is ignored.  
Elements that have more than one stable isotope also exhibit isotopic
splitting of their lines, and each isotope may 
have a slightly different set of
hyperfine splitting.  For our analyses, we adopted solar
isotopic ratios 
from meteoritic values (Anders \& Grevesse
1989).  
The wavelength of the hyperfine transitions and the relative 
$gf$ strengths were taken from Prochaska et al.\ (2000) for 
V, Co, Mn, Sc, and Cu, McWilliam (1998) for Ba, Kurucz (1995) 
for Eu, and Aoki et al.\ (2001) for La.

The software package MOOG (Sneden
1973) was used, in the synthesis mode $blends$, to derive abundances for
these elements.  

\subsubsection{Synthetic Spectra}

Three of the Eu lines and three of the La lines are in
portions of the spectrum where line blending makes it
impossible or at least unwise to measure simple equivalent
widths. For these lines, abundances were
found by comparison with synthetic spectra.  MOOG was used to
create a synthetic spectra of about 10~\AA~surrounding the line.  The
atomic and molecular line lists come from Kurucz (1995).  Hyperfine and
isotopic splitting were taken into account.  The stellar spectrum was
compared to the synthetic spectrum, and the abundance of the element was
changed until the best fit was found\footnote{The referee has commented
on the relatively poor quality of the fit for the $\lambda$4123 line
of europium. We did not change the $gf$ values of the adjacent
lines, which would have produced a better overall fit, but the europium line
itself is a good match.}.  
Figures~\ref{fig:syneu} and \ref{fig:synla} 
show examples of the synthetic spectra fit 
for one of our program stars using three different abundances.  
The (solid) middle line is the best fit to the stellar spectrum with the other lines 
representing changes of $\pm 0.1$~dex in the [La/H] and [Eu/H] abundances,
based on $\chi^{2}$ goodness-of-fit measures supplied by MOOG.

\subsubsection{Damping} 

Atomic line broadening by van der Waals damping 
was employed in our abundance analyses.
A correction factor of 3 was multiplied to the Uns\"{o}ld (1955)
approximation for the
damping constant for Ba as suggested by Holweger \& Muller (1974) 
because the fit to the line with synthetic spectra is better than 
with the uncorrected damping factor.  For all other elements 
the no multiplicative factor was included.

\subsection{Error Analysis}

There are several factors that affect the accuracy of our elemental 
abundances.  Uncertainties in $gf$ values produce random errors
in the results, but possible
systematic effects as well. Laboratory values may have
systematic effects that depend on the excitation potentials of the
lower energy levels, for example. However, such
should not alter the relative abundances of our own program stars, 
so the comparisons between our thin and thick disk program stars should be 
independent of these uncertainties.  The uncertainties in the 
atmospheric parameters and the equivalent width measurements affect 
the internal uncertainty of our elemental abundances and are a 
measure of the ability to compare our thin and thick disk stars. 
[Fe/H] is most strongly affected while element-to-iron ratios
are less sensitive, especially for elements whose ionization
potentials are similar to that for iron (7.90~eV).
We illustrate the effects of these uncertainties using
BD+34~927, a star in the middle
of the sample in both $T_{\rm eff}$ and metallicity. We calculated elemental
abundances using eight different atmospheric models.  
The temperature was varied by
$\pm$80~K, $\xi$ by $\pm$0.1~\kms, log g by $\pm$ 0.05, and [Fe/H] by
$\pm0.1$~dex. Table~\ref{tab:errors} summarizes the results.

Our abundance analyses, like essentially all of those discussed in
Section~4, are relative to solar values, which, as noted above,
were based on those derived from solar photospheric measures of
Grevesse \& Sauval (1998). In Table~\ref{tab:solar} we summarize
the results we obtained using our adopted $gf$ values, and
a solar model calculated using ATLAS9 with $T_{\rm eff}$ = 5770~K,
log~g = 4.44, and $\xi = 1.15$ \kms. Columns 2 and 3 list
the results from Grevesse \& Sauval, in units of log~$\epsilon_{\odot}$(X) =
log~$N_{\rm X}$ $-$ log~$N_{\rm H}$ + 12.00, and our derived values
of [X/H], relative to those values. Column~4 lists the individual
scatter {\em per line}, and column~5 provides the number of lines employed.
The solar spectrum was a lunar spectrum, obtained with the same
instrumentation as used for our program stars, and reduced in an
identical fashion. The agreement is excellent, except perhaps for
vanadium and manganese. This result was expected since we have
largely adopted $gf$ values from Prochaska et al.\ (2000), and they
found similar underabundances for these elements. This suggests
systematic errors may exist for our adopted $gf$ values for
these elements, or the manner in which we have corrected
for hyperfine splitting, or both. Fortunately, neither V nor Mn
are critical for the comparisons of the thin disk and thick
disk stellar populations.

The final elemental abundances for each star  are listed in 
Tables~\ref{tab:thindisk1} and \ref{tab:thickdisk3}.  We
include the standard error, per line, $\sigma$, and the number
of lines used, $N$. For plotting purposes we rely on the
mean error, defined as $\sigma/\sqrt{N}$. However,
uncertainty propagates into the determination
of $\sigma$ as well, especially when the number of
lines employed is small, and for plotting purposes, we have 
resorted to an alternative method to estimate $\sigma$.

The random uncertainty in the abundance obtained using a single line 
may be estimated from
from the uncertainty in the equivalent width measurement, following
Cayrel (1989).
The uncertainty in a measured equivalent width is estimated to be
\begin{equation}
\sigma(W_{\lambda})\approx \frac{\sqrt{FWHM \cdot \Delta x}}{S/N}
\end{equation} 
where $\Delta x$ is the dispersion in \AA\ pixels$^{-1}$ 
and FWHM is the width of the spectral line in \AA.  
For our spectra, $\Delta x \approx 0.05$ and FWHM ranges from 
0.15 to 0.20 depending upon the strength of the line.  
Using S/N = 75 for the blue spectra and S/N = 150 for the red 
spectra, we find $\sigma(W_{\lambda})$ = 2 m\AA\ and 1 m\AA\ respectively.  
Most of our line measurements are from the red spectra, 
but $\sigma(W_{\lambda})$ does not take into account the difficulties 
in continuum placement, so we use an uncertainty in our equivalent 
width of 2 m\AA\, which corresponds to an estimated abundance uncertainty of 
$\approx$ 0.05~dex. 

In the Figures we have therefore employed the larger
of 0.05/$\sqrt{N}$, following the discussion above, if N $<$ 5, or the
error of the mean of the results from the individual lines
if N $\geq$ 5. All
elements whose abundances are based on a single line therefore
have are therefore plotted with an error bar of 0.05 dex. This
obviously neglects uncertainties in the $gf$ values themselves,
which are hard to quantify.

\section{RESULTS}

The element-to-iron ratio of each
element is plotted as a function of [Fe/H]
in Figures~\ref{fig:alpha} - \ref{fig:rpros}, with
thick disk stars identified by filled squares and thin disk
stars by open squares.  The error bars for each point are the 
internal uncertainties based on the scatter in the abundance values from 
the individual absorption lines.  The abundance uncertainties due to 
uncertainties in atmospheric parameters are shown
in the upper right corner of each figure. 

\subsection{Alpha-elements} 

Oxygen abundances for 13 stars were found from the 7774 \AA~triplet.  The 
other 10 stars did not have spectral coverage in this region.
Abundances from the 7774 \AA~triplet are very sensitive to effective 
temperature and non-LTE effects. (See Kiselman 1993 for a discussion
of 1-d calculations and Kiselman \& Nordlund 1995 and Asplund 2004
for calculations involving 3-d models.)
Because our stars are fairly cool and cover a small range in 
atmospheric parameters, we assume that non-LTE effects would
affect all of our program stars to comparable levels. Thus
internal comparisons between the thin disk and thick disk stars
in our sample should be robust, and any derived differences
between the two populations should be real, even if the absolute
oxygen abundances are less reliable.
Figure~\ref{fig:alpha} shows the [O/Fe] abundances for our stars and a 
clear difference in the thin and thick disk patterns.  The thin 
disk oxygen abundances are on a plateau,
[O/Fe] $\approx$ +0.2, or perhaps rising slowly with declining [Fe/H],
for [Fe/H] $< -0.2$. At higher [Fe/H] values,
the thin disk [O/Fe] descends to solar abundances.
The thick disk abundances, 
on the other hand, are much higher, reaching
[O/Fe] values of $\approx$ +0.6 for [Fe/H] $< -0.5$, and at
higher [Fe/H], [O/Fe] declines toward solar values. At
[Fe/H] =$-$0.4 there is about a 0.2~dex difference 
in the thick and thin disk abundances.  

The abundances of Mg, Si, Ca, and Ti in Figure~\ref{fig:alpha} 
all show the same pattern.  
For all the $\alpha$-elements, the thick disk stars with 
[Fe/H] $<-$0.4 show enhancements over the thin disk stars of 
the same metallicity. In the range $-$0.4 $<$ [Fe/H] $< -0.2$, 
the [$\alpha$/Fe] abundances in the thick disk stars decline steeply
with increasing metallicity.  While this decline is most apparent 
in the Mg and Ti abundances, it is also present in Ca and Si.  
The decline in the $\alpha$-elements, including oxygen, 
is thought to signify the incorporation of the ejecta
from SNe~Ia into the interstellar medium of
the thick disk. These results therefore suggest that
the star formation rate in the thick disk was much higher
than that of the thin disk.

Note that BD+4~2696, at [Fe/H] = $-0.24$, does not follow the trend 
of the other thick disk stars, and that its abundance pattern is similar to the 
thin disk stars. We assume that our assignment of this star
to the thick disk population, based on its kinematics, were
inappropriate and that it represents a thin disk star whose kinematics
put it in the high velocity tails of the thin disk velocity
distributions. We have, nonetheless, shown the star as a thick disk
star in Figures 10-16 and 18-20. For clarity, we have omitted the
star from Figure 17. The star has [$\alpha$/Fe] = +0.05.

\subsection{Al \& Na}

The odd-Z, light elements Al and Na are thought to be produced primarily
in SNe~II (Woosley \& Weaver 1995) and, therefore, they should be
enhanced in stars born during the earliest stages of star
formation, presumably the stars with the lowest [Fe/H] values.
Figure~\ref{fig:light} shows that the behavior seen in the $\alpha$ elements
is repeated in the case of aluminum, but not in sodium,
as might have been expected. 
The thick disk stars show a plateau 
(at [Al/Fe] $\approx +0.3$) for [Fe/H] $< -0.3$,
or, possibly a slow rise in [Al/Fe] as [Fe/H] decreases.
The thin disk abundances remain near
solar.  The Al abundance of BD+4~2696 falls below the rest of the thick
disk stars, just as it does for the $\alpha$-elements.

Unlike the $\alpha$-elements, Na abundances, shown in 
Figure~\ref{fig:light}, remain near solar over the entire metallicity 
range for both the thick and thin
disk stars.
Our Na abundances show no trend with metallicity for either 
the thick or thin disk, though the average thick disk abundance is 
slightly higher (0.04~dex) than that of the thin disk stars. This is
quite surprising, and we have no explanation for the dissimilarity
in the behavior of Na from Al.

\subsection{Iron Peak Elements}

The even-Z iron peak elements Ni and Cr follow Fe in both the thick 
and thin disk stars as can be seen in Figure~\ref{fig:iron}.  
Unfortunately, the the abundances derived from ionized chromium lines 
are significantly higher than those obtained using neutral lines (by an 
average of 0.11~dex). Since significantly more neutral lines were measured,
we have relied solely on [Cr~I/Fe~I] values to obtain the [Cr/Fe]
abundance ratios\footnote{Cr~I has an ionization potential of 6.8 eV.}.
We do not understand the source of the disagreements between Cr~I and
Cr~II abundances, but our first guess would be that the $gf$ values
need to be re-determined.

Enhancements in [Sc/Fe] have 
been seen in metal-poor stars (Zhao \& Magain
1990), and Nissen et al.\ (2000) suggested that Sc is produced in the same
site as the $\alpha$-elements.  Our results are consistent
with this suggestion, in that [Sc/Fe] appears to behave 
similarly to the $\alpha$ elements, with higher abundances for the
thick disk stars compared to the thin disk. 
Bensby et al.\ (2003) had noted the similarity in the behavior of [Zn/Fe]
with the $\alpha$ elements, and we confirm that behavior, along
with that of [Co/Fe] (Figure~\ref{fig:iron2}).
For all three elements, BD+4~2696 continues to
behave chemically like a thin disk star. We are unaware of any
explanation as to why cobalt and zinc should show such behavior.
Vanadium may also show this behavior but with a reduced amplitude,
and the scatter in Figure~\ref{fig:iron} prohibits any firm
conclusions.

The supernovae yields of Mn seem to be
metallicity-dependent as studies of metal poor stars reveal
underabundances of Mn at low metallicities (Gratton 1989, Reddy et al.\
2003, Prochaska et al.\ 2000).  We observe [Mn/Fe] to decrease with
decreasing metallicity in both our thin and thick disk samples with no
distinction between the two populations.

[Cu/Fe] in the thick
disk is enhanced mildly for 
[Fe/H] $<-0.2$ at which point there appears to a 
small ($\approx$ 0.1 dex) step-like decrease to 
thin disk abundances.  The thin disk stars all appear to have
[Cu/Fe] abundance ratios at or slightly below the solar value.

\subsection{Heavy Elements}

The light neutron-capture elements Sr, Y, and Zr are thought to be significantly
produced in the $weak$ $s$-process, which 
has been proposed to occur in advanced
evolutionary phases of massive (M $>$ 10 M$_\odot$) stars (Lamb et al.\ 1977).
Because these elements are produced in massive stars, they should be
overabundant at early epochs,
though because the $weak$ $s$-process 
elements' production has been suggested to depend
on metallicity (Raiteri et
al.\ 1992), the overabundance may not be as great as that of the
$r$-process elements.

Because the stars in this study are metal-rich, several of the most
commonly used Sr II lines are saturated and heavily blended with Fe lines,
making them unusable.  This leaves one clean Sr I line and one fairly
clean Sr II line.  The Sr I lines gives consistently lower abundances 
(by $\approx$ 0.25~dex), than the Sr~II line. Indeed, this is seen
in our derived solar abundances (Table~\ref{tab:solar}).
Reddy et al.\ (2003) found similar behavior in that 
the Sr~I line yielded [Sr/Fe] 0.35~dex too low in their 
solar analysis. Therefore, we cite results in Tables~\ref{tab:thindisk1}
and \ref{tab:thickdisk3} and show abundances using only the Sr~II line in 
Figure~\ref{fig:spros}, though this line could not be measured in a 
few of the most metal-rich stars because of blending.   

Figure~\ref{fig:spros} shows no difference between the thick disk 
and thin disk abundances of Sr, Y, and Zr.  
The thick disk star BD+47~2491, at [Fe/H] = $-0.51$, shows a 
significant enhancement in all $s$-process elements we measured. While
this behavior could have been caused by mass transfer from a highly-evolved AGB
companion, no sign of orbital motion is seen in our
radial velocity monitoring (Table~\ref{tab:radvels}). 

Ba, La, and Ce are
thought to be formed during the
$main$ $s$-process.  The $main$ $s$-process accounts for the majority of the
$s$-process abundances in elements with Z $\ge$ 56, and is thought
to be occur primarily in low-mass
(M $<$ 4 M$_\odot$) AGB stars during He-shell burning (Busso, Gallino, \&
Wasserburg 1999). In solar system material, Burris et al.\ (2000)
estimated the $s$-process contributions to be 85\%, 75\%, and
81\% for these three elements, respectively.
Ba can be difficult to measure since its lines are often very
strong and exhibit both hyperfine and isotopic splitting.  
Small changes in equivalent widths and microturbulent velocity
may therefore result in differences
in the derived abundances.  Nonetheless, 
we find interesting differences between the thin disk and 
thick disk stars in the abundances of all three of these heavy 
$s$-process elements. Once again, BD+47~2491 is enhanced
in these $s$-process elements. Neglecting that star, it
is the thin disk stars that show enhanced abundance ratios
in [Ba/Fe], [La/Fe], and [Ce/Fe], at least for [Fe/H] $< -0.2$.
Again, it is clear that the chemical enrichment history
of the thin disk has been quite different than that of
the thick disk.
 
Nd is produced by both the $r$- and $s$-process, 
almost 50\% by each process
for solar system material (Burris et al.\ 2000).  
The abundances of this element therefore
represent a mixture of the two processes.  
Figure~\ref{fig:rpros} shows enhanced [Nd/Fe] ratios
at lower [Fe/H] for both populations, but, as in the case
of the other heavy $s$-process elements, Ba, La, and Ce,
[Nd/Fe] is slightly enhanced in the thin disk stars
compared to the thick disk stars for [Fe/H] $< -0.2$,
and, hence, behaves more like an $s$-process element
than one synthesized in the $r$-process.

Though the site of the $r$-process is not completely understood, 
it has been proposed to occur in SNe~II (Wasserburg \& Qian 2000).
As the only $r$-process element we able to measure, 
Eu values play an important role
in understanding the heavy element history of the thick and thin disks.  
(Its solar $r$-process contribution is estimated to be 91\%,
according to Burris et al.\ 2000.)
The [Eu/Fe] results in Figure~\ref{fig:rpros} show a similar 
pattern as the [$\alpha$/Fe] trends seen in Figure~\ref{fig:alpha},
which is consistent with also being
produced in SNe~II.
The thick disk [Eu/Fe] ratio is
$\approx$ +0.3~dex for
[Fe/H]$< -0.2$, at which point there 
appears to be a step-like, or at least very steep, decrease to solar
values.  The thin disk [Eu/Fe] vs.\ [Fe/H] trend is
smoother, declining steadily as [Fe/H] increases. Once again,
BD+4~2696 behaves like a thin disk star in its [Eu/Fe] ratio.

\section{COMPARISON WITH OTHER STUDIES}

For a variety of reasons, we must compare our results
with other disk abundance studies where 
stars from the thick and thin disk populations were chosen 
and analyzed in manners similar to those we have employed.
For example, while others have found that [$\alpha$/Fe] vs.\
[Fe/H] trends differ for thin disk and thick disk stars,
do Sc, Co, and Zn also obey these trends? Further, comparing
with other results can help assess the comparability of our
results, and combined samples have much greater power in
addressing the key question about the origin of the thick disk
population. In merging the samples, we have taken special
care with three critical issues. First, we have relied
on stars whose estimated ages are large, for reasons
discussed earlier. We have also relied on kinematics, 
especially using probabilities of thin disk or thick
disk membership. Finally, we have shifted all iron and
element-to-iron results to the solar abundance scale we have
adopted to reduce that source of systematic differences. We
have not re-analyzed the work of others for possible differences
in the treatment of isotopic or hyperfine splitting.

For the thick disk comparison, we use the abundances of our earlier
work (Prochaska et al.\ 2000). That study used stars selected
in the same manner as in this study: by kinematics 
and life expectancies.  These should be the best comparison for our 
stars as we used the same $gf$ values as did they. We have also
employed the thick disk stars identified and
studied by Edvardsson et al.\ (1993) and by 
Bensby et al.\ (2003, 2004a, 2005).
In all these cases, we relied on two primary criteria.
First, the published estimates for the ages of the stars 
must equal or exceed 8~Gyrs. Second, the probability of
thick disk membership, computed by Venn et al.\ (2004), must
equal or exceed 80\%. We note that
Bensby et al.\ (2004b)
found that those population assignments resulted in a very
large mean age, although there might be a trend to somewhat
lower ages at the higher metallicities. 

To be considered as part of
the thin disk, membership probability must equal or
exceed 80\%. We employed stars from the studies by
Edvardsson et al.\ (1993), Bensby et al.\ (2003, 2004a, 2005), 
and Reddy et al.\ (2003). We believe that the best comparison
between the thin and thick disk populations is with comparably
old stars.
We employed a slightly milder
age criterion in order to provide a comparable sample of
stars. Specifically, we included only those stars whose
ages estimated by Edvardsson et al.\ (1993),
Bensby et al.\ (2003), or Reddy et al.\ (2003) are at least
7 Gyrs. For the few stars without age estimates, we followed
our own practice and retained stars only if their life expectancies
exceed 10~Gyrs, following the same procedures described in Section~5.1.

We have excluded three ``interlopers" from the comparisons. HD~78747 has a high
probability of being a thin disk star according to Venn et al.\ (2004)
but chemically it behaves like a thick disk star, with [$<$Mg+Si+Ca+Ti$>$/Fe] = 
+0.26 at [Fe/H] = $-0.64$. We have already drawn attention 
to BD+4~2696 in our
own sample (see Section 7.1), and have excluded it. HD~4597 from 
Bensby et al.\ (2003) was assigned
to the thick disk by Venn et al.\ (2004) yet shows thin disk
chemical abundances ([$<$Mg+Si+Ca+Ti$>$/Fe] = +0.06 at [Fe/H] = $-0.34$.
Given our probability selection limit for population assignments
was $\geq 80$\%, we are not surprised by a few interlopers.
criterion, but again relied upon the population assignments
from Venn et al.\ (2004). 

We have supplemented the results for a subset of the
stars studied by Edvardsson et al.\ (1993)
with the Ba and Eu abundances obtained by
Koch \& Edvardsson (2002).

We have also employed results for the $s$-process and
$r$-process elements from Mashonkina \& Gehren (2000, 2001),
following their population designations. These population
assignments may differ in detail from those made by
Venn et al.\ (2004).

\subsection{Alpha Elements}

We have not made direct comparisons between our results 
for [O/Fe] and those
from other workers for reasons described earlier. Not only are
non-LTE effects potentially important, but some workers do not
include them (such as this paper), and others do attempt such
corrections. Even so, the corrections are not always handled
in the same manner, and it may be useful at a later time to
merge all the data and undertake a comprehensive re-analysis
of the data.

For Mg, Si, Ca, and Ti, careful inspection of 
Figure~\ref{fig:othersalphafe} shows good
agreement between our results and those from Prochaska et al.\ (2000),
as we would have expected given the similarity of the analytical
methods and basic data, as well as those from the other studies.
Of greater importance, however, is the confirmation of our finding
that the $\alpha$-elements indicate a rather different nucleosynthesis
history for the thick disk than for the thin disk. In general
the $\alpha$-elements  maintain a high value for the thick disk stars for
[Fe/H] $\leq\ -0.4$, above which they decline toward solar values
rather steeply. The thin disk stars show the same declines, but the
slopes are shallower, and begin at a much lower metallicity.
These trends are represented better in Figure~\ref{fig:mgsicatimeans},
where we have averaged the behaviors of the four
$\alpha$-elements measured in most of the dwarf and subgiant
stars: Mg, Si, Ca, and Ti. 
Please note that here we have excluded BD+4~2696, which, as we
have noted, does not appear to behave chemically like thick disk stars,
despite its high probability of thick disk membership. 
The solid lines represent formal fits to the data. For the thin
disk, the slope was computed using stars with $-0.8 \leq$\ [Fe/H] $\leq\ 0.0$.
For the thick disk, a straight mean was calculated for the stars
with $-1.1 \leq$\ [Fe/H] $\leq\ -0.4$, and a slope was computed
for the stars with $-0.4 \leq$\ [Fe/H] $\leq\ 0.0$.
The two stellar
populations obviously show different trends.  

\subsection{Iron Peak Elements}

We have already noted that some iron peak elements may behave
similarly to the $\alpha$-elements, implying contributions
to their abundances relative to iron from SNe~II. 
In fact, Prochaska et al.\ (2000) had already drawn attention
to the apparently high values of scandium, vanadium, cobalt,
and zinc in most or all of the thick disk stars they analyzed.

In Figure~\ref{fig:scvcozn} we show the behaviors of these four
elements. 
We do not show results for Mn, Cr, Ni, and Cu
since these elements show no statistical 
difference between the thick and thin disk populations.
Our results agree well with those of Prochaska et al.\ (2000)
and, in the case of zinc, with Bensby et al.\ (2003).
For Sc, V, Co, and Zn,
there are significant differences between the thin disk
and thick disk populations. 

\subsection{Heavy Elements}

In Figure~\ref{fig:yzrbalaeu} we show the results for the $s$-process
elements Y, Zr, Ba, and La, and the $r$-process element, Eu. 
We have noted already that while BD+47~2491 behaves like other thick disk stars
in [Eu/Fe], all of its $s$-process elements appear to be
enhanced substantially.

Y and Zr are representative of the
{\em weak} $s$-process, and [Y/Fe] and
[Zr/Fe] are only slightly different between the 
thick and thin disk stars, although there is 
significant scatter among the various studies.

Barium is the most widely studied of the heavy $s$-process elements,
and, like La, is produced in the {\em main} $s$-process.
As seen in Figure~\ref{fig:yzrbalaeu}, the thick disk has lower [Ba/Fe]
ratios than does the thin disk, at least for
[Fe/H] $<-$0.4.  (We use the solar-corrected [Ba/Fe] results
from Prochaska et al.\ 2000 since
they agree much better with other results.)  There is good
agreement between studies for the thick disk abundances, 
and the scatter seen among
the thick disk appears smaller than is seen among the thin 
disk stars.  The lower [Ba/Fe]
ratios seen in the thick disk agree with the idea that Ba production is
from low-mass stars, and these low mass stars began contributing to the 
thick disk at a higher metallicity than did those 
enriching the thin disk.  This is another 
indication of a high rate of star formation in the thick disk population.
Unfortunately, we have not found any other studies of La
in disk stars with which to
compare our data.

The Eu abundances in Figure~\ref{fig:yzrbalaeu} show very good 
agreement between studies. The thick disk stars from 
Mashonkina \& Gehren (2000, 2001), 
Prochaska et al.\ (2000), Bensby et al.\ (2005), and our study all show 
an enhancement of $\approx$0.4~dex for [Fe/H] $<-$0.4.  This
is perhaps not surprising given that the $\alpha$-elements are
thought to be synthesized primarily in SNe~II events, as are
the $r$-process elements.

In Figure~\ref{fig:othersstor} we present these results in
an alternative manner, following oft-used ratios that reflect
the $r$-process element Eu and the $s$-process elements
Y, Ba, and La, and plotted in such a way that
we anticipate high values when SNe~II are
the dominant sources of nucleosynthesis contributions. We note that
La has fewer potential
problems than does Ba in the abundance 
determinations since the La lines are weaker, and
form deeper in the stellar atmospheres, where non-LTE effects
are less likely to introduce systematic errors. 

If the light $s$-process element Y is created in 10~\msun\
stars (the {\em weak} $s$-process), [Eu/Y] should have
a roughly constant ratio, since Eu is likewise thought to
be created in the deaths of (even more) massive stars. While
the scatter is large, this expectation is borne out for the
thick disk stars. The thin disk stars, however, show a decline
from [Eu/Y] $\approx +0.2$ to 0.0 
as [Fe/H] rises from $-0.8$ to $-0.5$. The thick
disk stars maintain a high [Eu/Y] ration until [Fe/H] $\approx\ -0.2$,
at which point there appears to be a sharp decrease to solar
values, reminiscent of the behavior of [Cu/Fe].

Since the {\em main} $s$-process occurs on a much longer timescale than 
the $r$-process, [Eu/Ba] is expected to be enhanced 
early in a population's star formation history.  Figure~\ref{fig:othersstor} 
shows the [Eu/Ba] abundances, with the
pure $r$-process abundance ratio (Burris et al.\ 2002) shown
by a dotted line. The scatter in the [Eu/Ba] abundance ratios is
unfortunately rather large, but it is clear that the thick
disk stars show a larger enhancement of the $r$-process, at
least as represented by europium, than does the thin disk.
Indeed, comparison of Figure~\ref{fig:othersstor} with
Figure~\ref{fig:mgsicatimeans} shows very similar trends.
The thick disk stars show enhancement in [Eu/Fe] of
$\approx$ 0.45~dex for $-0.7 \leq$ [Fe/H] $<-$0.4.  
At [Fe/H] $\approx-$0.4, the [Eu/Ba] values for the
thick disk stars begin to 
decrease, though they are still higher than the thin disk 
values up to [Fe/H] $\approx-$0.2.
Combined with the $\alpha$-element abundances, this suggests that the
star formation timescale is 1 Gyr or longer for the thick disk,
at least in terms of producing stars with with [Fe/H] 
as high as $\approx\ -0.2$, based on
the theoretical time scale for SNe~Ia to begin enriching the ISM
is $\approx$ 1 Gyr (Matteucci \& Recchi 2001). Unfortunately,
we have too few [Eu/La] results from which to draw any
firm conclusions.

The $main$ to $weak$ $s$-process production may be compared
through [Ba/Y] abundances.  If the two elements
were created in AGB stars with significantly different
masses, as has seen suggested, we might see some structure
in the [Ba/Y] vs.\ [Fe/H] plane, for both the thin disk
and thick disk populations. The scatter is unfortunately
too large to enable us to draw any meaningful conclusions.

\section{THE HISTORY OF THE THICK DISK}

\subsection{Arguments in Favor of an Independent Origin for the Thick Disk}

We summarize here what we believe to be the key points
regarding an independent origin for the thick disk and
the implied accretion of its parent galaxy.

\begin{figure}
\epsscale{1}
\figurenum{1}
\plotone{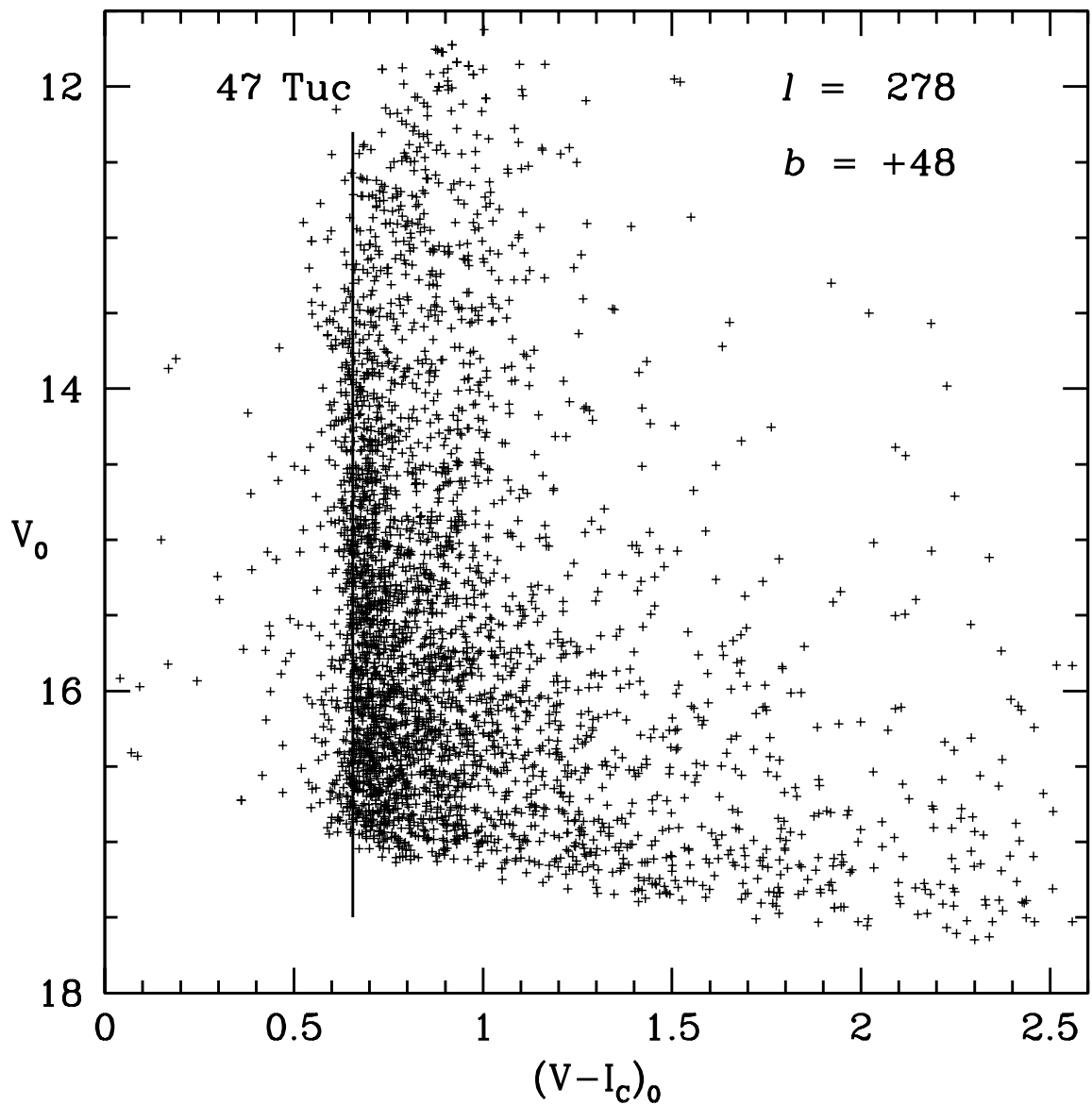}
\caption{$VI_{C}$ photometry of stars covering 7.9 square degrees
roughly centered on $\ell = 278$, $b = +48$, taken with the
Burrell Schmidt at Kitt Peak National Observatory. The photometry
has been corrected for interstellar reddening using the
maps of Schlegel et al.\ (1998). The de-reddened color of the
main sequence turn-off for the thick disk cluster 47~Tuc
has been shown as a vertical line for comparison.
\label{fig:ar0}}
\end{figure}

\begin{figure}
\epsscale{1}
\figurenum{2}
\plotone{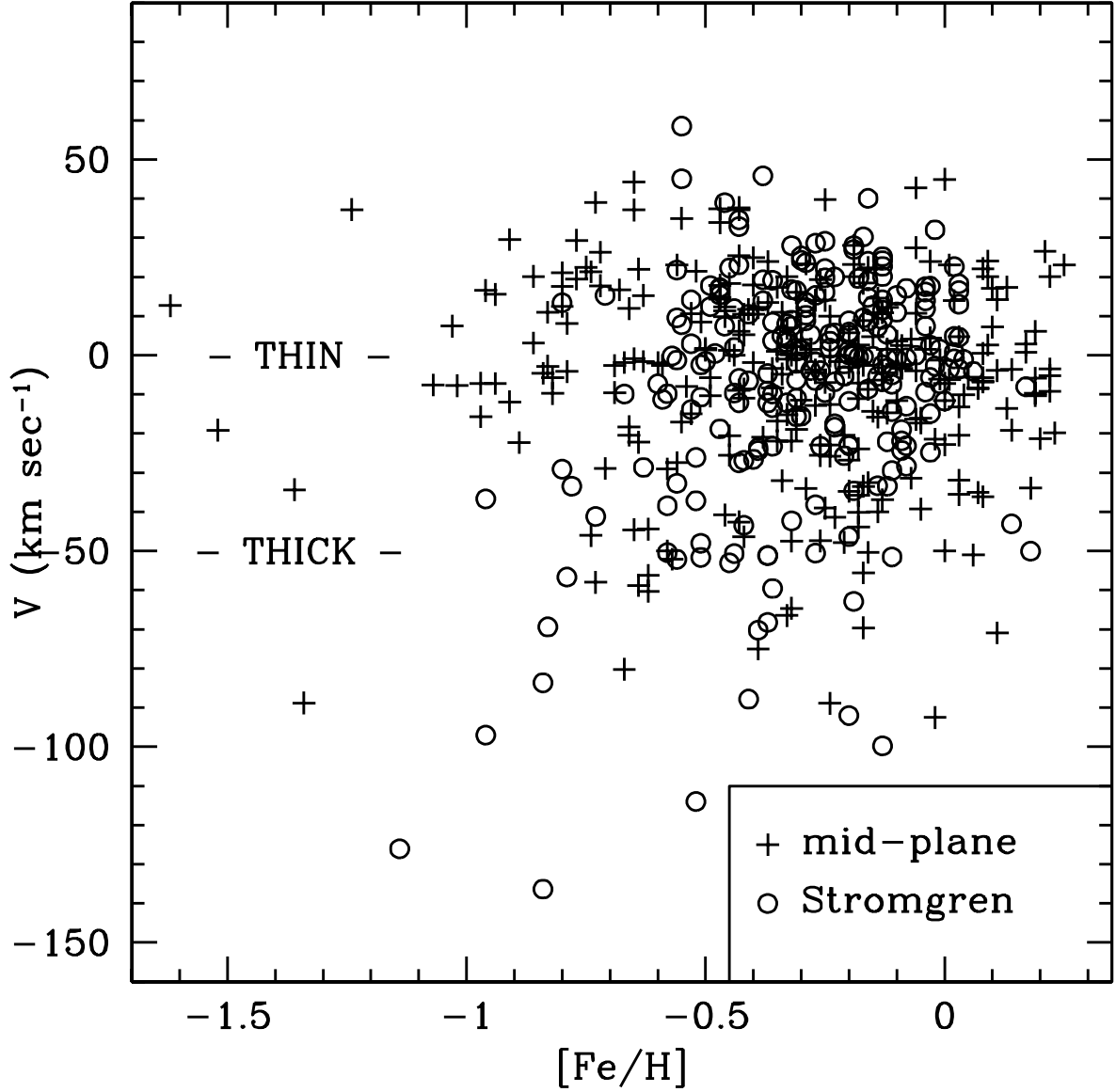}
\caption{Long-lived stars selected by Naoumov (1999) 
from his objective prism spectroscopy and from 
Str\"{o}mgren photometry of nearby field
stars published by Olsen (1993).
\label{fig:lsrv}}
\end{figure}

\begin{figure}
\epsscale{1}
\figurenum{3}
\plotone{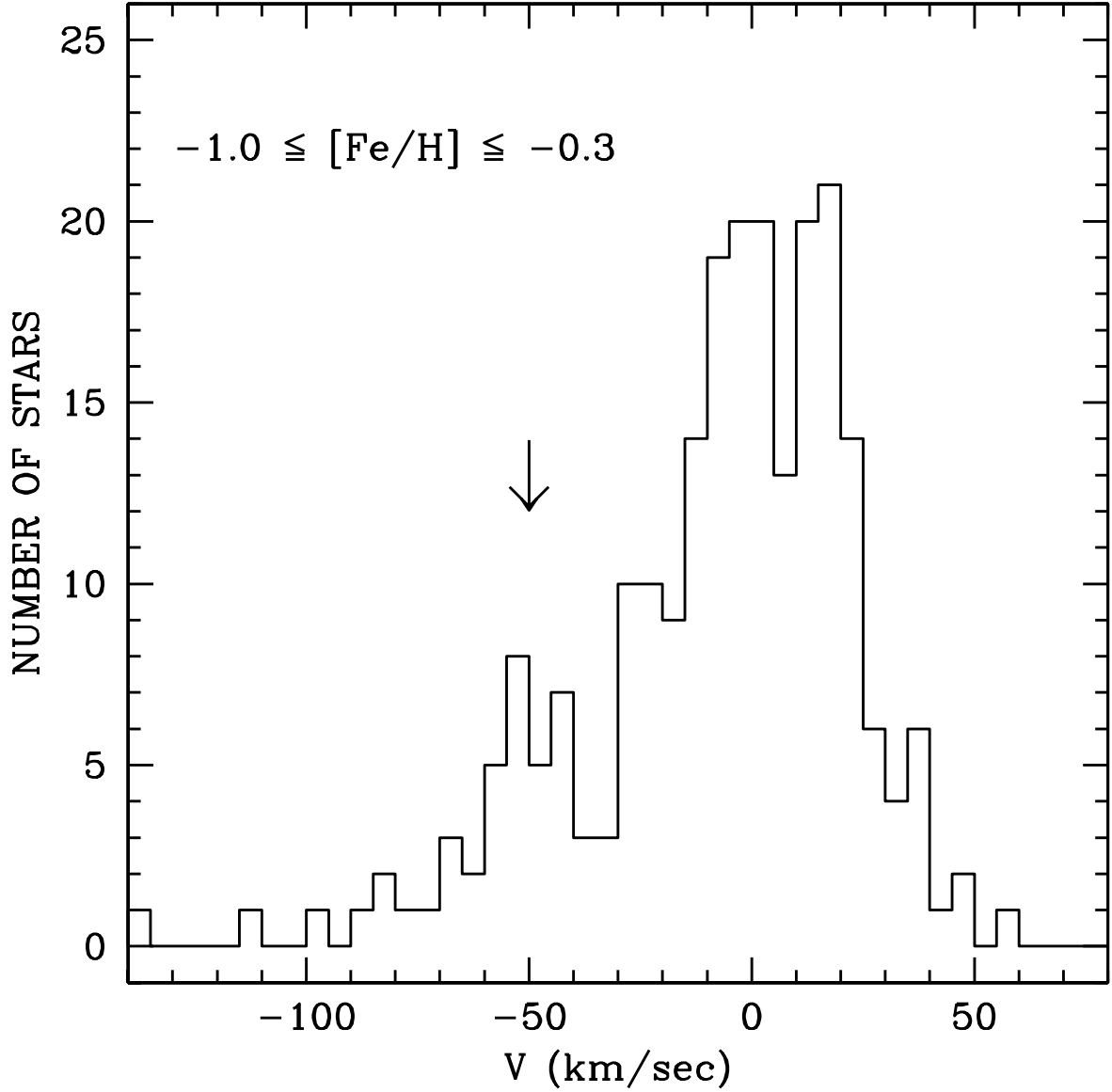}
\caption{The V velocity histogram for stars in
Figure~\ref{fig:lsrv} with $-1.0 \leq$\ [Fe/H\ $\leq\ -0.3$.
The asymmetric drift of the thick disk is indicated by the arrow.
\label{fig:vhistfem03m10}}
\end{figure}

\begin{figure}
\epsscale{1}
\figurenum{4}
\plotone{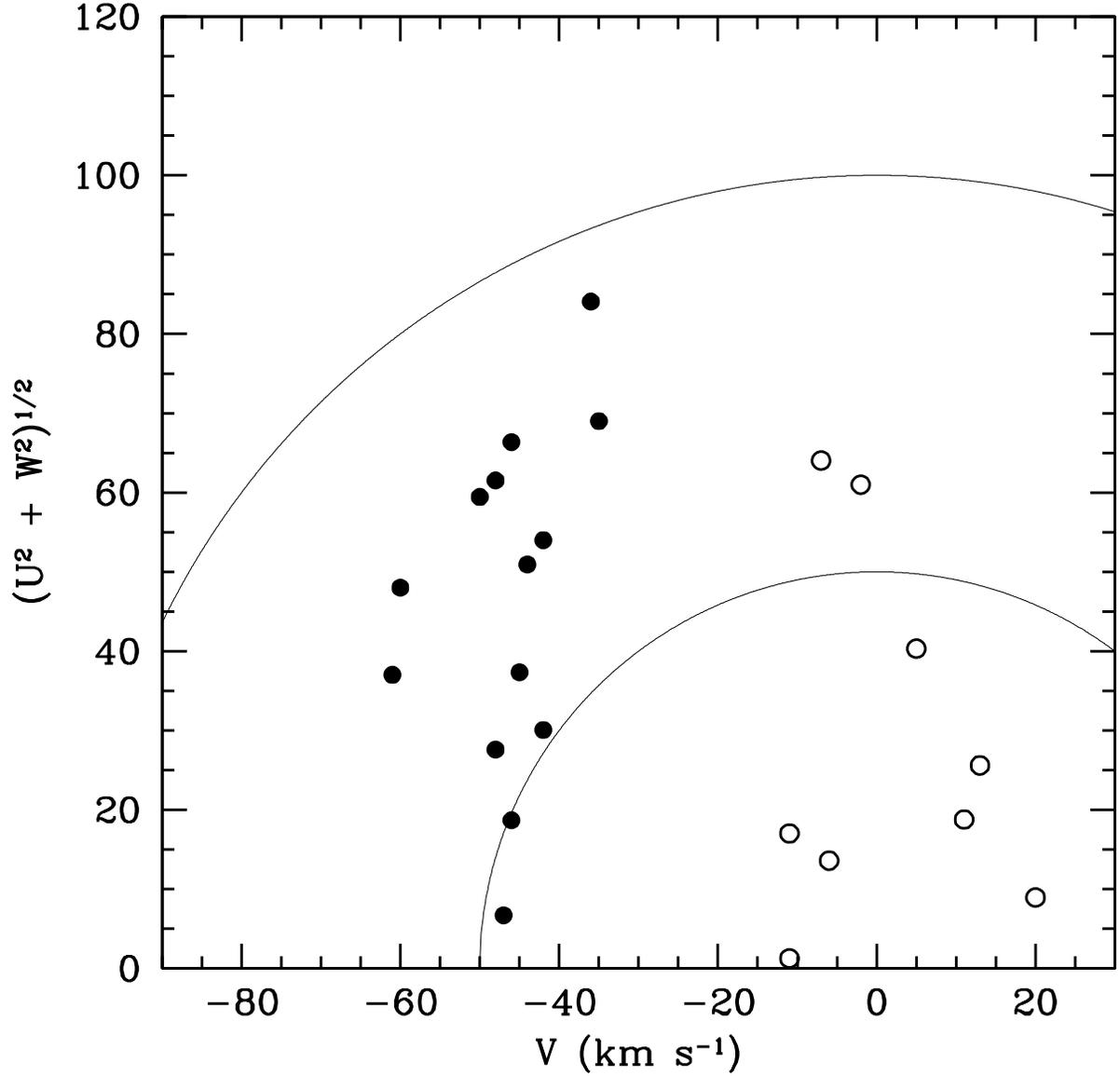}
\caption{The ``Toomre diagram" for the program stars. The
arcs represent kinetic energies of 50 and 100 \kms\ with
respect to the Local Standard of Rest. Stars we defined
to belong to the thin disk are plotted as open circles
while those belonging to the thick disk are filled circles.
Note that, as in Figure~\ref{fig:lsrv}, we have employed
the V velocity as a primary criterion.
\label{fig:toomre}}
\end{figure}

\begin{figure}
\epsscale{1}
\figurenum{5}
\plotone{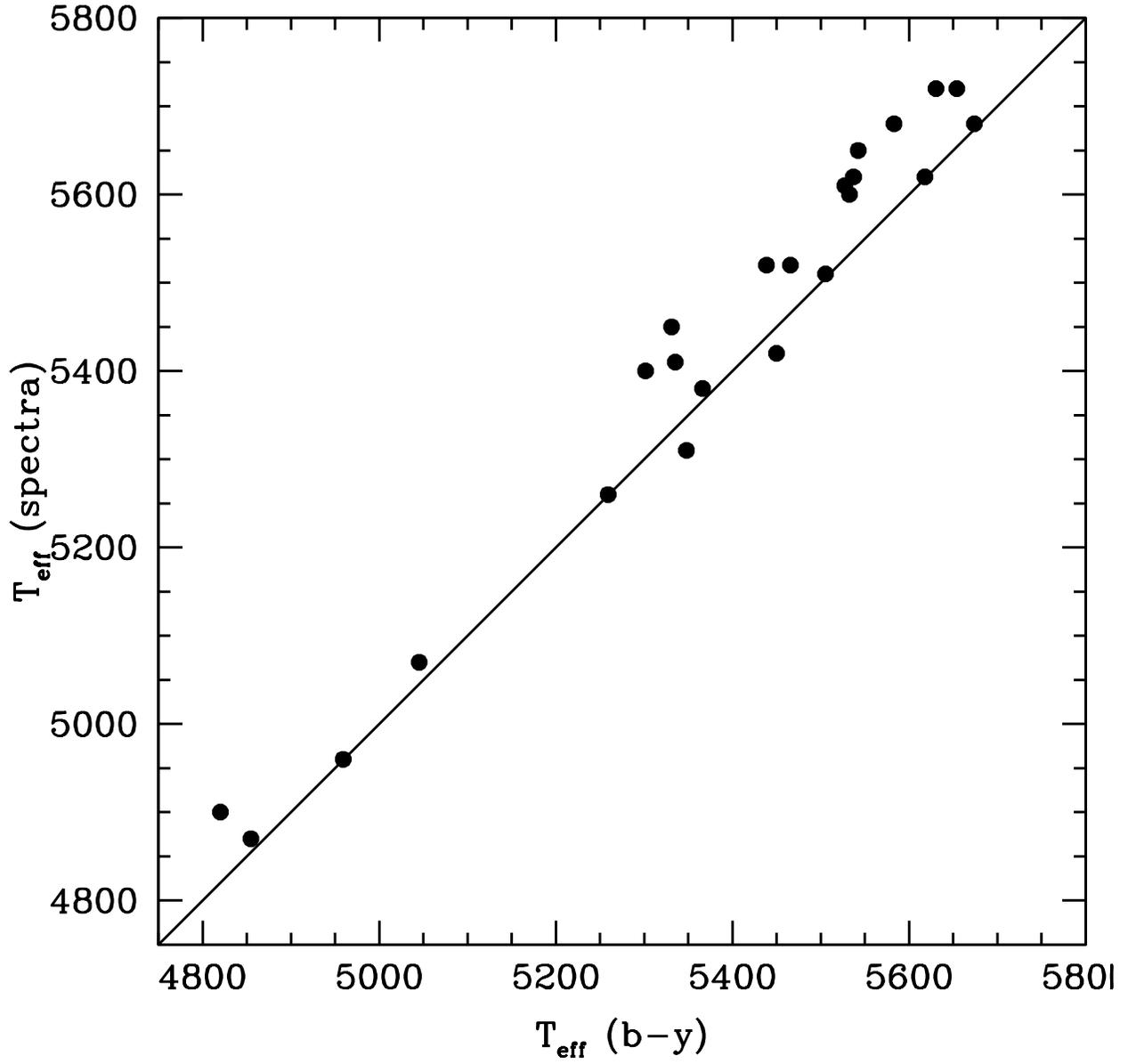}
\caption{Comparison between temperatures derived
from the color-temperature relations of Alonso et al.\ (1996)
and ($b-y$) 
Str\"{o}mgren photometry from Olsen (1993).
\label{fig:tby}} 
\end{figure}

\begin{figure}
\epsscale{1}
\figurenum{6}
\plotone{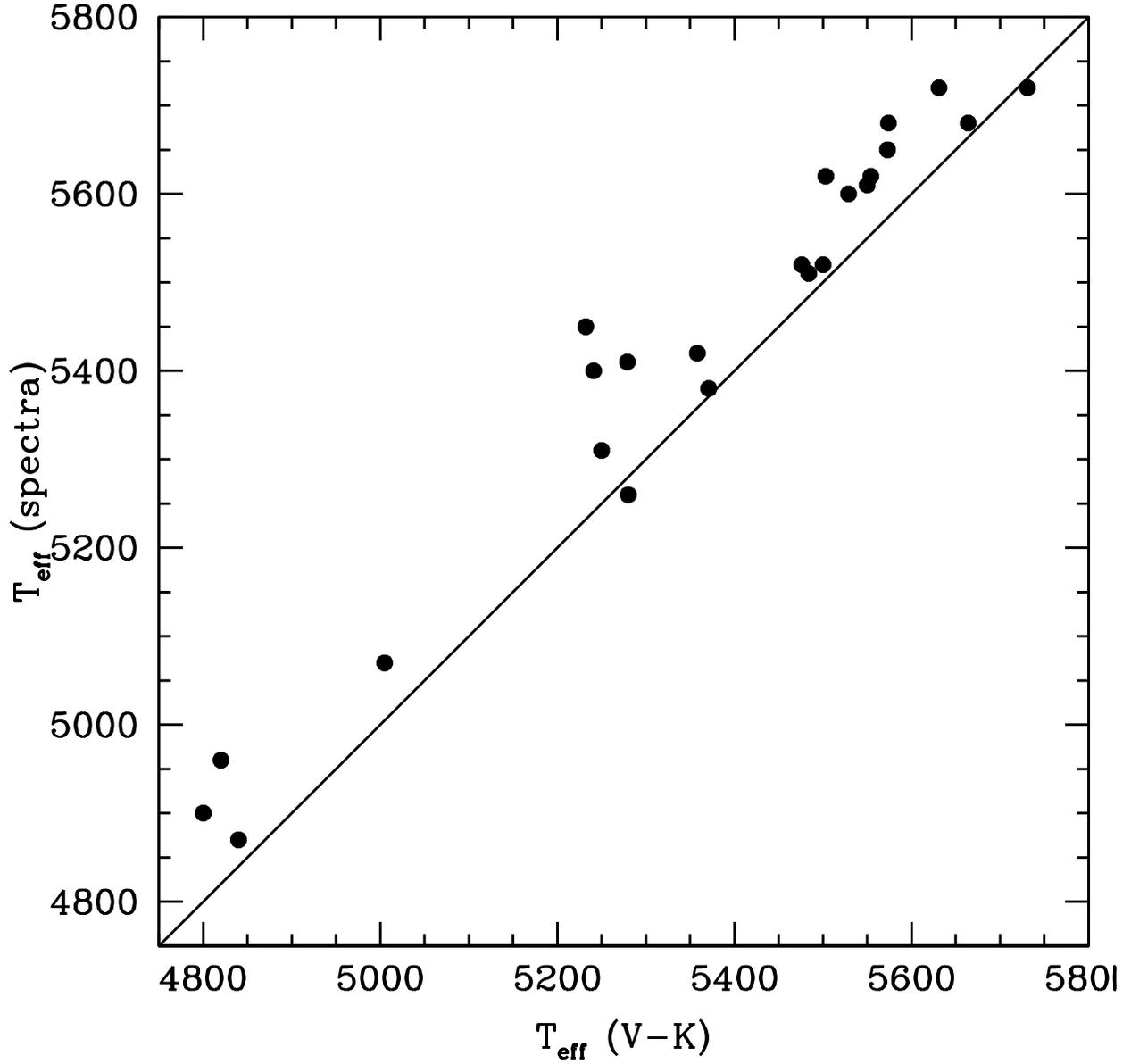}
\caption{Comparison between temperatures derived
from the color-temperature relations of Alonso et al.\ (1996)
and ($V-K$) 2MASS photometry
and our spectroscopic 
determinations. \label{fig:vkt}} 
\end{figure}

\begin{figure}
\epsscale{1}
\figurenum{7}
\plotone{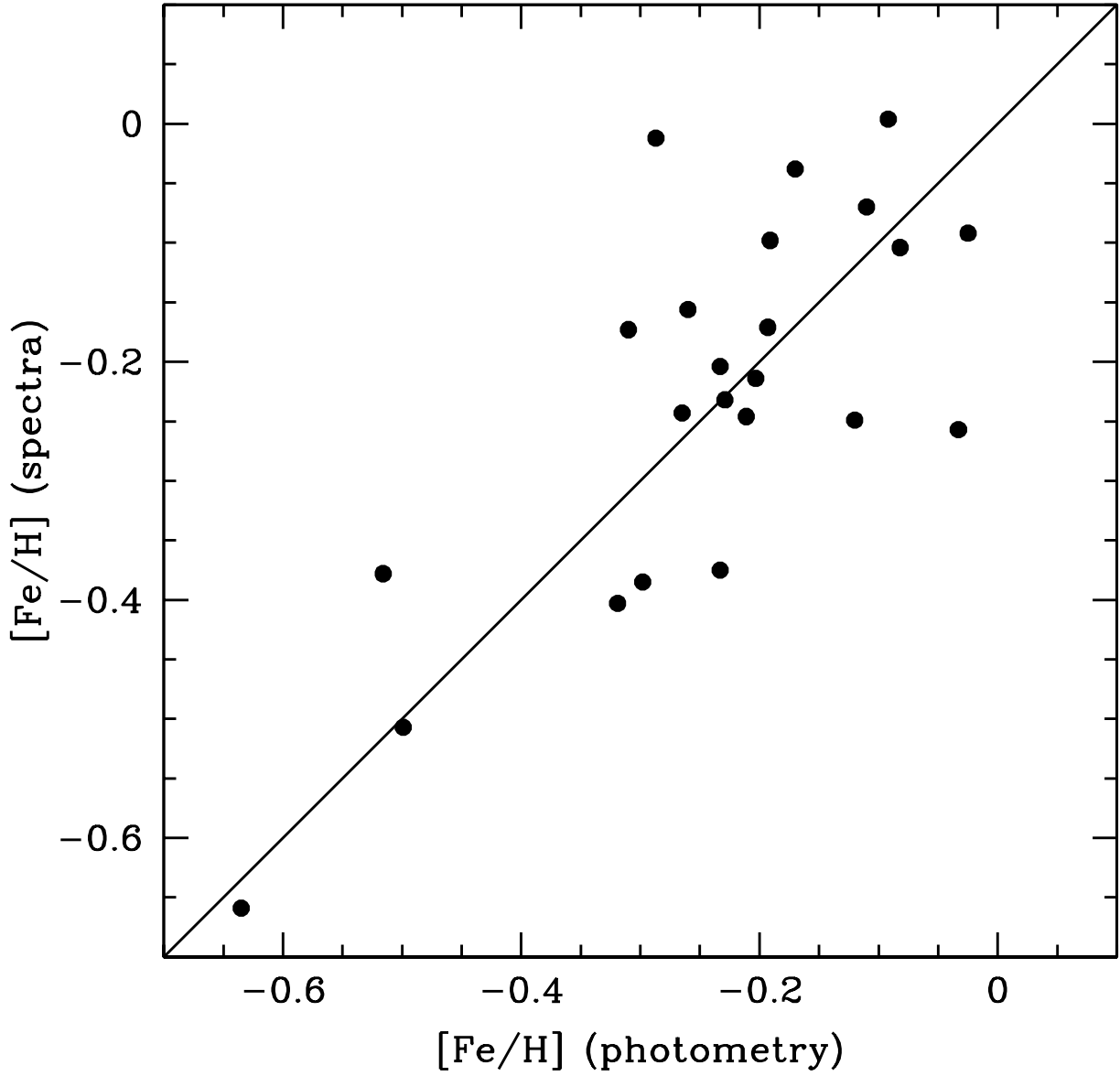}
\caption{Comparison between photometric and spectroscopic 
metallicities. The former were obtained from Str\"{o}mgren
photometry and the calibation of Schuster \& Nissen (1989). \label{fig:fecomp}}
\end{figure}

\begin{figure}
\epsscale{1}
\figurenum{8}
\plotone{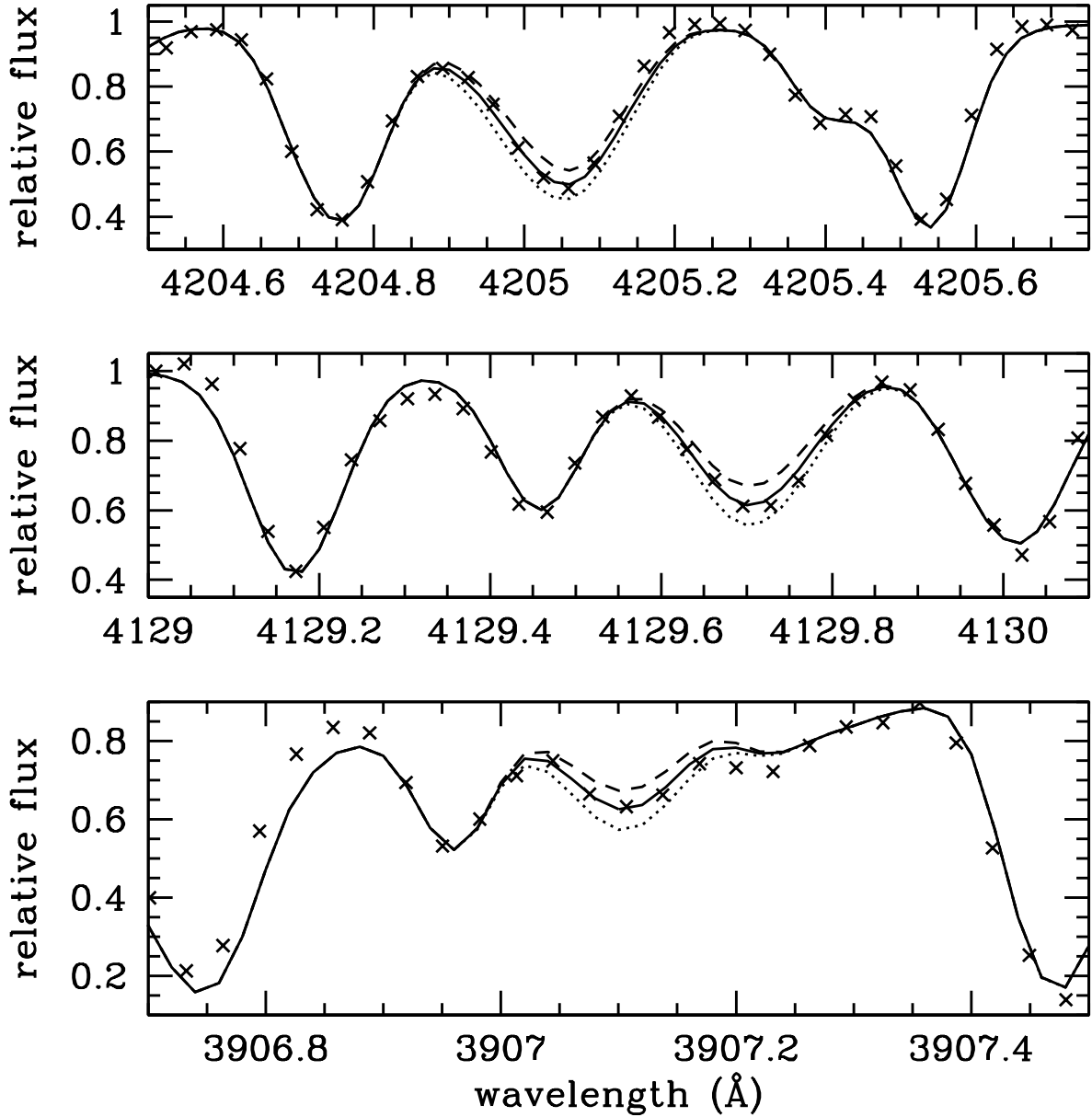}
\caption{Fits to synthetic spectra for the three Eu II lines at 
3907.1~\AA, 4129.7~\AA, and 4205.1~\AA\ for BD+34 927.
The solid lines are the best fits, while the dashed and dotted lines
represent changes of 0.1 dex in the Eu abundance.
\label{fig:syneu}}
\end{figure}

\begin{figure}
\epsscale{1}
\figurenum{9}
\plotone{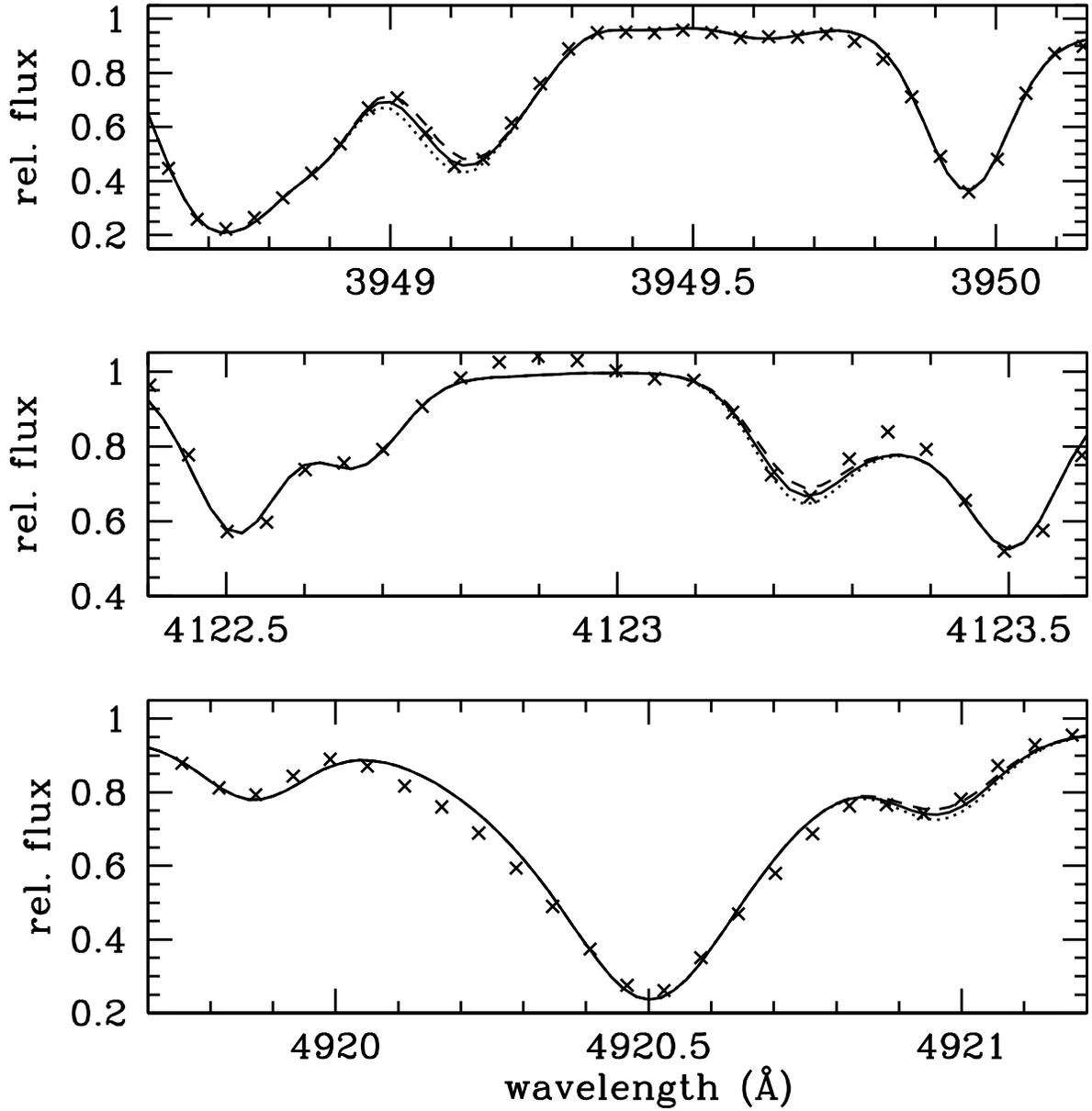}
\caption{Fits to synthetic spectra for the three La II lines at 
3949.1~\AA, 4123.2~\AA, and 4920.9~\AA~for BD+47~2491.
The solid lines are the best fits, while the dashed and dotted lines
represent changes of 0.1 dex in the La abundance.
\label{fig:synla}}
\end{figure}

\begin{figure}
\epsscale{1}
\figurenum{10}
\plotone{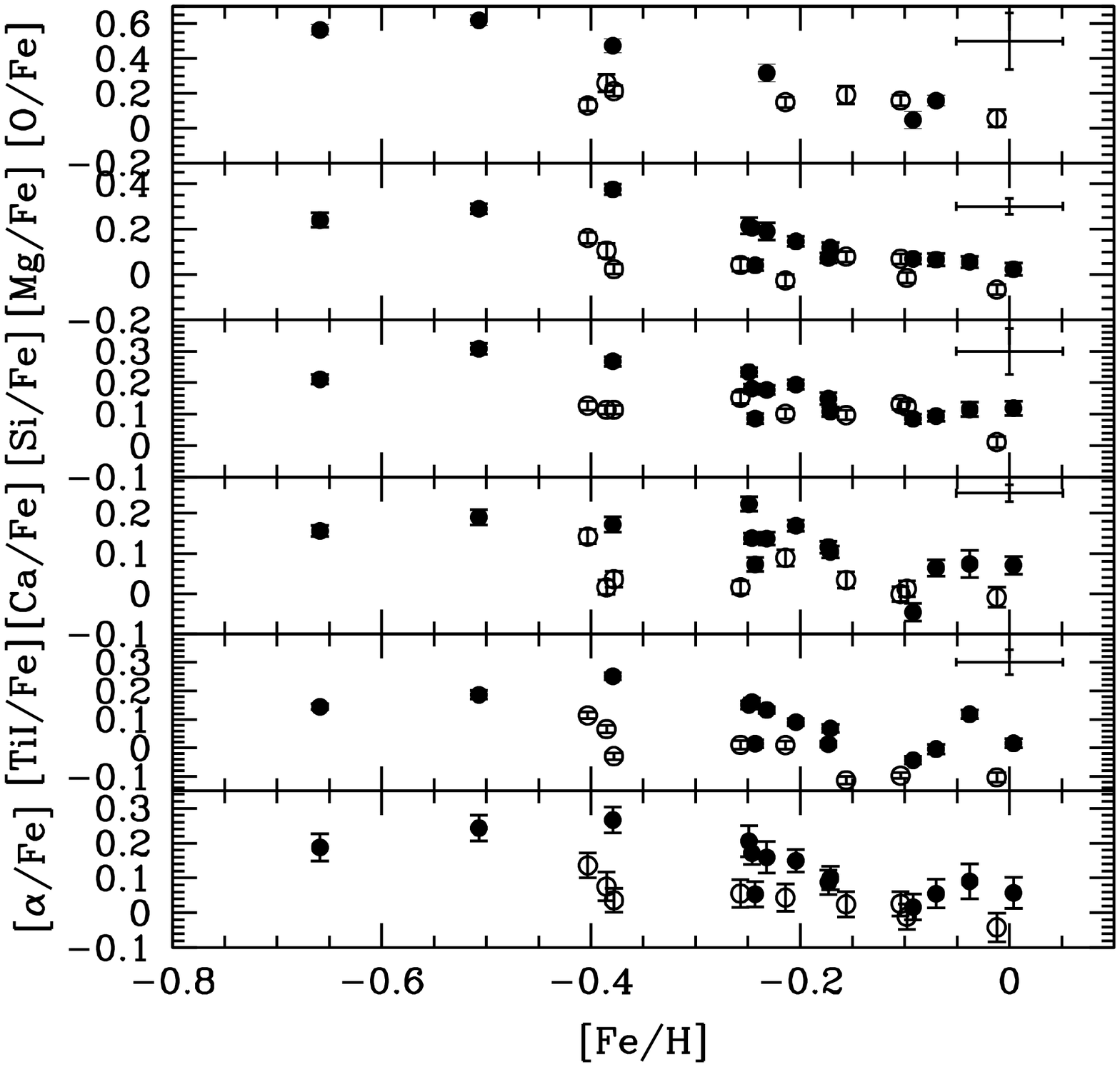}
\caption{Alpha-element abundances.
Filled circles are stars classifed as belonging
to the thick disk while open circles are thin
disk stars. 
The error bars
in the upper right corner are those derived from uncertainties
in the atmospheric parameters. 
\label{fig:alpha}} 
\end{figure}

\begin{figure}
\epsscale{1}
\figurenum{11}
\plotone{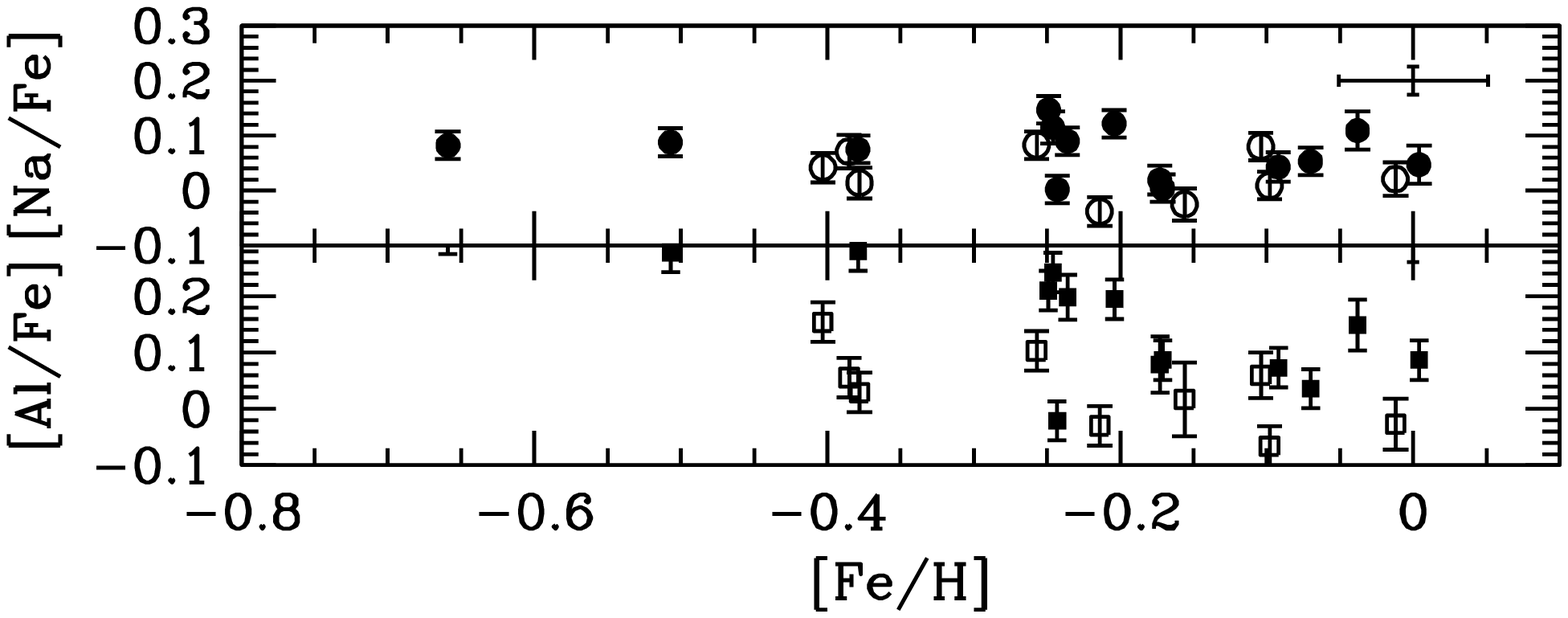}
\caption{Aluminum and sodium abundances. 
Filled circles are stars classifed as belonging
to the thick disk while open circles are thin
disk stars. 
The error bars
in the upper right corner are those derived from uncertainties
in the atmospheric parameters. 
\label{fig:light}} 
\end{figure}

\begin{figure}
\epsscale{1}
\figurenum{12}
\plotone{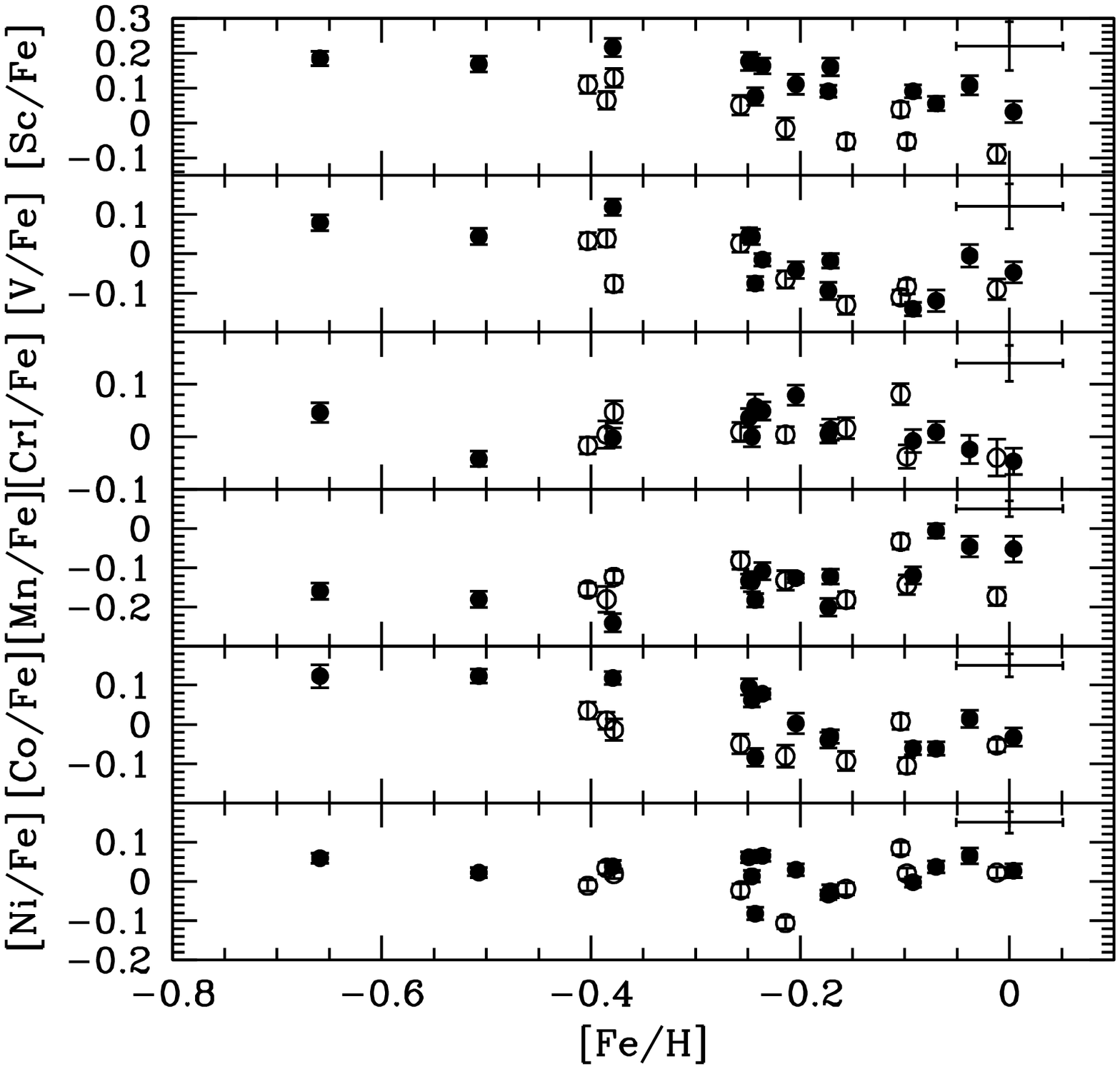}
\caption{Iron peak abundances. 
Filled circles are stars classifed as belonging
to the thick disk while open circles are thin
disk stars. 
The error bars
in the upper right corner are those derived from uncertainties
in the atmospheric parameters. 
\label{fig:iron}} 
\end{figure}

\begin{figure}
\epsscale{1}
\figurenum{13}
\plotone{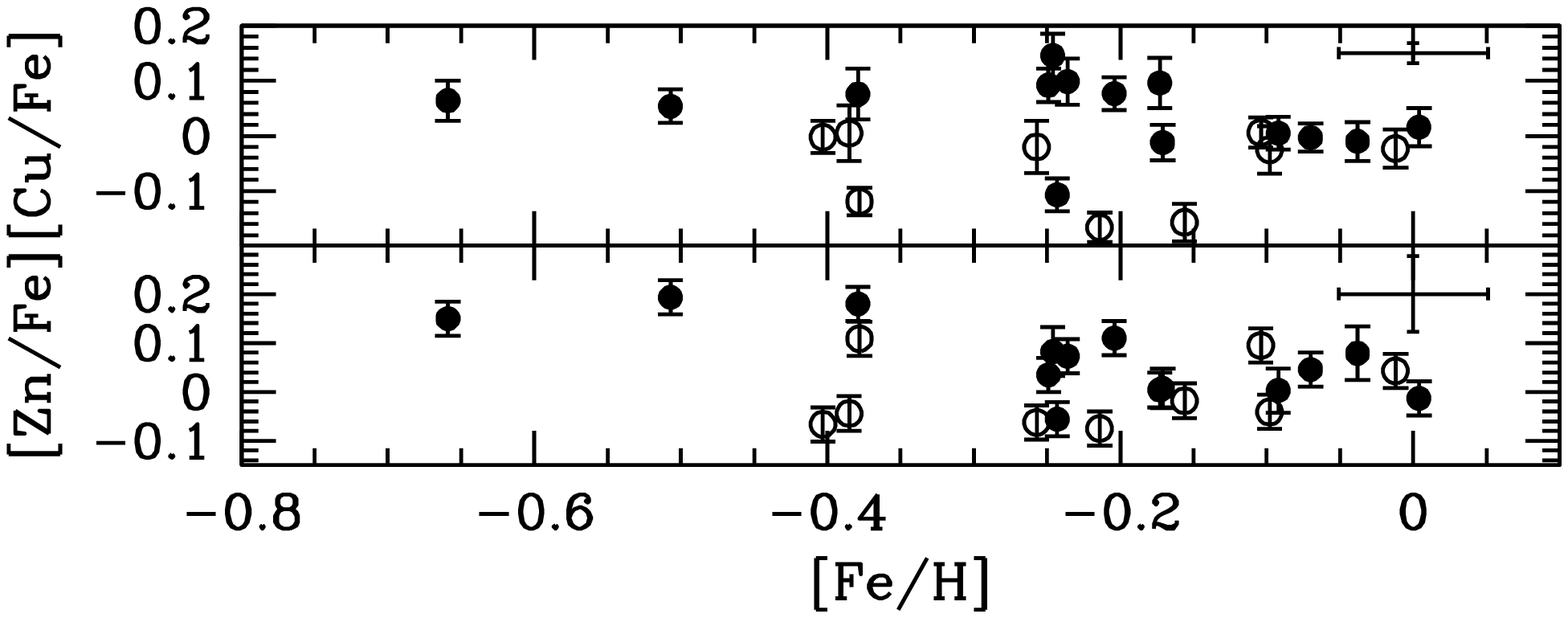}
\caption{Copper and zinc abundances.
Filled circles are stars classifed as belonging
to the thick disk while open circles are thin
disk stars. 
The error bars
in the upper right corner are those derived from uncertainties
in the atmospheric parameters. 
\label{fig:iron2}} 
\end{figure}

\begin{figure}
\epsscale{1}
\figurenum{14}
\plotone{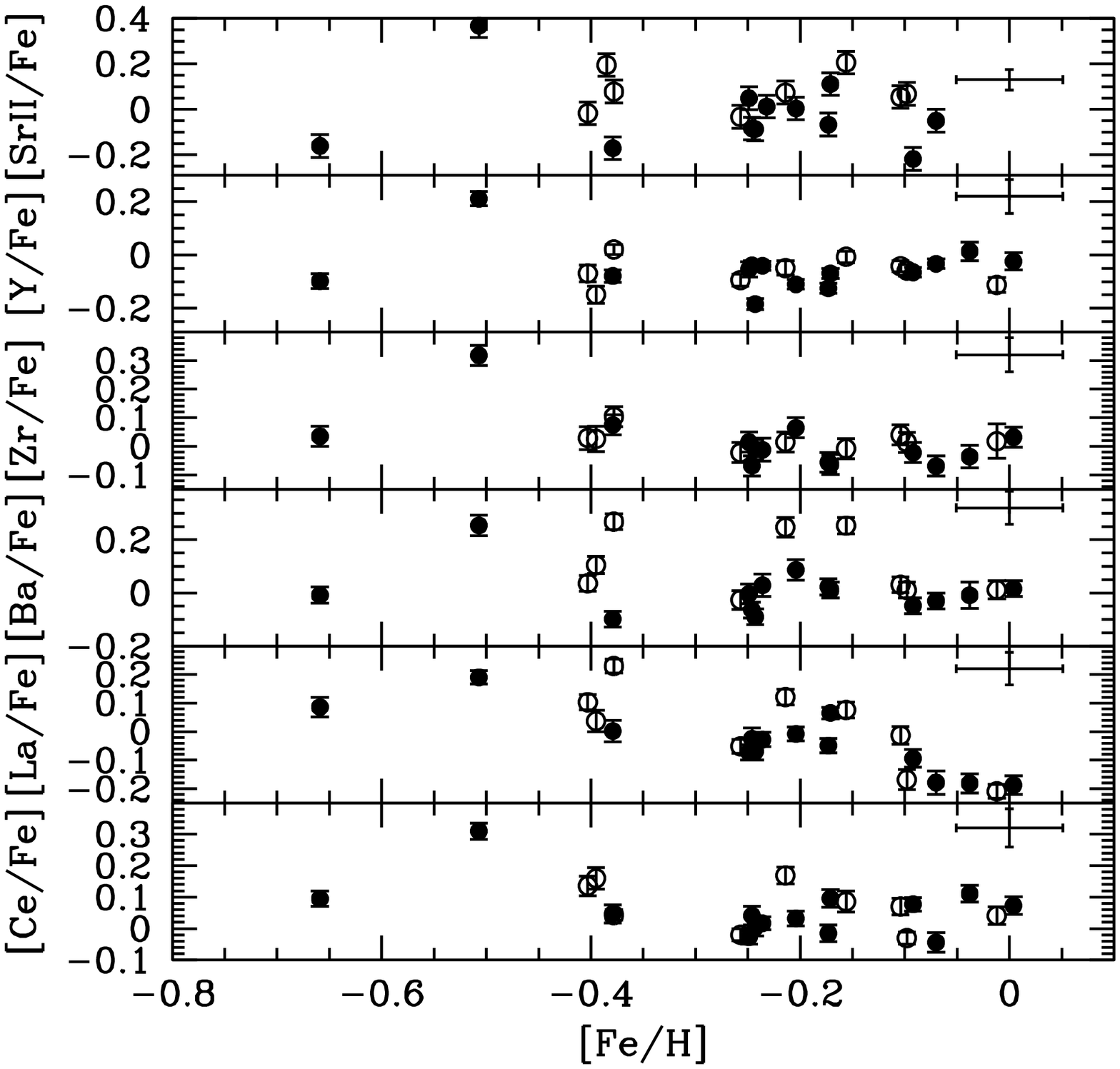}
\caption{$s$-process elemental abundances. 
Filled circles are stars classifed as belonging
to the thick disk while open circles are thin
disk stars. 
The error bars
in the upper right corner are those derived from uncertainties
in the atmospheric parameters. 
\label{fig:spros}} 
\end{figure}

\begin{figure}
\epsscale{1}
\figurenum{15}
\plotone{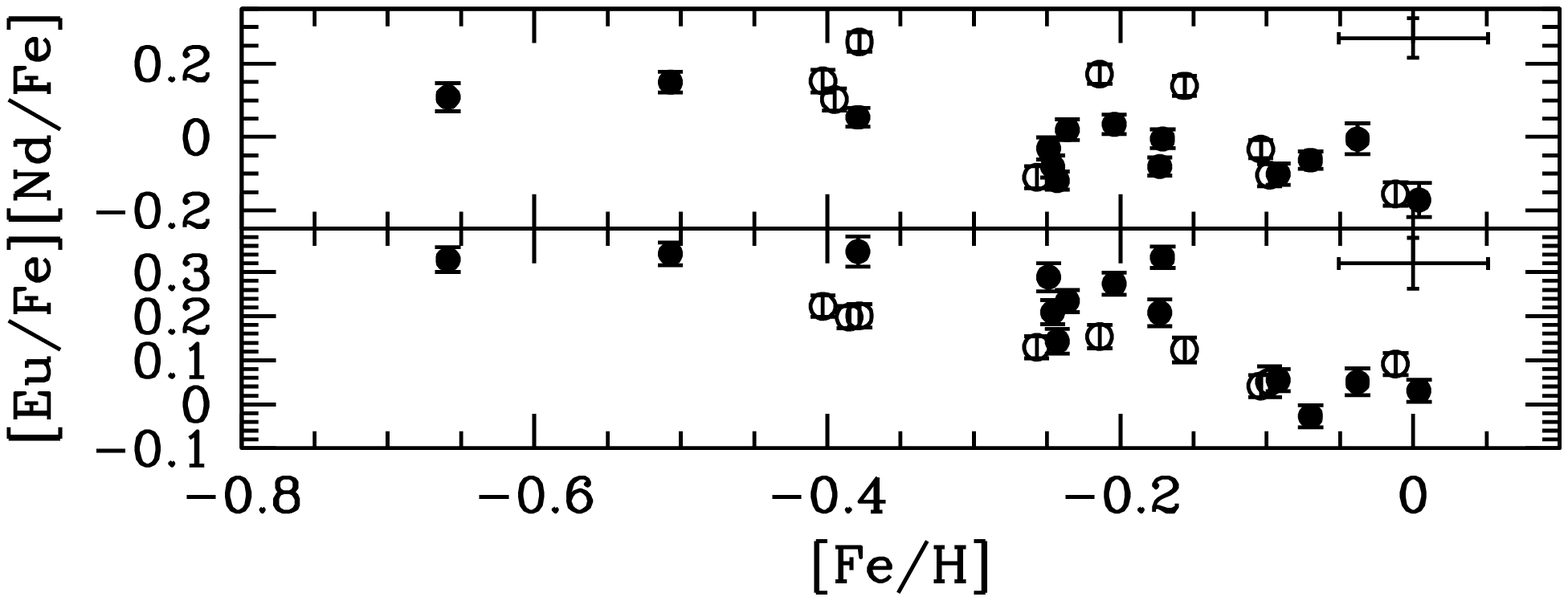}
\caption{$s$-process and $r$-process elemental abundances.
Filled circles are stars classifed as belonging
to the thick disk while open circles are thin
disk stars. 
The error bars
in the upper right corner are those derived from uncertainties
in the atmospheric parameters. 
\label{fig:rpros}} 
\end{figure}

\begin{figure}
\epsscale{1}
\figurenum{16}
\plotone{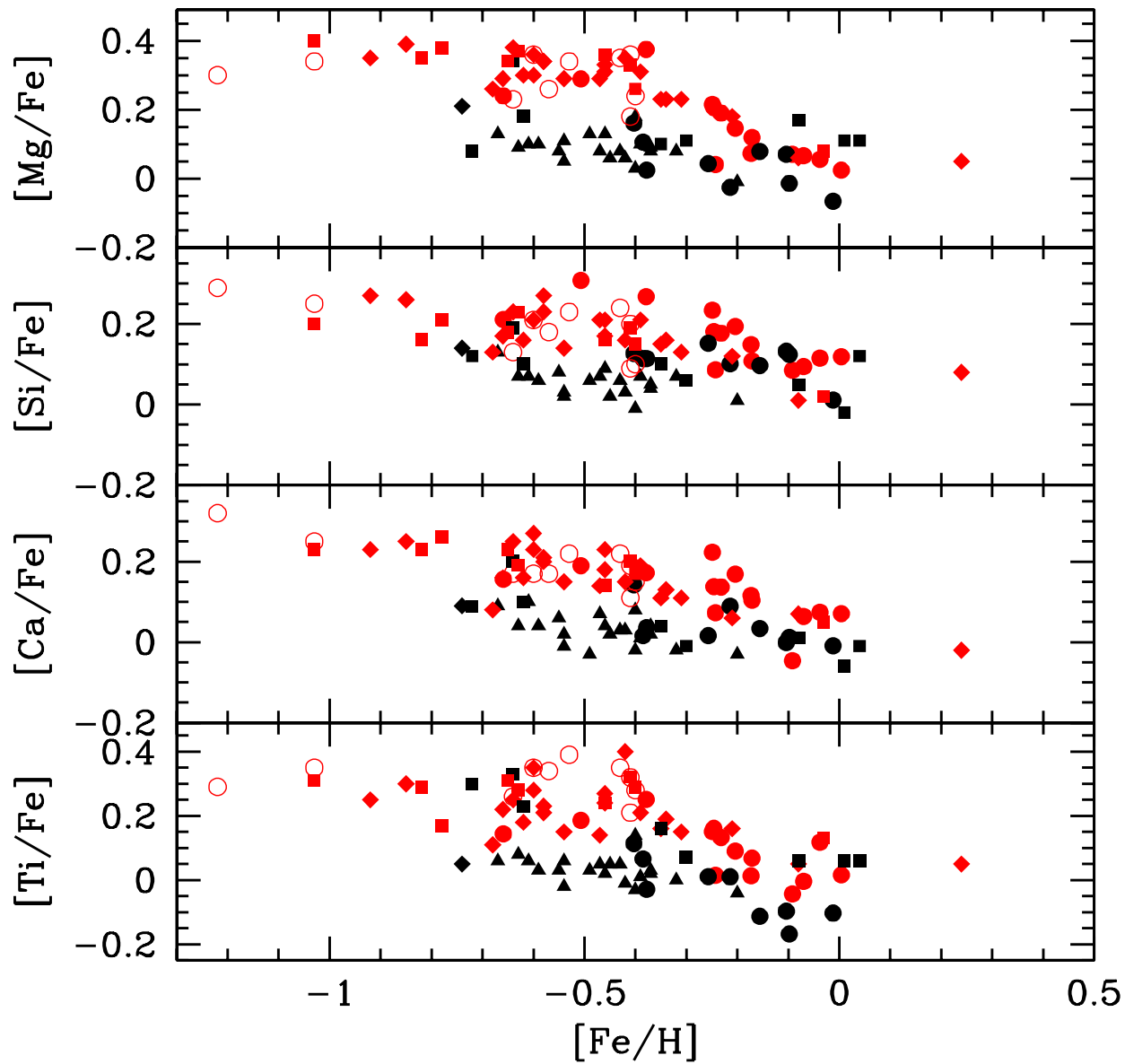}
\caption{Comparison of O, Mg, Si, Ca, and Ti abundances with other 
studies. Red indicates thick disk stars; black thin disk stars.
Our results are shown as filled circles; those from Prochaska
et al.\ (2000) as open circles; Edvardsson et al.\ (1993) as
filled squares; Bensby et al.\ (2003, 2004a) as filled
diamonds; and Reddy et al.\ (2003) as filled triangles.
\label{fig:othersalphafe}}
\end{figure}

\begin{figure}
\epsscale{1}
\figurenum{17}
\plotone{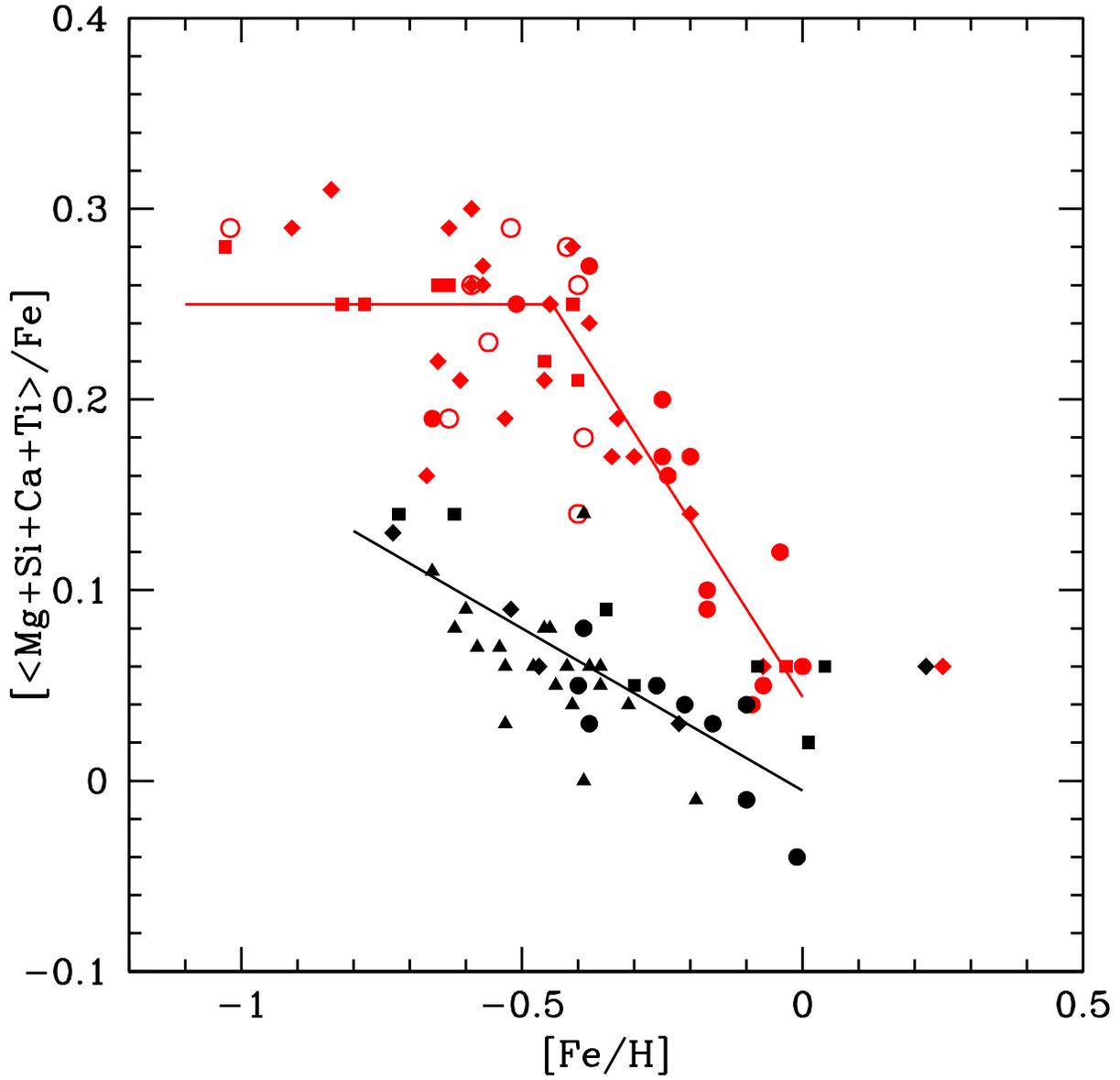}
\caption{A comparison of average $\alpha$-element (Mg, Si, Ca, Ti) abundances with 
other studies. Same symbols as Figure~\ref{fig:othersalphafe}. 
The sloped lines are linear least squares fits, as described in Section~8.1. 
For thick disk stars with [Fe/H] $\leq\ -0.4$, a straight mean for
all stars has been shown.
\label{fig:mgsicatimeans}}
\end{figure}

\begin{figure}
\epsscale{1}
\figurenum{18}
\plotone{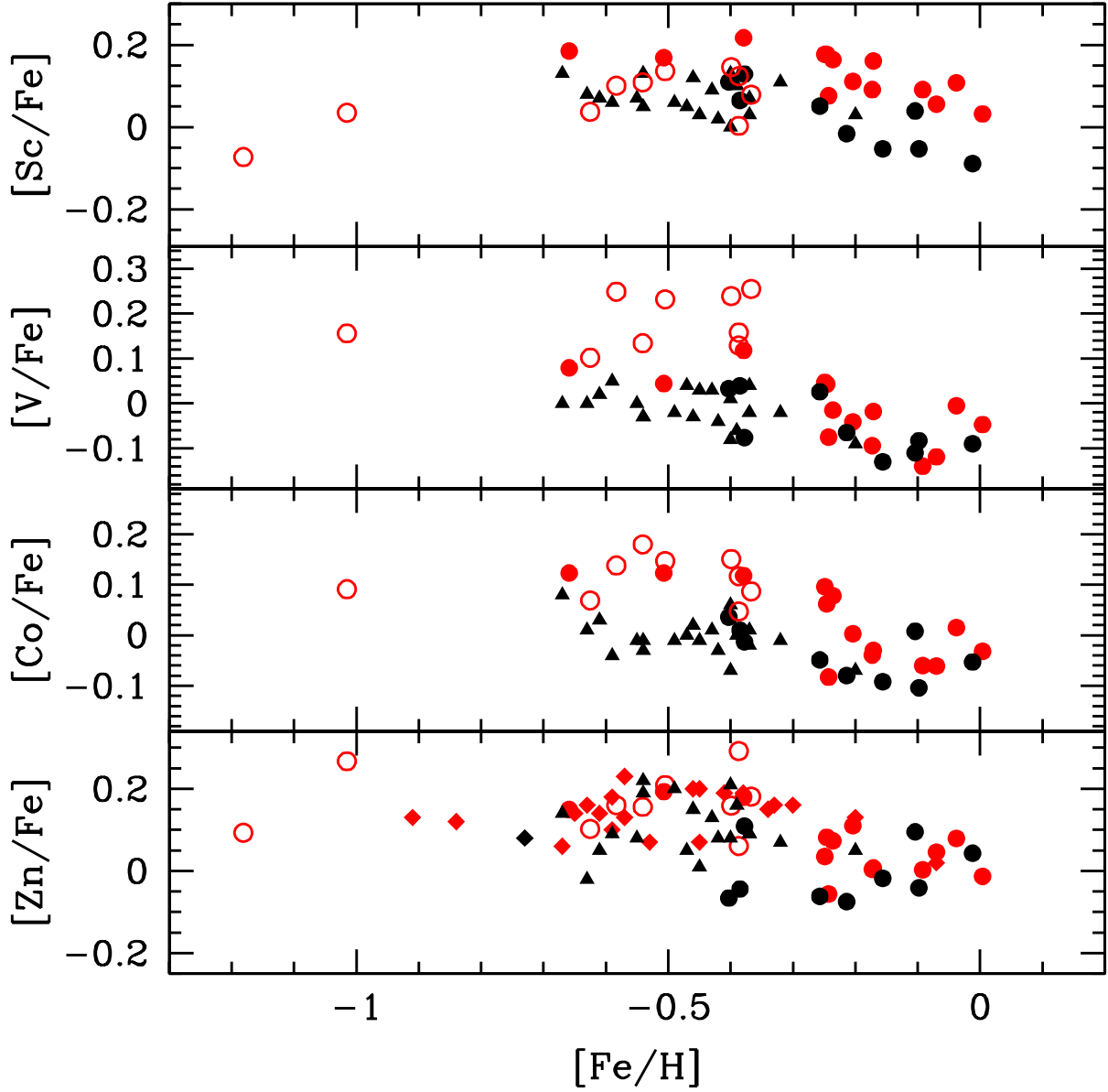}
\caption{Comparison of Sc, V, Co, and Zn abundances with other 
studies. Same symbols as Figure~\ref{fig:othersalphafe}.
\label{fig:scvcozn}}
\end{figure}

\begin{figure}
\epsscale{1}
\figurenum{19}
\plotone{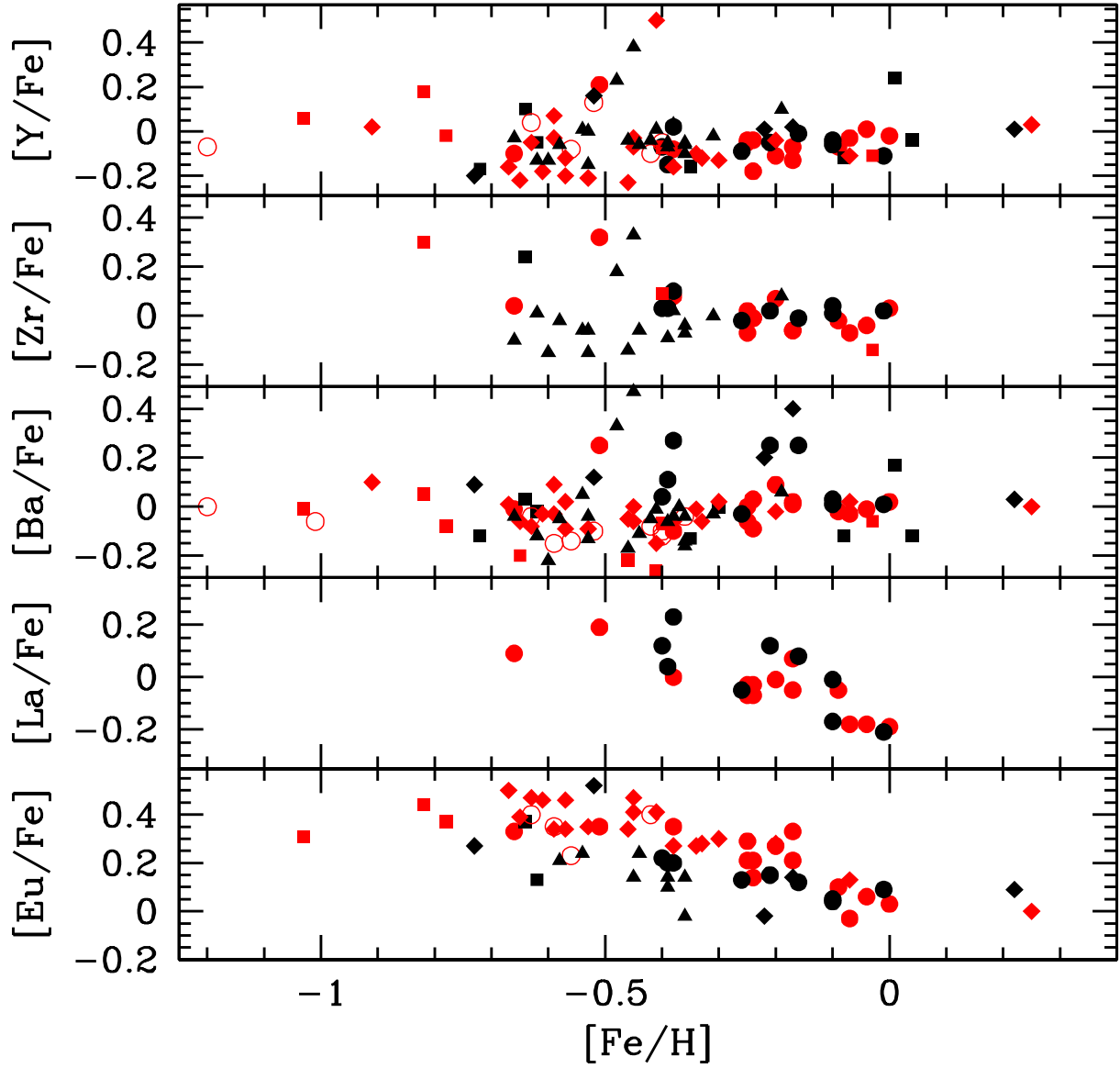}
\caption{Comparison with other studies of heavy element 
abundances. Same symbols 
as Figure~\ref{fig:othersalphafe} with inverted filled triangles
representing data from 
Mashonkina \& Gehren (2000, 2001). \label{fig:yzrbalaeu}}
\end{figure}

\begin{figure}
\epsscale{1}
\figurenum{20}
\plotone{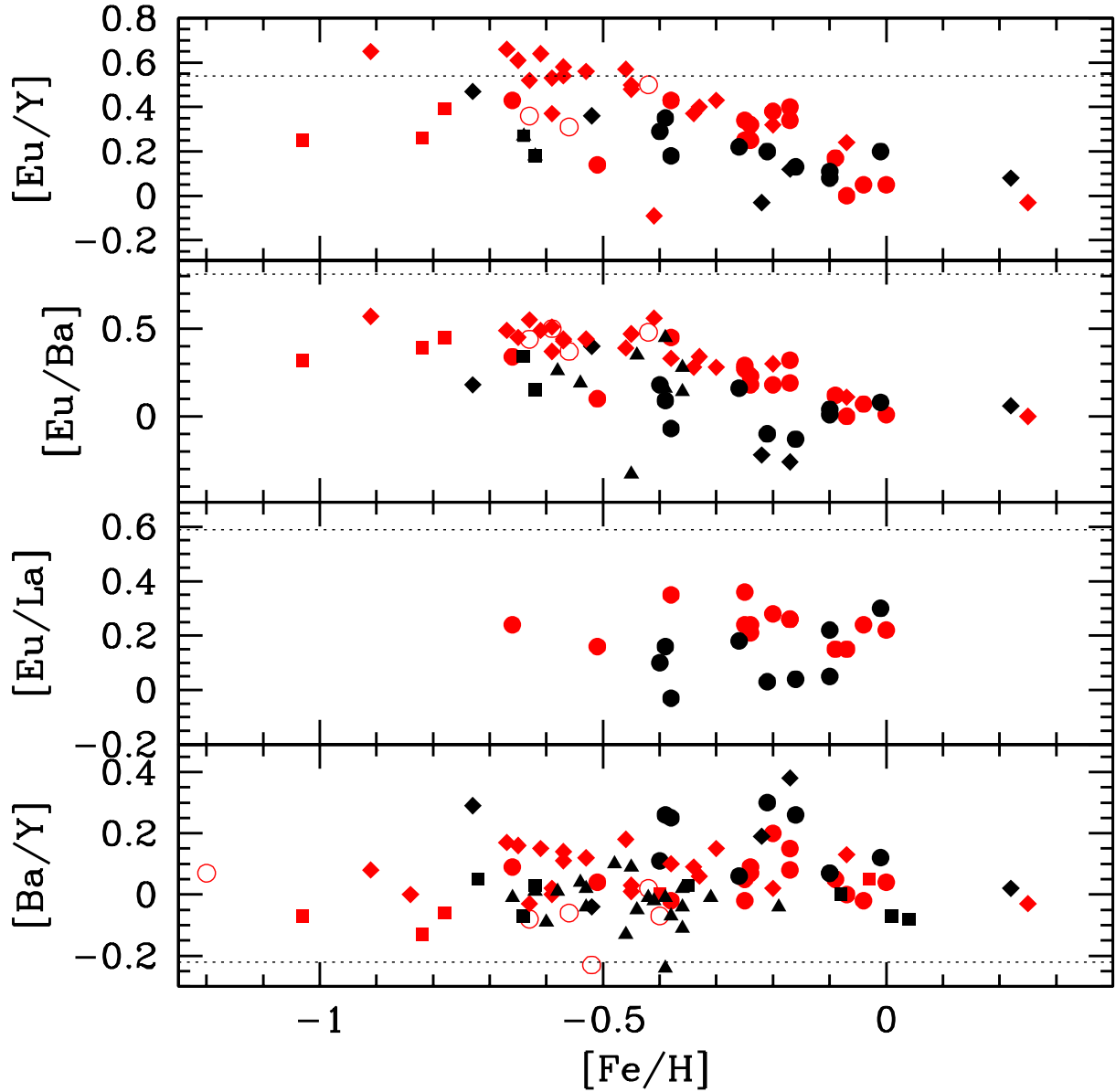}
\caption{Comparison with other studies of $r$-process to
$s$-process elemental abundance ratios. 
The symbols are the same 
as Figure~\ref{fig:yzrbalaeu}. The dotted lines represent the
predicted ratio for production of both elements only via
the $r$-process, using results from Burris
et al.\ (2000). \label{fig:othersstor}}
\end{figure}

First, the thick disk stars, at least in the solar
neighborhood, are old (see Figure~\ref{fig:ar0} and the
discussion in Section~1). Second, the thick disk appears
to have the same mean metallicity, independent of distance
from the plane. There is no vertical metallicity gradient (Carney
et al.\ 1989; Gilmore et al.\ 1995). Third, the limited evidence
appears to show that the thin and thick disk populations maintain
distinctive [Fe/H] vs.\ angular momentum trends (Figure~\ref{fig:lsrv}).
Finally, our results (Figures~\ref{fig:alpha}-\ref{fig:rpros})
and those of others (Figures~\ref{fig:othersalphafe}-\ref{fig:othersstor})
clearly show that the chemical enrichment histories of the thin
and thick disk stellar populations have been very different,
with the most prominent signature being that star formation
proceeded more rapidly in the thick disk, such that [Fe/H]
had reached levels as high as [Fe/H] $\approx -0.4$ before
contributions from SNe~Ia became significant, whereas
in the thin disk star formation was slower, and SNe~Ia contributions
began to appear when [Fe/H] $\approx -1.0$. 

We concur with Venn et al.\ (2004) that the abundance patterns
seen among the dwarf galaxies that have been studied cannot
be reconciled with those seen in among the thick disk stars,
even when attention is restricted to those galaxies that completed
their star formation in one primary, if possibly prolonged, burst.
We say ``prolonged" because the low [$\alpha$/Fe] values suggest
SNe~Ia events were contributing to the pollution of the
interstellar media of these galaxies even when the [Fe/H] levels
were still very low. Indeed, the low mean metallicities of these
galaxies suggests that they were unable to retain the bulk of their
gas and did not evolve as ``closed box" systems.

The thick disk, on the other hand, has a high mean metallicity.
If the thick disk arose due to the accretion of a small galaxy
by the Milky Way, $<$[Fe/H]$> \approx -0.5$ suggests a fairly massive
parent galaxy, comparable in mass to the Magellanic Clouds (see
Grebel, Gallagher, \& Harbeck 2003 for a good discussion of the
relationships between dwarf galaxies' mean metallicities 
and luminosities.) Unfortunately, direct comparison with elemental
abundances and abundance ratios in thick disk stars and
stars in the Magellanic Clouds is compromised by the continued
star formation in both Clouds, and in the bulk of the available
analyses referring to stars with younger ages. We would be
caught in the same problems to which we have referred before
were we to attempt such comparisons.

\subsection{What Was the Progenitor Galaxy for the Thick Disk?}

There are three remaining questions we may ask nonetheless.

First, what type of galaxy might have converted most of its
gas into stars rapidly, so that we could see a behavior
from its remnants within the Milky Way consistent with
Figure~\ref{fig:ar0}? The answer does not appear to involve
dwarf galaxies like the Magellanic Clouds nor the dwarf
spheroidal galaxies. But it could have been a more compact
dwarf elliptical galaxy.

Second, what is known about the abundance patterns in dwarf
elliptical galaxies? Unfortunately, not very much, given their
large distances. M32 is an interesting test case, and the
literature is full of contradictory claims about the mean
age and mean metallicity of its central regions, as well as
about the bulk of its stellar population. The situation is
finally becoming clearer (Schiavon, Caldwell, \& Rose 2004; 
Rose et al.\ 2004; Worthey 2004). In the core of the galaxy, $<$[Fe/H]$>$
is higher than solar, and [Mg/Fe] is subsolar. From a
chemical evolution perspective, this suggests that the core
of M32 had a prolonged star formation history, enabling SNe~Ia
to contribute very significantly to the chemical enrichment.
Schiavon et al.\ (2004), in fact, argue that the luminosity-weighted
age of the stars in the core of M32 is only a few Gyrs,
consistent with this view. The situation changes outside the
core, with $<$[Fe/H]$>$ declining to $\approx -0.25$ at
one effective radius. The weighted age rises as well, by
about 3 Gyrs, although that remains very young in comparison
to the age of the Galaxy's thick disk stellar population.
As would be expected, [Mg/Fe] is somewhat higher at one
effective radius, roughly $-0.1$ compared to $-0.25$ in the
core, but it remains subsolar and consistent with prolonged
star formation. We conclude that M32, in its present state,
is, like the other surviving dwarf galaxies, 
not a good template for the progenitor of the Galactic
thick disk, although it is possible that had it merged with
the Milky Way very early in its chemical evolution, it might
have resulted in what we see today. We cannot distinguish
the detailed chemical enrichment history of M32, unfortunately.

Third, what lessons may we learn from the wealth
of elemental abundance ratios we and our colleagues have determined
from careful studies of thick disk stars? There appear to be
several answers, although not all are as firmly established
as we desire.

\begin{itemize}

\item The ``traditional" $\alpha$-elements O, Mg, Ca, and Si
are enhanced relative to iron, for thick disk stars from
[Fe/H] values as high as $-0.4$, at which point the
abundance ratios begin to decline toward solar values.
This can be understood, as we have stressed, if those elements
arise in the rapid evolution of massive stars, and that such
stars were able to enrich the interstellar medium of the thick
disk's progenitor galaxy much more rapidly than did the thin
disk. SNe~Ia begin to appear in the thin disk when [Fe/H] $\approx -1.0$
while the thick disk was able to reach [Fe/H $\approx -0.4$.

\item Apparently other elements in the thick disk stars were also
created primarily in rapidly-evolving massive stars, including 
Ti, Sc, V, Co, Zn, and Eu. The latter element's comparable behavior
is not at all surprising since in the solar system it is thought
to have been created primarily in the $r$-process.

\item The uniformity of the [Ba/Y] ratio in thick disk stars,
and its similarity to values in seen among thin disk stars,
at least for [Fe/H] $< -0.3$, suggests that only one of the
two processes, the {\em weak} or the {\em main} $s$-process,
were working. Perhaps neither $s$-process was working: ytrrium and
barium both have contributions from the $r$-process.
Qualitatively, we would expect enhanced [Eu/Y] and enhanced
[Eu/Ba] ratios, as seen in Figure~\ref{fig:othersstor}.
Qantitatively, however, the actual ratios appear to fall
short of ``pure" $r$-process predictions. And it is also
possible, even likely, that we have too few data, and that the
scatter is too large, to discern the appearance or change
in the relative contributions of the two $s$-processes.

\item There is almost a ``step function" or at least a
very steep slope signifying rapid changes in the abundances
of [Eu/Y] and [Eu/Fe], beginning at
[Fe/H] $\approx -0.2$. Either a new nucleosynthesis source
began to appear in the thick disk interstellar medium, or,
possibly, its interstellar medium was being diluted by
that of the Galactic thin disk. This could be a signal of
the merger event itself.

\end{itemize}

\subsection{Hierarchical Assembly of the Milky Way}

In the above discussions we have assumed that the progenitor
of the thick disk was a separate galaxy, which underwent its
own star formation history, and which subsequently was captured
and absorbed by the Milky Way, as opposed to the idea that
the disk itself has experienced rather distinct episodes of
star formation, for reasons that are not well understood.
There is, perhaps, an attractive alternative that blends
aspects of both metaphors. In essence, if galaxies are
assembled in a hierarchical fashion, such that small units
merge to make larger ones, out of which eventually emerge
the galaxies we see today, the thick disk might have resulted
from a merger of a moderately large such galaxy, but at an
early time in the formation of the Milky Way itself. The
thick disk progenitor would have still experienced a different
chemical enrichment history, although perhaps not for very long.
Brook et al.\ (2004, 2005) have developed chemodynamical
evolutionary models that explain the observed features of
the thick disk vs.\ the thin disk, including the differences
in kinematics as well as [$\alpha$/Fe] vs.\ [Fe/H]. We certainly
look forward to even more detailed models that explore as well
the heavy $s$-process and $r$-process elements.

\section{SUMMARY}

We conclude that the thick disk does appear to have undergone
a chemical enrichment history that was separate from that of the
thin disk, most probably in a nearby dwarf galaxy. The star formation
history was, however, sufficiently prolonged that SNe~Ia were
contributing to the galaxy's interstellar medium by the
time that the mean metallicity level had reached [Fe/H] = $-0.4$
(or, alternatively, [$\alpha$/H] $\approx -0.1$), where we now we
could include Sc, V, Co, Zn, and Eu in this broad mix as elements
created in the same mass range of stars, albeit in different sites,
as the more ``traditional" $\alpha$-elements. When $<$[Fe/H]$>$ in
the thick disk parent galaxy reached $\approx -0.2$, however,
its stars {\em and} gas appear to have merged with the Milky Way.

We are very grateful to the National Science Foundation for support
through grants AST-9988156 and AST-030541 to the University of
North Carolina. We also thank the anonymous referee for a challenging
but ultimately very constructive review that improved this paper
considerably.

\clearpage

\begin{deluxetable}{lrrrrrrrr}
\tabletypesize{\footnotesize}
\tablenum{1}
\tablewidth{0pc}
\footnotesize
\tablecaption{Radial Velocities \label{tab:radvels}}
\tablehead{\colhead{Name}& 
  \colhead{N} & \colhead{Span}  &
  \colhead{$V_{\rm rad}$} & \colhead{$\sigma$} & \colhead{E} &
  \colhead{I} & \colhead{E/I} & \colhead{P($\chi^{2}$)} 
  }

\startdata
 
BD+65~3 & 5 & 802 & $-60.3$ & 0.2 & 0.2 & 0.4 & 0.7 & 0.776245 \\
BD+69~238  & 5 & 1363 & $-13.4$ & 0.2 & 0.4 & 0.4 & 1.0 & 0.393702 \\ 
BD+34~927 & 12 & 3050 & 39.0 & 0.2 & 0.3 & 0.6 & 0.6 & 0.949663 \\
BD+17~1145 & 12 & 3021 & 5.0 & 0.2 & 0.7 & 0.5 & 1.3 & 0.151076 \\
BD+01~1600 & 7 & 2289 & $-8.1$ & 0.2 & 0.4 & 0.5 & 0.9 & 0.489412 \\
BD+30~1423 & 4 & 791 & 20.0 & 0.2 & 0.4 & 0.4 & 0.8 & 0.554606 \\ 
BD+05~1611 & 8 & 2341 & 21.7 & 0.2 & 0.6 & 0.6 & 1.0 & 0.337620 \\
BD+13~1655 & 4 & 791 & 38.5 & 0.2 & 0.4 & 0.4 & 0.9 & 0.447799 \\
BD+46~1590 & 20 & 2547 & 33.7 & 0.1 & 3.9 & 0.5 & 7.2 & 0.000000\tablenotemark{a} \\
BD+06~2398 & 31 & 7316 & 19.0 & 0.1 & 0.7 & 0.5 & 1.4 & 0.000001\tablenotemark{b} \\
BD+18~2542 & 28 & 1838 & 15.4 & 0.3 & 3.1 & 0.7 & 4.8 & 0.000000\tablenotemark{c} \\
BD+11~2439 & 9 & 3166 & $-24.6$ & 0.1 & 0.3 & 0.4 & 0.7 & 0.867265 \\
BD+02~2585 & 18 & 3186 & 2.1 & 0.2 & 0.7 & 0.6 & 1.2 & 0.311593 \\
BD+04~2696 & 7 & 3635 & $-56.5$ & 0.3 & 0.6 & 0.7 & 0.9 & 0.461471 \\
BD+09~2736 & 11 & 3186 & $-7.0$ & 0.2 & 0.6 & 0.6 & 1.0 & 0.443047 \\
BD+15~2658 & 41 & 3679 & 8.3 & 0.1 & 22.5 & 0.5 & 47.1 & 0.000000\tablenotemark{d} \\
BD+23~2747 & 11 & 2197 & $-26.9$ & 0.1 & 0.4 & 0.4 & 0.9 & 0.649191 \\
BD+26~2677 & 50 & 3665 & $-29.8$ & 0.1 & 0.7 & 0.5 & 1.5 & 0.000000\tablenotemark{e} \\
BD+47~2491 & 9 & 2521 & $-84.2$ & 0.2 & 0.4 & 0.5 & 0.9 & 0.534524 \\ 
BD+45~2684 & 9 & 2015 & $-64.9$ & 0.2 & 0.5 & 0.4 & 1.3 & 0.112319 \\
BD+15~4026 & 8 & 2013 & 20.0 & 0.2 & 0.4 & 0.4 & 0.9 & 0.679412 \\
BD+43~4116 & 11 & 2447 & 4.9 & 0.2 & 0.5 & 0.5 & 1.1 & 0.087191 \\
BD+40~4912 & 9 & 2419 & $-70.7$ & 0.2 & 0.4 & 0.4 & 1.0 & 0.450598 \\ 

\enddata
\tablenotetext{a}{A single-lined spectroscopic binary, with P = 1402 days.}
\tablenotetext{b}{A possible single-lined spectroscopic binary, with 
an undetermined orbital period.}
\tablenotetext{c}{A single-lined spectroscopic binary, with P = 742 days.}
\tablenotetext{d}{A double-lined spectroscopic binary, with P = 23.58 days.}
\tablenotetext{e}{A single-lined spectroscopic binary, with an undetermined
but long orbital period.}
\end{deluxetable}

\clearpage

\begin{deluxetable}{llrrrrrrrr}
\tablenum{2}
\tabletypesize{\footnotesize}
\tablewidth{0pc}
\footnotesize
\tablecaption{Photometric Data \label{tab:photometry}}
\tablehead{\colhead{BD}& \colhead{HD}  & 
  \colhead{$V$} & \colhead{$b-y$} & 
  \colhead{$m_{1}$} & \colhead{$c_{1}$}  &
  \colhead{$\beta$} & \colhead{$K$} &
  \colhead{$J-K$} & \colhead{$V-K$}
  }
\startdata
 
BD+65 3 &   404 & 8.62 & 0.516 & 0.425 & 0.326 & \nodata & 6.52 & 0.51 & 2.10 \\
BD+69 238 & 25665 & 7.70 & 0.541 & 0.497 & 0.238 & \nodata & 5.46 & 0.53 & 2.24 \\
BD+34 927 & 31501 & 8.16 & 0.453 & 0.257 & 0.303 & 2.554 & 6.30 & 0.44 & 1.86 \\
BD+17 1145 & 42160 & 8.48 & 0.417 & 0.205 & 0.309 & 2.584 & 6.88 & 0.36 & 1.60 \\
BD+1 1600 & 51219 & 7.40 & 0.431 & 0.236 & 0.326 & 2.599 & 5.72 & 0.43 & 1.68 \\
BD+30 1423 & 53927 & 8.32 & 0.522 & 0.417 & 0.264 & 2.544 & 6.06 & 0.59 & 2.26  \\
BD+5 1611 & 56202 & 8.42 & 0.400 & 0.187 & 0.318 & 2.609 & 6.93 & 0.32 & 1.49 \\
BD+13 1655 & 57901 & 8.18 & 0.556 & 0.530 & 0.285 & 2.532 & 5.89 & 0.57 & 2.29  \\
BD+46 1590 & 87899 & 8.88 & 0.423 & 0.193 & 0.259 & \nodata & 7.21 & 0.38 & 1.60 \\
BD+6 2398 & 95980 & 8.25 & 0.402 & 0.199 & 0.338 & 2.592 & 6.72 & 0.37 & 1.53 \\
BD+18 2542 & 103419 & 9.34 & 0.437 & 0.261 & 0.298 & 2.598 & 7.57 & 0.42 & 1.77 \\
BD+11 2439 & 106210 & 7.57 & 0.421 & 0.206 & 0.327 & 2.587 & 5.94 & 0.40 & 1.63 \\
BD+2 2585 & 111515 & 8.12 & 0.435 & 0.205 & 0.247 & 2.558 & 6.36 & 0.45 & 1.76 \\
BD+4 2696 & 114094 & 9.67 & 0.434 & 0.229 & 0.278 & 2.562 & 8.02 & 0.42 & 1.65 \\
BD+9 2736 & 115231 & 8.42 & 0.423 & 0.210 & 0.319 & 2.581 & 6.80 & 0.40 & 1.62 \\
BD+15 2658 & 122676 & 7.13 & 0.458 & 0.242 & 0.314 & \nodata & 5.25 & 0.47 & 1.88 \\
BD+23 2747 & 131042 & 7.50 & 0.411 & 0.188 & 0.309 & 2.577 & 5.90 & 0.33 & 1.60 \\
BD+26 2677 & 136274 & 7.99 & 0.458 & 0.264 & 0.289 & \nodata & 6.12 & 0.41 & 1.87 \\
BD+47 2491 & 159062 & 7.22 & 0.458 & 0.258 & 0.238 & \nodata & 5.39 & 0.41 & 1.83 \\
BD+45 2684 & 168009 & 6.30 & 0.410 & 0.203 & 0.344 & 2.597 & 4.76 & 0.36 & 1.54 \\
BD+15 4026 & 190067 & 7.15 & 0.452 & 0.233 & 0.287 & \nodata & 5.32 & 0.45 & 1.83 \\
BD+43 4116 & 209393 & 7.97 & 0.417 & 0.234 & 0.262 & 2.581 & 6.32 & 0.39 & 1.65 \\
BD+40 4912 & 215942 & 8.05 & 0.411 & 0.203 & 0.329 & 2.587 & 6.44 & 0.38 & 1.61 \\

\enddata
\end{deluxetable}

\clearpage

\begin{deluxetable}{lrrrrrrrr}
\tablenum{3}
\tabletypesize{\footnotesize}
\tablewidth{0pc}
\footnotesize
\tablecaption{Photometric Temperatures \& Kinematics \label{tab:stars}}
\tablehead{\colhead{Name}& \colhead{[Fe/H]$_{phot}$}  & 
\colhead{$T_{\rm eff}$} &\colhead{$T_{\rm eff}$} & \colhead{U}  &
  \colhead{V} & \colhead{W} & Thin & Thick \\
  & & ($b-y$) & ($V-K$) &\colhead{\kms} &\colhead{\kms} &\colhead{\kms}
  & (\%) & (\%) }

\startdata
 
BD+65 0003 &   $-$0.09 &5045  & 5005  &$-$30 &$-$42 &    2  & 0.48 & 0.51 \\
BD+34 0927 &   $-$0.23 &5348  & 5250   &   44 &$-$48 &   43 & 0.09 & 0.88 \\
BD+17 1145 &   $-$0.23 &5542  & 5573 &$-$19 &$-$48 &   20 & 0.31 & 0.67 \\
BD+01 1600 & $-$0.03 &5505    & 5484 &$-$57 &$-$46 &$-$34 & 0.15 & 0.82 \\
BD+13 1655 & $-$0.17 &4820    & 4800 &    6 &$-$47 &    3 & 0.42 & 0.57 \\
BD+04 2696 &  $-$0.26 &5439 & 5500 &   71 &$-$36 &$-$45 & 0.11 & 0.90 \\
BD+11 2439 & $-$0.23 &5532  & 5529&$-$58 &$-$54 &$-$42 & 0.05 & 0.90 \\
BD+02 2585 &  $-$0.63 &5366 & 5371&$-$50 &$-$69 &$-$35 & 0.03 & 0.92 \\
BD+09 2736 & $-$0.19 &5527 & 5550&   48 &$-$60 & 0 & 0.17 & 0.81 \\
BD+23 2747~&  $-$0.31 &5583 & 5574&   35 &$-$45 &   13 & 0.38 & 0.60 \\
BD+26 2677 & $-$0.21 &5301  & 5241&   29 &$-$61 &   23 & 0.12 & 0.85 \\
BD+47 2491 & $-$0.50 &5259  & 5280&   17 &$-$44 &$-$48 & 0.09 & 0.88 \\
BD+45 2684 &  $-$0.11 &5630 & 5631& $-$5 &$-$46 &$-$18 & 0.37 & 0.62 \\
BD+40 4912 &  $-$0.12 &5618  & 5554& $-$58 &$-$50 &   13 & 0.24 & 0.73 \\
BD+69 0238 & $-$0.29 &4855  & 4840& $-$4 & $-$6 &$-$13 & 0.84 & 0.16 \\
BD+30 1423 &$-$0.30 &4959  & 4820 &   13 &$-$11 &$-$11 & 0.82 & 0.18 \\
BD+05 1611 &  $-$0.26 &5654  & 5731&    61 & $-$2 &    1 & 0.81 & 0.19 \\
BD+46 1590 &  $-$0.52 &5466  & 5476&   37 &    5 &   16 & 0.84 & 0.15 \\
BD+06 2398 &  $-$0.08 &5674  & 5664&   64 & $-$7 & $-$2 & 0.77 & 0.22 \\
BD+18 2542 & $-$0.03 &5450  & 5358&   16 &   13 &   20 & 0.86 & 0.13 \\
BD+15 2658 & $-$0.19 &5331 & 5232& $-$8 &   11 &   17 & 0.87 & 0.12 \\
BD+15 4026 & $-$0.32 &5335  & 5279&$-$63 & $-$1 &$-$11 & 0.79 & 0.21 \\
BD+43 4116 &  $-$0.20 &5537  & 5503& $-$4 &   20 &    8 & 0.91 & 0.09 \\

\enddata
\end{deluxetable}

\clearpage

\begin{deluxetable}{lll}
\tablenum{4}
\tablewidth{0pc}
\tablecaption{Observations \label{tab:observations}}

\tablehead{\colhead{Date} &
  \colhead{Telescope} &
  \colhead{$\lambda$ range (\AA)}  }

\startdata

January 14$-$19, 2000& KPNO& 5000-7400 \\
July 9$-$13, 2001& KPNO& 3900-5500 \\
January 28$-$31, 2002& KPNO& 3700-5200 \\
May 25$-30$, 2002 & KPNO& 3750-5300 \\
July 25$-$29, 2002& KPNO& 5420-7840 \\
November 28$-$December 1, 2002& McDonald & 3700-8600 \\
January 12$-$14, 2003& McDonald & 3800-8700 \\

\enddata
\end{deluxetable}

\clearpage

\begin{deluxetable}{lccccccccc}
\tablenum{5}
\tabletypesize{\footnotesize}
\tablewidth{0pc}
\tablecaption{Observational Data \label{tab:spectra}}

\tablehead{\colhead{Name} & \colhead{R.\ A.} & 
  \colhead{DEC.} & \colhead{$V$} & \colhead{Red} &
  \colhead{Exp} & \colhead{S/N\tablenotemark{a}} & \colhead{Blue }& 
  \colhead{Exp} & \colhead{S/N\tablenotemark{a}} \\
  & \colhead{(J2000)} & \colhead{(J2000)} & &
  \colhead{Date} & \colhead{(min)} & &
  \colhead{Date} & \colhead{(min)} &   }
\startdata
BD+65 0003    & 00 08 56.90 & +65 38 47.0 & 8.61 & 01/2000  & 30 & 240 & 07/2001  & 65 & 135\\
BD+34 0927  & 04 57 59.37 & +34 16 04.9 & 8.16 & 01/2000  & 35 & 410 & 11/2002  & 60 &  120\\
BD+17 1145 & 06 10 01.07 & +17 56 03.2 & 8.48 & 01/2000  & 30 & 380 & 11/2002  & 60 &  150\\
BD+01 1600 & 06 56 34.19 & +01 09 43.5 & 7.39 & 01/2000  & 10 & 280 & 11/2002  & 20 & 130\\
BD+13 1655 & 07 23 47.07 & +12 57 53.0 & 8.18 & 01/2000  & 15 & 220 & 01/2002  & 30 & 120\\
BD+04 2696 & 13 08 13.16 & +03 46 36.6 & 9.67 & 01/2000  & 30 & 210 & 05/2002   & 60 & 140\\
BD+11 2439 & 12 13 13.12 & +10 49 18.0 & 7.57 & 01/2000  & 15 & 420 & 05/2002   & 20 & 250\\
BD+02 2585 & 12 49 44.83 & +01 11 16.9 & 8.11 & 05/2003   & 30 & 490 & 01/2003  & 30 & 180\\
BD+09 2736 & 13 15 36.97 & +09 00 57.7 & 8.42 & 01/2000  & 20 & 280 & 05/2002   & 40 & 210\\
BD+23 2747 & 14 50 40.98 & +22 54 27.4 & 7.50 & 01/2000  & 10 & 300 & 01/2002  & 15 & 115\\
BD+26 2677 & 15 18 59.06 & +25 41 30.1 & 7.99 & 01/2000  & 15 & 350 & 05/2002   & 20 & 170\\
BD+47~2491 & 17 30 16.43 & +47~24 07.9 & 7.22 & 07/2002  & 10 & 400 & 05/2002   & 40 & 280\\
BD+45 2684 & 18 15 59.06 & +45 12 33.5 & 6.31 & 07/2002  & 10 & 650 & 07/2001  & 30 & 255\\
BD+40 4912 & 22 48 13.95 & +41 31 57.0 & 8.05 & 01/2000  & 27 & 240 & 07/2001  & 75 &  165\\
BD+69 0238  & 04 09 35.04 & +69 32 29.0 & 7.70 & 01/2000  & 15 & 330 & 11/2002  & 60 & 100\\ 
BD+30 1423 & 07 08 04.24 & +29 50 04.2 & 8.32 & 01/2000  & 40 & 230 & 11/2002  & 40 &  150\\
BD+05 1611 & 07 16 18.55 & +05 04 33.9 & 8.41 & 01/2000  & 40 & 250 & 11/2002  & 30 &  190\\
BD+46 1590 & 10 09 14.20 & +46 17 02.3 & 8.88 & 05/2003   & 10 & 200 & 01/2003  & 60 & 120\\ 
           &             &             &              &            &    &     & 01/2002  & 30 & 100\\
BD+06 2398 & 11 04 18.93 & +05 47~44.5 & 8.25 & 05/2003   & 10 & 280 & 01/2003  & 60 & 180\\ 
           &             &             &      &                   &    &      & 01/2002  & 15 & 130\\
BD+18 2542 & 11 54 33.78 & +17 52 50.0 & 9.34 & 05/2000  & 90 & 250 & 05/2002   & 80 & 130\\ 
BD+15 2658 & 14 02 56.86 & +14 58 31.2 & 7.13 & 01/2000  & 15 & 410 & 07/2001  & 25 & 115\\ 
BD+15 4026 & 20 02 34.12 & +15 35 31.5 & 7.15 & 07/2002  & 10 & 400 & 05/2002   & 20 & 180\\
BD+43 4116 & 22 02 05.39 & +44 20 35.4 & 7.97 & 07/2002  & 25 & 480 & 05/2002   & 20 & 240\\
\enddata
\tablenotetext{a}{The S/N level cited is per resolution element. A spectroscopic
resolution element covers two pixels. In the red, the reference wavelength is
6645 \AA, and in the blue it is 4130 \AA.}
\end{deluxetable}

\clearpage

\begin{deluxetable}{lcrccrccrcc}
\tablenum{6}
\tabletypesize{\footnotesize}
\tablewidth{0pc}
\tablecaption{Atmospheric Parameters \label{tab:parameters}}
\tablehead{
   \colhead{Star} & \colhead{$T_{\rm eff}$} & \colhead{[M/H]}  & 
   \colhead{$\xi$(km/s)} & \colhead{log g} &
   \colhead{[FeI/H]} & \colhead{N} &
   \colhead{$\sigma$} &
   \colhead{[FeII/H]} & \colhead {N} & \colhead{$\sigma$}  }

\startdata

BD+65 3     &  5070 &     0.0   & 1.07 &  4.40 &      0.004 &   39    & 0.107 &     0.012   &   10    &  0.105 \\
BD+34 927   &  5310 &  $-$0.4   & 1.00 &  4.60 &   $-$0.379 &   78    & 0.089 &  $-$0.375   &   14    &  0.097 \\
BD+17 1145  &  5650 &  $-$0.2   & 1.03 &  4.40 &   $-$0.232 &   85    & 0.085 &  $-$0.237   &   19    &  0.075 \\
BD+01 1600  &  5510 &  $-$0.1   & 1.13 &  4.30 &   $-$0.092 &   73    & 0.080 &  $-$0.096   &   17    &  0.094 \\
BD+13 1655  &  4900 &     0.0   & 1.07 &  4.50 &   $-$0.038 &   43    & 0.090 &  $-$0.047~  &   11    &  0.050 \\
BD+04 2696  &  5520 &  $-$0.3   & 1.15 &  4.45 &   $-$0.243 &   75    & 0.087 &  $-$0.247~  &   17    &  0.075 \\
BD+11 2439  &  5600 &  $-$0.2   & 1.10 &  4.35 &   $-$0.204 &   62    & 0.081 &  $-$0.198   &   17    &  0.094 \\
BD+02 2585  &  5380 &  $-$0.7   & 0.95 &  4.50 &   $-$0.659 &  103    & 0.072 &  $-$0.664   &   24    &  0.072 \\
BD+09 2736  &  5610 &  $-$0.2   & 1.27 &  4.40 &   $-$0.171 &   72    & 0.086 &  $-$0.172   &   14    &  0.088 \\
BD+23 2747~ &  5680 &  $-$0.2   & 1.13 &  4.35 &   $-$0.173 &   70    & 0.073 &  $-$0.173   &   15    &  0.064 \\
BD+26 2677  &  5400 &  $-$0.3   & 1.10 &  4.50 &   $-$0.246 &   65    & 0.089 &  $-$0.254   &   14    &  0.090 \\
BD+47~2491  &  5260 &  $-$0.5   & 1.05 &  4.45 &   $-$0.507 &   68    & 0.095 &  $-$0.500   &   14    &  0.095 \\
BD+45 2684  &  5720 &  $-$0.1   & 1.15 &  4.20 &   $-$0.070 &   70    & 0.078 &  $-$0.060   &   15    &  0.038 \\
BD+40 4912  &  5620 &  $-$0.3   & 1.10 &  4.30 &   $-$0.249 &   74    & 0.082 &  $-$0.257   &   14    &  0.081 \\
BD+69 238   &  4870 &     0.0   & 0.85 &  4.40 &   $-$0.012 &   73    & 0.099 &  $-$0.016   &   13    &  0.081 \\
BD+30 1423  &  4960 &  $-$0.4   & 0.90 &  4.60 &   $-$0.385 &   78    & 0.094 &  $-$0.390   &   17    &  0.079 \\
BD+05 1611  &  5720 &  $-$0.2   & 1.35 &  4.45 &   $-$0.156 &   69    & 0.073 &  $-$0.159   &   17    &  0.087 \\
BD+46 1590  &  5520 &  $-$0.4   & 1.03 &  4.50 &   $-$0.378 &  120    & 0.082 &  $-$0.385   &   25    &  0.076 \\
BD+06 2398  &  5680 &  $-$0.1   & 1.12 &  4.25 &   $-$0.104 &  119    & 0.089 &  $-$0.098   &   25    &  0.088 \\   
BD+18 2542  &  5420 &  $-$0.3   & 1.30 &  4.40 &   $-$0.257 &   69    & 0.099 &  $-$0.256   &   13    &  0.076 \\
BD+15 2658  &  5450 &  $-$0.1   & 1.02 &  4.30 &   $-$0.098 &   67    & 0.072 &  $-$0.090   &   12    &  0.053 \\
BD+15 4026  &  5410 &  $-$0.4   & 1.15 &  4.60 &   $-$0.403 &   76    & 0.089 &  $-$0.403   &   17    &  0.079 \\
BD+43 4116  &  5620 &  $-$0.2   & 1.30 &  4.55 &   $-$0.214 &   78    & 0.091 &  $-$0.217   &   13    &  0.077 \\

\enddata
\end{deluxetable}

\clearpage

\begin{deluxetable}{lrrrrrrrr}
\tabletypesize{\scriptsize}
\tablenum{7}
\tablewidth{0pc}
\scriptsize
\tablecaption{Equivalent Widths for Thin Disk Stars \label{tab:ew}}
\tablehead{\colhead{$\lambda$(\AA)} & \colhead{$\chi$(eV)} & 
  \colhead{log $gf$} &\colhead{ref} & \colhead{BD+69~238} & 
  \colhead{BD+30~1423} & \colhead{BD+5~1611} & \colhead{BD+46~1590} &
  \colhead{BD+6~2398} }

\startdata
      &        &      &       &      & O I  &    &   \\
\hline
\\
7771.954& 9.14&  0.360&    62&  \nodata &  \nodata &  \nodata &  44.5 & \nodata \\
7774.177& 9.14&  0.210&    62&  \nodata &  \nodata &  \nodata &  36.8 & \nodata \\
7775.395& 9.14& $-0.010$&    62&  11.9&  11.4&  43.2&  28.8 & \nodata \\
\hline
\\
      &        &      &       &      & Na I  &    &   \\
\hline
\\
5682.65 & 2.10 & $-0.890$ & 99 & 149.9 & 122.1 & 89.6 & 77.9 & 92.9 \\
5688.210 & 2.10 & $-0.580$ & 99 & \nodata & \nodata & 112.1 & 110.8 & 114.1 \\
6154.230 & 2.10 & $-1.570$ & 48 & 77.9 & 51.6 & 30.2 & 25.6 & 40.1 \\
6160.753 & 2.10 & $-1.270$ & 48 & 94.7 & 70.7 & 41.2 & 37.8 & 57.1 \\
\\
\enddata
\end{deluxetable}

\clearpage

\begin{deluxetable}{lrrrrrrrrr}
\tablenum{8}
\tabletypesize{\footnotesize}
\tablewidth{0pc}
\tablecaption{Uncertainties in Atmospheric Parameters \label{tab:errors}}

\tablehead{\colhead{Ion} &  \colhead{$\Delta\xi$} & 
  \colhead{$\Delta\xi$} & \colhead{$\Delta T_{\rm eff}$} & 
  \colhead{$\Delta T_{\rm eff}$} &
  \colhead{$\Delta$log g} & \colhead{$\Delta$log g} & 
  \colhead{$\Delta$[M/H]} & \colhead{$\Delta$[M/H]} \\ 
  & \colhead{+0.10} & \colhead{$-$0.10} & 
  \colhead{+80K} & \colhead{$-$80K} & \colhead{+0.05} & 
  \colhead{$-$0.05} & \colhead{+0.1} & 
  \colhead{$-$0.1} & \colhead{$\sigma$}}

\startdata

FeI/H  &$-$0.022&    0.024&    0.042& $-$0.040& $-$0.003&    0.005&    0.016& $-$0.014&  0.051\\
FeII/H &$-$0.016&    0.022& $-$0.032&    0.040&    0.020& $-$0.018&    0.035& $-$0.034&  0.061\\
OI/Fe  &   0.012& $-$0.024& $-$0.158&    0.146&    0.013& $-$0.025& $-$0.016&    0.004&  0.163\\
NaI/Fe &   0.022& $-$0.010& $-$0.003&    0.006& $-$0.005&    0.007& $-$0.009&    0.011&  0.026\\
MgI/Fe &   0.018& $-$0.020& $-$0.026&    0.022& $-$0.005&    0.000& $-$0.008&    0.002&  0.034\\
AlI/Fe &   0.022& $-$0.004& $-$0.014&    0.026&    0.003&    0.005& $-$0.016&    0.019&  0.039\\  
SiI/Fe &   0.018& $-$0.016& $-$0.067&    0.070&    0.008& $-$0.005&    0.004& $-$0.007&  0.073\\
CaI/Fe &   0.006& $-$0.008&    0.005& $-$0.013& $-$0.014&    0.009& $-$0.006&    0.002&  0.021\\
ScII/Fe&   0.012& $-$0.008& $-$0.062&    0.061&    0.023& $-$0.023&    0.017& $-$0.017&  0.070\\
TiI/Fe &   0.004& $-$0.006&    0.034& $-$0.040& $-$0.001&    0.000& $-$0.018&    0.015&  0.044\\
TiII/Fe&   0.002& $-$0.006& $-$0.062&    0.051&    0.022& $-$0.026&    0.015& $-$0.018&  0.070\\
VI/Fe  &   0.014& $-$0.014&    0.043& $-$0.051&    0.000& $-$0.003& $-$0.020&    0.017&  0.057\\
CrI/Fe &  0.001&    0.002&    0.024& $-$0.032& $-$0.007&    0.004& $-$0.008&    0.006&  0.034\\
CrII/Fe&   0.002& $-$0.004& $-$0.085&    0.086&    0.023& $-$0.021&    0.010& $-$0.008&  0.090\\
MnI/Fe &   0.010& $-$0.008&    0.016& $-$0.016& $-$0.002&    0.002& $-$0.006&    0.006&  0.020\\
CoI/Fe &   0.018& $-$0.022& $-$0.014&    0.014&    0.009& $-$0.012& $-$0.003&    0.000&  0.029\\
NiI/Fe &   0.004& $-$0.006& $-$0.026&    0.026&    0.003& $-$0.005&    0.005& $-$0.006&  0.027\\
CuI/Fe &   0.008& $-$0.012& $-$0.010&    0.010&    0.006& $-$0.009&    0.004& $-$0.003&  0.018\\
ZnI/Fe &$-$0.008&    0.006& $-$0.070&    0.073&    0.013& $-$0.015&    0.014& $-$0.016&  0.077\\
SrI/Fe &$-$0.018&    0.016&    0.034& $-$0.038& $-$0.007&    0.005& $-$0.016&    0.014&  0.046\\
SrII/Fe&   0.008& $-$0.018& $-$0.078&    0.074&    0.023& $-$0.025&    0.004& $-$0.016&  0.085\\
YII/Fe &   0.008& $-$0.010& $-$0.054&    0.046&    0.017& $-$0.022&    0.019& $-$0.023&  0.064\\
ZrII/Fe&   0.002& $-$0.004& $-$0.046&    0.050&    0.023& $-$0.025&    0.019& $-$0.016&  0.059\\
BaII/Fe&$-$0.026&    0.016& $-$0.042&    0.030&    0.010& $-$0.018&    0.027& $-$0.033&  0.062\\
LaII/Fe&   0.022& $-$0.020& $-$0.042&    0.042&    0.023& $-$0.025&    0.019& $-$0.016&  0.057\\
CeII/Fe&   0.014& $-$0.024& $-$0.046&    0.038&    0.023& $-$0.025&    0.016& $-$0.020&  0.061\\
NdII/Fe&   0.018& $-$0.018& $-$0.037&    0.038&    0.024& $-$0.026&    0.021& $-$0.018&  0.054\\
EuII/Fe&   0.012& $-$0.014& $-$0.043&    0.034&    0.020& $-$0.030&    0.014& $-$0.019&  0.058\\
\enddata

\end{deluxetable}

\clearpage

\begin{deluxetable}{lrrrr}
\tablenum{9}
\tabletypesize{\footnotesize}
\tablewidth{0pc}
\tablecaption{Derived Solar Abundances \label{tab:solar}}

\tablehead{ \colhead{Ion} &
  \colhead{log n$_{\odot}$(X)} &
  \colhead{[X/H]}  &
  \colhead{$\sigma$} &
  \colhead{N} }

\startdata

Fe I & 7.51 & $-0.006$ & 0.067 & 91 \\
Fe II & 7.51 & $-0.007$ & 0.081 & 22 \\
O I & 8.83 & +0.036 & \nodata & 1 \\
Na I & 6.33 & +0.004 & 0.017 & 4 \\
Mg I & 7.58 & $-0.037$ & 0.080 & 4 \\
Al I & 6.47 & $-0.021$ & 0.043 & 4 \\
Si I & 7.55 & $+0.026$ & 0.035 & 10 \\
Ca I & 6.36 & +0.013 & 0.060 & 14 \\
Sc II & 3.17 & +0.008 & 0.032 & 6 \\
Ti I & 5.02 & $-0.087$ & 0.061 & 49 \\
Ti II & 5.02 & $-0.071$ & 0.092 & 15 \\
V I & 4.00 & $-0.137$ & 0.053 & 16 \\
Cr I & 5.67 & $-0.073$ & 0.065 & 17 \\
Cr II & 5.67 & +0.090 & 0.058 & 5 \\
Mn I & 5.39 & $-0.164$ & 0.071 & 15 \\
Co I & 4.92 & $-0.032$ & 0.087 & 10 \\
Ni I & 6.25 & +0.013 & 0.077 & 35 \\
Cu I & 4.21 & $-0.003$ & 0.072 & 3 \\
Zn I & 4.60 & $-0.074$ & 0.028 & 2 \\
Sr I & 2.97 & $-0.260$ & \nodata & 1 \\
Sr II & 2.97 & 0.000 & \nodata & 1 \\
Y II & 2.24 & $-0.070$ & 0.052 & 10 \\
Zr II & 2.60 & $-0.029$ & 0.035 & 2 \\
Ba II & 2.13 & +0.023 & 0.021 & 3 \\
La II & 1.17 & $-0.069$ & 0.095 & 9 \\
Ce II & 1.58 & $-0.018$ & 0.076 & 8 \\
Nd II & 1.50 & $-0.054$ & 0.097 & 18 \\
Eu II & 0.51 & +0.068 & 0.055 & 4 \\
\enddata

\end{deluxetable}

\clearpage

\begin{deluxetable}{lrrrrrrrrrrrrrrr}
\tablenum{10}
\tabletypesize{\scriptsize}
\tablewidth{0pt}
\scriptsize
\tablecaption{Abundances - Thin Disk Stars \label{tab:thindisk1}}
\tablehead{ &  \colhead{BD+69} & & & \colhead{BD+30} & & & \colhead{BD+5} 
 & & & \colhead{BD+46} & & & \colhead{BD+6} \\
 & \colhead{238} & \colhead{$\sigma$} & \colhead{N} &
 \colhead{1423} & \colhead{$\sigma$} & \colhead{N} &
 \colhead{1611} & \colhead{$\sigma$} & \colhead{N} &
 \colhead{1590} & \colhead{$\sigma$} & \colhead{N} &
 \colhead{2398} & \colhead{$\sigma$} & \colhead{N} }
\startdata
             
$[$O/Fe$]$ & 0.058 & \nodata & 1 & 0.260 & \nodata & 1 & 0.192 & \nodata & 1 & 0.214 & 0.010 & 3 & 0.160 & 0.017 & 3 \\
$[$Na/Fe$]$ & 0.021 & 0.035 & 3 & 0.071 & 0.036 & 3 & $-0.025$ & 0.057 & 4 & 0.014 & 0.055 & 4 & 0.080 & 0.034 & 4 \\
$[$Mg/Fe$]$ & $-0.066$ & 0.030 & 5 & 0.106 & 0.076 & 6 & 0.079 & 0.051 & 5 & 0.024 & 0.040 & 5 & 0.070 & 0.038 & 5 \\
$[$Al/Fe$]$ & $-0.027$ & 0.064 & 2 & 0.056 & 0.049 & 2 & 0.017 & 0.092 & 2 & 0.029 & 0.007 & 2 & 0.060 & 0.057 & 2 \\
$[$Si/Fe$]$ & 0.011 & 0.048 & 9 & 0.114 & 0.047 & 12 & 0.097 & 0.032 & 11 & 0.114 & 0.049 & 11 & 0.204 & 0.062 & 13 \\
$[$Ca/Fe$]$ & $-0.009$ & 0.070 & 8 & 0.016 & 0.066 & 14 & 0.034 & 0.083 & 17 & 0.036 & 0.083 & 18 & $-0.001$ & 0.075 & 18 \\
$[$Sc/Fe$]$ & $-0.089$ & 0.075 & 8 & 0.065 & 0.076 & 9 & $-0.053$ & 0.060 & 8 & 0.129 & 0.084 & 10 & 0.039 & 0.065 & 10 \\
$[$TiI/Fe$]$& $-0.103$ & 0.107 & 46 & 0.066 & 0.098 & 42 & $-0.113$ & 0.082 & 40 & $-0.029$ & 0.091 & 56 & $-0.097$ & 0.070 & 56 \\
$[$TiII/Fe$]$& $-0.125$ & 0.077 & 12 & 0.012 & 0.054 & 13 & $-0.019$ & 0.071 & 12 & 0.081 & 0.082 & 20 & 0.012 & 0.091 & 19 \\ 
$[$V/Fe$]$ & $-0.090$ & 0.102 & 16 & 0.039 & 0.087 & 15 & $-0.130$ & 0.090 & 16 & $-0.076$ & 0.090 & 18 & $-0.110$ & 0.077 & 18 \\
$[$CrI/Fe$]$ & $-0.040$ & 0.120 & 12 & 0.004 & 0.102 & 15 & 0.016 & 0.084 & 18 & 0.047 & 0.092 & 19 & 0.081 & 0.090 & 20 \\
$[$CrII/Fe$]$& 0.084 & 0.059 & 5 & 0.055 & 0.081 & 5 & 0.204 & 0.055 & 5 & 0.164 & 0.026 & 5 & 0.120 & 0.061 & 4 \\
$[$Mn/Fe$]$ & $-0.173$ & 0.072 & 10 & $-0.180$ & 0.120 & 13 & $-0.181$ & 0.082 & 15 & $-0.123$ & 0.064 & 16 & $-0.033$ & 0.075 & 15 \\
$[$FeI/H$]$ & $-0.012$ & 0.099 & 73 & $-0.385$ & 0.094 & 74 & $-0.156$ & 0.073 & 69 & $-0.378$ & 0.082 & 120 & $-0.104$ & 0.089 & 113 \\
$[$FeII/H$]$& $-0.016$ & 0.081 & 13 & $-0.390$ & 0.079 & 17 & $-0.159$ & 0.087 & 17 & $-0.385$ & 0.076 & 25 & $-0.098$ & 0.088 & 25 \\
$[$Co/Fe$]$ & $-0.053$ & 0.040 & 10 & 0.010 & 0.069 & 10 & $-0.092$ & 0.079 & 11 & $-0.103$ & 0.089 & 11 & 0.008 & 0.070 & 12 \\
$[$Ni/Fe$]$ & 0.022 & 0.081 & 34 & 0.034 & 0.093 & 38 & $-0.019$ & 0.094 & 38 & 0.019 & 0.077 & 37 & 0.084 & 0.086 & 28 \\
$[$Cu/Fe$]$ & $-0.023$ & 0.049 & 2 & 0.005 & 0.071 & 2 & $-0.157$ & 0.059 & 3 & $-0.119$ & 0.010 & 4 & 0.006 & 0.054 & 4 \\
$[$Zn/Fe$]$ & 0.043 & 0.035 & 2 & $-0.044$ & 0.021 & 2 & $-0.018$ & 0.014 & 2 & 0.109 & 0.021 & 2 & 0.095 & 0.007 & 2 \\
$[$SrII/Fe$]$ & 0.002 & \nodata & 1 & 0.195 & \nodata & 1 & 0.206 & \nodata & 1 & 0.078 & \nodata & 1 & 0.054 & \nodata & 1 \\
$[$Y/Fe$]$ & $-0.112$ & 0.078 & 8 & $-0.149$ & 0.082 & 6 & $-0.007$ & 0.062 & 8 & 0.021 & 0.059 & 10 & $-0.042$ & 0.065 & 10 \\
$[$Zr/Fe$]$ & 0.018 & 0.060 & 2 & 0.026 & 0.064 & 2 & $-0.008$ & 0.028 & 2 & 0.104 & 0.028 & 2 & 0.040 & 0.028 & 2 \\
$[$Ba/Fe$]$ & 0.012 & 0.060 & 3 & 0.105 & 0.056 & 3 & 0.253 & 0.032 & 3 & 0.268 & 0.020 & 3 & 0.031 & 0.023 & 3 \\
$[$La/Fe$]$ & $-0.209$ & 0.073 & 9 & 0.037 & 0.112 & 9 & 0.076 & 0.072 & 9 & 0.229 & 0.073 & 9 & $-0.013$ & 0.094 & 9 \\
$[$Ce/Fe$]$ & 0.041 & 0.088 & 10 & 0.160 & 0.090 & 7 & 0.086 & 0.094 & 8 & 0.043 & 0.054 & 10 & 0.070 & 0.086 & 10 \\
$[$Nd/Fe$]$ & $-0.155$ & 0.097 & 9 & 0.103 & 0.095 & 10 & 0.140 & 0.083 & 9 & 0.260 & 0.097 & 14 & $-0.033$ & 0.095 & 14 \\
$[$Eu/Fe$]$ & 0.092 & 0.039 & 4 & 0.198 & 0.049 & 4 & 0.124 & 0.056 & 4 & 0.201 & 0.052 & 4 & 0.042 & 0.043 & 4 \\
\enddata

\end{deluxetable}

\clearpage

\begin{deluxetable}{lrrrrrrrrrrrrrrr}
\tablenum{10}
\tabletypesize{\scriptsize}
\tablewidth{0pt}
\scriptsize
\tablecaption{Abundances - Thin Disk Stars \label{tab:thindisk2}}
\tablehead{ &  \colhead{BD+18} & & & \colhead{BD+15} & & & \colhead{BD+15} 
 & & & \colhead{BD+43}  \\
 & \colhead{2542} & \colhead{$\sigma$} & \colhead{N} &
 \colhead{2658} & \colhead{$\sigma$} & \colhead{N} &
 \colhead{4026} & \colhead{$\sigma$} & \colhead{N} &
 \colhead{4116} & \colhead{$\sigma$} & \colhead{N} }
\startdata
             
$[$O/Fe$]$  & \nodata & \nodata & \nodata & \nodata & \nodata & \nodata & 0.132 & 0.057 & 3 & 0.150 & 0.046 & 3 \\
$[$Na/Fe$]$ & 0.083 & 0.014 & 4 & 0.009 & 0.048 & 4 & 0.042 & 0.053 & 4 & $-0.038$ & 0.051 & 4 \\
$[$Mg/Fe$]$ & 0.043 & 0.056 & 4 & $-0.014$ & 0.051 & 5 & $-0.161$ & 0.048 & 5 & $-0.026$ & 0.057 & 5 \\
$[$Al/Fe$]$ & 0.103 & \nodata & 1 & $-0.066$ & 0.014 & 2 & 0.154 & 0.049 & 2 & $-0.030$ & 0.028 & 2 \\
$[$Si/Fe$]$ & 0.152 & 0.059 & 11 & 0.124 & 0.036 & 11 & 0.127 & 0.045 & 11 & 0.101 & 0.055 & 11 \\
$[$Ca/Fe$]$ & 0.016 & 0.060 & 14 & 0.012 & 0.082 & 18 & 0.142 & 0.073 & 16 & 0.089 & 0.080 & 16 \\
$[$Sc/Fe$]$ & 0.051 & 0.083 & 9 & $-0.053$ & 0.060 & 9 & 0.110 & 0.074 & 9 & $-0.016$ & 0.089 & 8 \\
$[$TiI/Fe$]$ & 0.011 & 0.098 & 41 & $-0.168$ & 0.087 & 43 & 0.114 & 0.079 & 43 & 0.010 & 0.088 & 39 \\
$[$TiII/Fe$]$& 0.050 & 0.084 & 8 & $-0.085$ & 0.078 & 11 & 0.090 & 0.088 & 14 & 0.077 & 0.098 & 14 \\ 
$[$V/Fe$]$ &  0.026 & 0.101 & 15 & $-0.083$ & 0.066 & 14 & 0.033 & 0.077 & 14 & $-0.065$ & 0.086 & 15 \\
$[$CrI/Fe$]$ & 0.009 & 0.071 & 16 & $-0.038$ & 0.092 & 18 & $-0.017$ & 0.069 & 19 & 0.004 & 0.066 & 19 \\
$[$CrII/Fe$]$& 0.217 & 0.091 & 5 & 0.104 & 0.057 & 5 & 0.059 & 0.064 & 5 & 0.106 & 0.067 & 5 \\
$[$Mn/Fe$]$ & $-0.082$ & 0.079 & 13 & $-0.143$ & 0.094 & 14 & $-0.155$ & 0.056 & 13 & $-0.132$ & 0.089 & 13 \\
$[$FeI/H$]$ & $-0.257$ & 0.099 & 69 & $-0.098$ & 0.072 & 67 & $-0.403$ & 0.089 & 76 & $-0.214$ & 0.091 & 78 \\
$[$FeII/H$]$ & $-0.256$ & 0.076 & 13 & $-0.090$ & 0.053 & 12 & $-0.403$ & 0.079 & 17 & $-0.217$ & 0.077 & 13 \\
$[$Co/Fe$]$ & $-0.049$ & 0.072 & 8 & $-0.104$ & 0.064 & 10 & 0.036 & 0.070 & 11 & $-0.080$ & 0.089 & 10 \\
$[$Ni/Fe$]$ & $-0.023$ & 0.099 & 33 & 0.020 & 0.082 & 35 & $-0.011$ & 0.082 & 35 & $-0.106$ & 0.092 & 37 \\
$[$Cu/Fe$]$ & $-0.020$ & 0.081 & 3 & $-0.025$ & 0.075 & 3 & $-0.002$ & 0.058 & 4 & $-0.166$ & 0.053 & 4 \\
$[$Zn/Fe$]$ & $-0.062$ & 0.035 & 2 & $-0.041$ & 0.021 & 2 & $-0.066$ & 0.021 & 2 & $-0.075$ & 0.021 & 2 \\
$[$SrII/Fe$]$ & $-0.033$ & \nodata & 1 & 0.068 & \nodata & 1 & $-0.017$ & \nodata & 1 & 0.074 & \nodata & 1 \\
$[$Y/Fe$]$ & $-0.094$ & 0.057 & 7 & $-0.058$ & 0.067 & 9 & $-0.069$ & 0.090 & 8 & $-0.049$ & 0.075 & 8 \\
$[$Zr/Fe$]$ & $-0.022$ & 0.035 & 2 & 0.014 & 0.042 & 2 & 0.029 & 0.057 & 6 & 0.015 & 0.021 & 2 \\
$[$Ba/Fe$]$ & $-0.026$ & 0.060 & 3 & 0.011 & 0.021 & 3 & 0.036 & 0.035 & 3 & 0.247 & 0.064 & 3 \\
$[$La/Fe$]$ & $-0.051$ & 0.072 & 9 & $-0.168$ & 0.104 & 9 & 0.103 & 0.082 & 9 & 0.121 & 0.085 & 9 \\
$[$Ce/Fe$]$ & $-0.020$ & 0.039 & 7 & $-0.030$ & 0.042 & 7 & 0.136 & 0.07 & 8 & 0.169 & 0.079 & 9 \\
$[$Nd/Fe$]$ & $-0.110$ & 0.084 & 8 & $-0.104$ & 0.089 & 9 & 0.153 & 0.103 & 11 & 0.172 & 0.081 & 10 \\
$[$Eu/Fe$]$ & 0.130 & 0.050 & 4 & 0.051 & 0.069 & 4 & 0.223 & 0.047 & 4 & 0.154 & 0.054 & 4 \\
\enddata
\end{deluxetable}

\clearpage

\begin{deluxetable}{lrrrrrrrrrrrrrrr}
\tablenum{11}
\tabletypesize{\scriptsize}
\tablewidth{0pt}
\scriptsize
\tablecaption{Abundances - Thick Disk Stars \label{tab:thickdisk1}}
\tablehead{ &  \colhead{BD+65} & & & \colhead{BD+34} & & & \colhead{BD+17} 
 & & & \colhead{BD+1} & & & \colhead{BD+13} \\
 & \colhead{3} & \colhead{$\sigma$} & \colhead{N} &
 \colhead{927} & \colhead{$\sigma$} & \colhead{N} &
 \colhead{1145} & \colhead{$\sigma$} & \colhead{N} &
 \colhead{1600} & \colhead{$\sigma$} & \colhead{N} &
 \colhead{1655} & \colhead{$\sigma$} & \colhead{N} }
\startdata
             
$[$O/Fe$]$  &   \nodata  & \nodata & \nodata & 0.475 & 0.061 & 3 & 0.318 & 0.057 & 2 & 0.048 & 0.078 & 2 &  \nodata & \nodata & \nodata \\
$[$Na/Fe$]$ &  0.082 & 0.021 & 2 & 0.075 & 0.037 & 4 & 0.090 & 0.031 & 4 & 0.043 & 0.054 & 4 & 0.109 & 0.035 & 2 \\
$[$Mg/Fe$]$ & 0.024 & 0.053 & 4 & 0.375 & 0.058 & 6 & 0.190 & 0.092 & 6 & 0.070 & 0.033 & 5 & 0.056 & 0.055 & 5 \\
$[$Al/Fe$]$ & 0.310 & 0.035 & 2 & 0.280 & 0.021 & 2 & 0.198 & 0.057 & 2 & 0.073 & 0.035 & 2 & 0.149 & 0.064 & 2 \\
$[$Si/Fe$]$ & 0.131 & 0.065 & 8 & 0.268 & 0.053 & 12 & 0.177 & 0.033 & 12 & 0.176 & 0.042 & 10 & 0.215 & 0.066 & 11 \\
$[$Ca/Fe$]$ & 0.071 & 0.063 & 8 & 0.172 & 0.075 & 15 & 0.137 & 0.065 & 17 & $-0.046$ & 0.089 & 16 & 0.074 & 0.077 & 5 \\
$[$Sc/Fe$]$ & 0.032 & 0.093 & 9 & 0.217 & 0.077 & 9 & 0.164 & 0.068 & 9 & 0.091 & 0.053 & 8 & 0.108 & 0.079 & 8 \\
$[$TiI/Fe$]$ & 0.016 & 0.103 & 41 & 0.251 & 0.093 & 51 & 0.133 & 0.086 & 50 & $-0.043$ & 0.092 & 53 & 0.118 & 0.083 & 30 \\
$[$TiII/Fe$]$& $-$0.058 & 0.098 & 9 & 0.314 & 0.072 & 13 & 0.150 & 0.042 & 13 & $-$0.051 & 0.082 & 14 & 0.039 & 0.105 & 8  \\ 
$[$V/Fe$]$ & $-0.047$ & 0.100 & 14 & 0.118 & 0.088 & 17 & $-0.015$ & 0.062 & 16 & $-0.140$ & 0.067 & 16 & $-$0.005 & 0.107 & 14 \\
$[$CrI/Fe$]$ & $-0.047$ & 0.095 & 14 & $-$0.002 & 0.076 & 17 & 0.049 & 0.076 & 19 & $-$0.008 & 0.091 & 17 &$-$0.024 & 0.088 & 11 \\
$[$CrII/Fe$]$& 0.160 & 0.074 & 4 & 0.193 & 0.058 & 5 & 0.110 & 0.059 & 5 & 0.110 & 0.068 & 5 & 0.116 & 0.057 & 5  \\
$[$Mn/Fe$]$ & $-$0.052 & 0.100 & 9 & $-$0.240 & 0.091 & 16 & $-0.108$ & 0.088 & 16 & $-$0.119 & 0.082 & 14 & $-$0.046 & 0.079 & 9 \\
$[$FeI/H$]$ & 0.004 & 0.107 & 39 & $-$0.379 & 0.089 & 78 & $-$0.236 & 0.085 & 85 & $-$0.092 & 0.080 & 73 & $-$0.038 & 0.090 & 43 \\
$[$FeII/H$]$ & 0.012 & 0.105 & 10 & $-$0.375 & 0.097 & 14 & $-$0.235 & 0.075 & 19 & $-$0.096 & 0.094 & 17 & $-$0.047 & 0.050 & 11 \\
$[$Co/Fe$]$ & $-$0.032 & 0.073 & 10 & 0.118 & 0.053 & 11 & 0.078 & 0.044 & 11 & $-$0.060 & 0.053 & 11 & 0.015 & 0.068 & 10 \\
$[$Ni/Fe$]$ & 0.027 & 0.103 & 31 & 0.038 & 0.095 & 38 & 0.065 & 0.083 & 38 & $-$0.002 & 0.081 & 38 & 0.065 & 0.108 & 29 \\
$[$Cu/Fe$]$ & 0.016 & 0.028 & 2 & 0.076 & 0.080 & 3 & 0.099 & 0.072 & 3 & 0.005 & 0.035 & 3 & $-$0.010 & 0.028 & 2 \\
$[$Zn/Fe$]$ & $-$0.013 & 0.035 & 2 & 0.180 & 0.064 & 2 & 0.073 & 0.049 & 2 & 0.003 & 0.064 & 2 & 0.079 & 0.078 & 2 \\
$[$SrII/Fe$]$ & 0.086 & \nodata & 1 & $-0.171$ & \nodata & 1 & 0.012 & \nodata & 1 & $-0.218$ & \nodata & 1 & 0.198 & \nodata & 1 \\
$[$Y/Fe$]$ & $-$0.024 & 0.085 & 7 & $-$0.079 & 0.065 & 8 & $-$0.040 & 0.052 & 9 & $-$0.065 & 0.046 & 8 & 0.014 & 0.093 & 7 \\
$[$Zr/Fe$]$ & 0.032 & 0.071 & 2 & 0.075 & 0.028 & 2 & $-$0.012 & 0.057 & 2 & $-$0.022 & 0.028 & 2 & $-$0.036 & 0.057 & 2 \\
$[$Ba/Fe$]$ & 0.016 & 0.044 & 3 & $-$0.098 & 0.049 & 3 & 0.029 & 0.075 & 3 & $-$0.048 & 0.052 & 3 & $-$0.009 & 0.085 & 3 \\
$[$La/Fe$]$ & $-$0.187 & 0.094 & 8 & 0.002 & 0.112 & 9 & $-$0.028 & 0.076 & 9 & $-$0.094 & 0.094 & 9 & $-$0.181 & 0.102 & 9 \\
$[$Ce/Fe$]$ & 0.073 & 0.078 & 8 & 0.047 & 0.091 & 10 & 0.017 & 0.064 & 10 & 0.077 & 0.063 & 9 & 0.111 & 0.071 & 7 \\
$[$Nd/Fe$]$ & $-$0.172 & 0.123 & 7 & 0.054 & 0.087 & 12 & 0.020 & 0.092 & 11 & $-$0.101 & 0.108 & 14 & $-$0.005 & 0.126 & 9 \\
$[$Eu/Fe$]$ & 0.031 & 0.026 & 4 & 0.347 & 0.068 & 4 & 0.206 & 0.043 & 4 & 0.103 & 0.050 & 4 & 0.060 & 0.046 & 3 \\
\enddata
\end{deluxetable}

\clearpage

\begin{deluxetable}{lrrrrrrrrrrrrrrr}
\tablenum{11}
\tabletypesize{\scriptsize}
\tablewidth{0pt}
\scriptsize
\tablecaption{Abundances - Thick Disk Stars (continued) \label{tab:thickdisk2}}
\tablehead{ &  \colhead{BD+4} & & & \colhead{BD+11} & & & \colhead{BD+2} 
 & & & \colhead{BD+9} & & & \colhead{BD+23} \\
 & \colhead{2696} & \colhead{$\sigma$} & \colhead{N} &
 \colhead{2439} & \colhead{$\sigma$} & \colhead{N} &
 \colhead{2585} & \colhead{$\sigma$} & \colhead{N} &
 \colhead{2736} & \colhead{$\sigma$} & \colhead{N} &
 \colhead{2747} & \colhead{$\sigma$} & \colhead{N} }
\startdata
             
$[$O/Fe$]$  & \nodata & \nodata & \nodata & \nodata & \nodata & \nodata & 0.565 & 0.035 & 3 & \nodata &  \nodata & \nodata & \nodata & \nodata \\
$[$Na/Fe$]$ & 0.002 & 0.038 & 4 & 0.122 & 0.049 & 4 & 0.082 & 0.049 & 4 & 0.004 & 0.044 & 4 & 0.019 & 0.052 & 4 \\
$[$Mg/Fe$]$ & 0.041 & 0.054 & 5 & 0.146 & 0.038 & 5 & 0.240 & 0.061 & 4 & 0.119 & 0.049 & 5 & 0.073 & 0.043 & 5 \\
$[$Al/Fe$]$ & $-0.021$ & 0.014 & 2 & 0.195 & 0.035 & 2 & 0.310 & 0.049 & 2 & 0.087 & 0.014 & 2 & 0.079 & 0.071 & 2 \\
$[$Si/Fe$]$ & 0.086 & 0.041 & 10 & 0.194 & 0.049 & 11 & 0.211 & 0.055 & 14 & 0.108 & 0.044 & 11 & 0.149 & 0.053 & 14 \\
$[$Ca/Fe$]$ & 0.073 & 0.069 & 16 & 0.169 & 0.056 & 16 & 0.156 & 0.057 & 17 & 0.104 & 0.058 & 16 & 0.116 & 0.062 & 17 \\
$[$Sc/Fe$]$ & 0.076 & 0.076 & 9 & 0.111 & 0.086 & 9 & 0.185 & 0.066 & 10 & 0.161 & 0.077 & 9 & 0.091 & 0.047 \\
$[$TiI/Fe$]$ & 0.015 & 0.093 & 43 & 0.090 & 0.080 & 42 & 0.144 & 0.065 & 57 & 0.068 & 0.097 & 50 & 0.013 & 0.071 & 39 \\
$[$TiII/Fe$]$& 0.045 & 0.075 & 13 & 0.165 & 0.063 & 13 & 0.231 & 0.059 & 20 & 0.160 & 0.066 & 14 & 0.152 & 0.076 & 12 \\ 
$[$V/Fe$]$ & $-0.075$ & 0.067 & 15 & $-0.041$ & 0.077 & 14 & 0.079 & 0.085 & 18 & $-0.018$ & 0.069 & 15 & $-0.094$ & 0.084 & 15 \\
$[$CrI/Fe$]$ & 0.058 & 0.097 & 18 & 0.079 & 0.086 & 20 & 0.046 & 0.080 & 18 & 0.014 & 0.083 & 17 & 0.005 & 0.076 & 20 \\
$[$CrII/Fe$]$& 0.065 & 0.009 & 5 & 0.128 & 0.051 & 5 & 0.135 & 0.031 & 5 & 0.069 & 0.019 & 5 & 0.137 & 0.068 & 5 \\
$[$Mn/Fe$]$ & $-0.182$ & 0.067 & 15 & $-0.126$ & 0.041 & 15 & $-0.159$ & 0.082 & 16 & $-0.122$ & 0.073 & 15 & $-0.200$ & 0.084 & 15 \\
$[$FeI/H$]$ & $-0.243$ & 0.087 & 75 & $-0.204$ & 0.081 & 62 & $-0.659$ & 0.072 & 103 & $-0.171$ & 0.086 & 72 & $-0.173$ & 0.073 & 70 \\
$[$FeII/H$]$ & $-0.243$ & 0.075 & 17 & $-0.198$ & 0.094 & 17 & $-0.664$  & 0.072 & 24 & $-0.172$ & 0.088 & 14 & $-0.172$ & 0.064 & 15 \\
$[$Co/Fe$]$ & $-0.083$ & 0.070 & 10 & 0.003 & 0.081 & 10 & 0.123 & 0.096 & 11 & $-0.030$ & 0.058 & 11 & $-0.039$ & 0.065 & 11 \\
$[$Ni/Fe$]$ & $-0.082$ & 0.094 & 35 & 0.030 & 0.091 & 35 & 0.059 & 0.082 & 38 & $-0.025$ & 0.093 & 34 & $-0.034$ & 0.072 & 35 \\
$[$Cu/Fe$]$ & $-0.107$ & 0.036 & 3 & 0.077 & 0.015 & 3 & 0.064 & 0.072 & 4 & $-0.012$ & 0.055 & 3 & 0.096 & 0.080 & 3 \\
$[$Zn/Fe$]$ & $-0.056$ & 0.021 & 2 & 0.110 & 0.042 & 2 & 0.150 & 0.007 & 2 & 0.007 & 0.057 & 2 & 0.004 & 0.049 & 2 \\
$[$SrII/Fe$]$ & $-0.087$ & \nodata & 1 & 0.004 & \nodata & 1 & $-0.161$ & \nodata & 1 & 0.111 & \nodata & 1 & $-0.067$ & \nodata & 1 \\
$[$Y/Fe$]$ & $-0.184$ & 0.059 & 9 & $-0.110$ & 0.034 & 8 & $-0.098$ & 0.086 & 10 & $-0.069$ & 0.058 & 9 & $-0.125$ & 0.049 & 8 \\
$[$Zr/Fe$]$ & $-0.011$ & 0.000 & 2 & 0.065 & 0.007 & 2 & 0.035 & 0.028 & 2 & $-0.063$ & 0.042 & 2 & $-0.056$ & 0.021 & 2 \\
$[$Ba/Fe$]$ & $-0.090$ & 0.047 & 3 & 0.087 & 0.068 & 3 & $-0.008$ & 0.021 & 3 & 0.011 & 0.044 & 3 & 0.023 & 0.044 & 3 \\
$[$La/Fe$]$ & $-0.069$ & 0.090 & 9 & $-0.008$ & 0.073 & 9 & 0.086 & 0.101 & 9 & 0.065 & 0.060 & 9 & $-0.049$ & 0.074 & 9 \\
$[$Ce/Fe$]$ & 0.004 & 0.077 & 8 & 0.032 & 0.069 & 9 & 0.095 & 0.076 & 10 & 0.096 & 0.085 & 9 & $-0.015$ & 0.071 & 7 \\
$[$Nd/Fe$]$ & $-0.119$ & 0.080 & 11 & 0.035 & 0.092 & 13 & 0.109 & 0.121 & 10 & $-0.005$ & 0.092 & 13 & $-0.081$ & 0.076 & 9 \\
$[$Eu/Fe$]$ & 0.143 & 0.055 & 4 & 0.274 & 0.036 & 4 & 0.329 & 0.055 & 4 & 0.334 & 0.022 & 4 & 0.208 & 0.047 & 4 \\
\enddata
\end{deluxetable}

\clearpage

\begin{deluxetable}{lrrrrrrrrrrrrrrr}
\tablenum{11}
\tabletypesize{\scriptsize}
\tablecaption{Abundances - Thick Disk Stars (continued) \label{tab:thickdisk3}}
\tablehead{ &  \colhead{BD+26} & & & \colhead{BD+47} & & & \colhead{BD+45} 
 & & & \colhead{BD+40}  \\
 & \colhead{2677} & \colhead{$\sigma$} & \colhead{N} &
 \colhead{2491} & \colhead{$\sigma$} & \colhead{N} &
 \colhead{2684} & \colhead{$\sigma$} & \colhead{N} &
 \colhead{4912} & \colhead{$\sigma$} & \colhead{N} }
\startdata
             
\tablewidth{0pt}
\scriptsize
$[$O/Fe$]$  & \nodata & \nodata & \nodata & 0.620 & 0.038 & 3 & 0.059 & 0.015 & 3 &  \nodata & \nodata & \nodata \\
$[$Na/Fe$]$ & 0.115 & 0.057 & 4 & 0.088 & 0.040 & 4 & 0.053 & 0.036 & 4 & 0.147 & 0.040 & 4 \\
$[$Mg/Fe$]$ & 0.206 & 0.044 & 5 & 0.289 & 0.029 & 5 & 0.066 & 0.060 & 5 & 0.215 & 0.071 & 4 \\
$[$Al/Fe$]$ & 0.242 & 0.014 & 2 & 0.278 & 0.035 & 2 & 0.036 & 0.014 & 2 & 0.210 & 0.007 & 2\\
$[$Si/Fe$]$ & 0.181 & 0.039 & 11 & 0.308 & 0.058 & 10 & 0.094 & 0.053 & 11 & 0.234 & 0.034 & 12 \\
$[$Ca/Fe$]$ & 0.138 & 0.049 & 15 & 0.190 & 0.075 & 16 & 0.064 & 0.082 & 17 & 0.223 & 0.070 & 16 \\
$[$Sc/Fe$]$ & 0.177 & 0.060 & 8 & 0.169 & 0.070 & 9 & 0.056 & 0.063 & 9 & 0.177 & 0.077 & 9 \\
$[$TiI/Fe$]$ & 0.161 & 0.089 & 40 & 0.186 & 0.100 & 44 & $-0.004$ & 0.106 & 41 & 0.151 & 0.087 & 44 \\
$[$TiII/Fe$]$& 0.163 & 0.092 & 11 & 0.216 & 0.091 & 11 & $-0.012$ & 0.074 & 11 & 0.177 & 0.083 & 11 \\ 
$[$V/Fe$]$ & 0.043 & 0.073 & 15 & 0.044 & 0.080 & 15 & $-0.119$ & 0.101 & 14 & 0.047 & 0.077 & 9 \\
$[$CrI/Fe$]$ & 0.000 & 0.079 & 18 & $-0.042$ & 0.065 & 18 & 0.009 & 0.086 & 19 & 0.036 & 0.082 & 20 \\
$[$CrII/Fe$]$& 0.050 & 0.052 & 5 & 0.099 & 0.018 & 5 & 0.156 & 0.070 & 5 & 0.231 & 0.060 & 5 \\
$[$Mn/Fe$]$  &$-0.135$ & 0.091 & 13 & $-0.180$ & 0.080 & 14 & $-0.006$ & 0.070 & 15 & $-0.133$ & 0.070 & 15 \\
$[$FeI/H$]$  & $-0.246$ & 0.089 & 65 & $-0.507$ & 0.095 & 68 & $-0.070$ & 0.078 & 70 & $-0.249$ & 0.082 & 74 \\
$[$FeII/H$]$ & $-0.254$ & 0.090 & 14 & $-0.500$ & 0.095 & 14 & $-0.060$ & 0.038 & 15 & $-0.257$ & 0.081 & 14 \\
$[$Co/Fe$]$  & 0.062 & 0.058 & 11 & 0.123 & 0.056 & 10 & $-0.061$ & 0.058 & 11 & 0.096 & 0.067 & 10 \\
$[$Ni/Fe$]$ & 0.013 & 0.085 & 34 & 0.022 & 0.078 & 37 & 0.037 & 0.086 & 34 & 0.061 & 0.081 & 33 \\
$[$Cu/Fe$]$  & 0.146 & 0.069 & 3 & 0.054 & 0.050 & 3 & $-0.003$ & 0.039 & 4 & 0.092 & 0.050 & 3 \\
$[$Zn/Fe$]$  & 0.082 & 0.071 & 2 & 0.193 & 0.014 & 2 & 0.046 & 0.028 & 2 & 0.035 & 0.028 & 2 \\
$[$SrII/Fe$]$ & $-0.084$ & \nodata & 1 & 0.367 & \nodata & 1 & $-0.050$ & \nodata & 1 & 0.049 & \nodata & 1 \\
$[$Y/Fe$]$ & $-0.039$ & 0.053 & 9 & 0.211 & 0.082 & 9 & $-0.033$ & 0.051 & 8 & $-0.054$ & 0.083 & 8 \\
$[$Zr/Fe$]$ & $-0.068$ & 0.042 & 2 & 0.318 & 0.049 & 2 & $-0.069$ & 0.021 & 2 & 0.015 & 0.014 & 2 \\
$[$Ba/Fe$]$ & $-0.064$ & 0.026 & 3 & 0.254 & 0.065 & 3 & $-0.030$ & 0.026 & 3 & $-0.004$ & 0.064 & 3 \\
$[$La/Fe$]$ & $-0.025$ & 0.115 & 9 & 0.190 & 0.074 & 9 & $-0.178$ & 0.123 & 9 & $-0.072$ & 0.081 & 9 \\
$[$Ce/Fe$]$ & 0.041 & 0.085 & 8 & 0.310 & 0.073 & 8 & $-0.044$ & 0.088 & 8  & $-0.028$ & 0.058 & 7 \\
$[$Nd/Fe$]$ & $-0.081$ & 0.098 & 11 & 0.150 & 0.091 & 10 & $-0.063$ & 0.064 & 8 & $-0.031$ & 0.095 & 10 \\
$[$Eu/Fe$]$ & 0.209 & 0.053 & 4 & 0.347 & 0.051 & 4 & $-0.027$ & 0.035 & 3 & 0.289 & 0.063 & 4 \\
\enddata
\end{deluxetable}

{}

\clearpage

\clearpage

\clearpage

\clearpage

\clearpage

\clearpage

\clearpage

\clearpage

\clearpage 

\clearpage

\clearpage

\clearpage

\clearpage

\clearpage

\clearpage

\clearpage

\clearpage

\clearpage

\clearpage

\clearpage

\clearpage

\begin{thebibliography}{}

\bibitem[\protect\citeauthoryear{{Alonso}, {Arribas}, \&
  {Martinez-Roger}}{{Alonso} et~al.}{1996}]{1996A&A...313..873A}
{Alonso}, A., {Arribas}, S.,  \& {Martinez-Roger}, C.\ 1996, A\&A, 313, 873

\bibitem[\protect\citeauthoryear{{Anders} \& {Grevesse}}{{Anders} \&
  {Grevesse}}{1989}]{1989GeCoA..53..197A}
{Anders}, E.,  \& {Grevesse}, N.\ 1989, Geochim.\ Cosmochim.\ Acta, 53, 197

\bibitem[Aoki et~al.\ (2001)]{Aoki01}
Aoki, W., Ryan, S.\ G., Norris, J.\ E., Beers, T.\ C., Ando, H.,
Iwamoto, N., Kajino, T., Mathews, G.\ J., \& Fujimoto, M.\ Y.\ 2001, \apj, 561, 346

\bibitem[Armandroff (1989)]{armandroff89} Armandroff, T.\ E.\ 1989, \aj, 97, 375

\bibitem[Asiain et al.\ (1999)]{asiain99} Asiain, R., Figueras, F., \& Torra, J.\
1999, \aap, 350, 434

\bibitem[Asplund (2004)]{asplund2004} Asplund, M.\ 2004, \aap, 417, 751

\bibitem[Bensby et al.\ (2003)]{bensy03} Bensby, T., Feltzing, S., \&
Lundstr\"{o}m, I.\ 2003, \aap, 410, 527

\bibitem[Bensby et al.\ (2004a)]{bensy04a} Bensby, T., Feltzing, S., \&
Lundstr\"{o}m, I.\ 2004a, \aap, 415, 155

\bibitem[Bensby et al.\ (2004b)]{bensy04b} Bensby, T., Feltzing, S., \&
Lundstr\"{o}m, I.\ 2004b, \aap, 421, 969

\bibitem[Bensby et al.\ (2005)]{bensy05} Bensby, T., Feltzing, S., 
Lundstr\"{o}m, I., \& Ilyin, I.\  2005, \aap, 433, 185

\bibitem[Brook et al.\ (2005)]{brook2005} Brook, C.\ B., Gibson, B.\ K.,
Martel, H., \& Kawata, D.\ 2005, astro-ph/0503273

\bibitem[Brewer \& Carney (2004)]{brewer04} Brewer, M.-M., \& Carney, B.\ W.\ 2004,
\pasa, 21, 134

\bibitem[\protect\citeauthoryear{{Brown} \& {Wallerstein}}{{Brown} \&
  {Wallerstein}}{1992}]{1992AJ....104.1818B}
{Brown}, J.~A.,  \& {Wallerstein}, G.\ 1992, AJ, 104, 1818

\bibitem[\protect\citeauthoryear{{Burkert}, {Truran}, \& {Hensler}}{{Burkert}
  et~al.}{1992}]{1992ApJ...391..651B}
{Burkert}, A., {Truran}, J.~W.,  \& {Hensler}, G.\ 1992, ApJ, 391, 651

\bibitem[\protect\citeauthoryear{{Burris} et~al.}{{Burris}
  et~al.}{2000}]{2000ApJ...544..302B}
{Burris}, D.~L., {Pilachowski}, C.~A., {Armandroff}, T.~E., {Sneden}, C.,
  {Cowan}, J.~J.,  \& {Roe}, H.\ 2000, ApJ, 544, 302

\bibitem[\protect\citeauthoryear{{Busso}, {Gallino}, \& {Wasserburg}}{{Busso}
  et~al.}{1999}]{1999ARA&A..37..239B}
{Busso}, M., {Gallino}, R.,  \& {Wasserburg}, G.~J.\ 1999, ARA\&A, 37, 239

\bibitem[\protect\citeauthoryear{{Carney}, {Latham}, \& {Laird}}{{Carney}
  et~al.}{1989}]{1989AJ.....97..423C}
{Carney}, B.~W., {Latham}, D.~W.,  \& {Laird}, J.~B.\ 1989, AJ, 97, 423

\bibitem[\protect\citeauthoryear{{Carretta} et~al.}{{Carretta}
  et~al.}{2004}]{2004A&A...416..925C}
{Carretta}, E., {Gratton}, R.~G., {Bragaglia}, A., {Bonifacio}, P.,  \&
  {Pasquini}, L.\ 2004, A\&A, 416, 925

\bibitem[Carretta et al.\ (2004)]{carretta04} Carretta, E., Gratton, R.\ G.,
Bragaglia, A., Bonifacio, P., \& Pasquini, L.\ 2004, \aap, 416, 925

\bibitem[Carretta et al.\ (2000)]{carretta00} Carretta, E., Gratton, R.\ G.,
\& Sneden, C.\ 2000, \aap, 356, 238

\bibitem[\protect\citeauthoryear{{Chen} et~al.}{{Chen}
  et~al.}{2001}]{2001ApJ...553..184C}
{Chen}, B., Stoughton, C., Smith, J.\ A., Uomoto, A., Pier, J.\ R.,
Yanny, B., Ivezi\'{c}, \v{Z}., York, D.\ G., Anderson, J.\ E., Annis, J.,
Brinkmann, J., Csabai, I., Fukugita, M., Hindsley, R., Lupton, R., \& Munn, J.\ A.\
2001, ApJ, 553, 184

\bibitem[\protect\citeauthoryear{{Chen} et~al.}{{Chen}
  et~al.}{2000}]{2000A&AS..141..491C}
{Chen}, Y.~Q., {Nissen}, P.~E., {Zhao}, G., {Zhang}, H.~W.,  \& {Benoni}, T.\
  2000, A\&AS, 141, 491

\bibitem[Chen et al.\ 2003]{chen03} Chen, Y.\ Q., Zhao, G., Nissen, P.\ E.,
Bai, G.\ S., \& Qiu, H.\ M.\ 2003, \apj, 591, 925

\bibitem[\protect\citeauthoryear{{Chiba} \& {Beers}}{{Chiba} \&
  {Beers}}{2000}]{2000AJ....119.2843C}
{Chiba}, M.,  \& {Beers}, T.~C.\ 2000, AJ, 119, 2843

\bibitem[\protect\citeauthoryear{{Cohen} et~al.}{{Cohen}
  et~al.}{1999}]{1999ApJ...523..739C}
{Cohen}, J.~G., {Gratton}, R.~G., {Behr}, B.~B.,  \& {Carretta}, E.\ 1999, ApJ,
  523, 739

\bibitem[\protect\citeauthoryear{{Cowan} et~al.}{{Cowan}
  et~al.}{2002}]{2002ApJ...572..861C}
{Cowan}, J.~J., Sneden, C., Burles, S., Ivans, I.\ I., Beers, T.\ C.,
Truran, J.\ W., Lawler, J.\ E., Primas, F., Fuller, G.\ M., Pfeiffer, B.,
\& Kratz, K.-L.\ 2002, ApJ, 572, 861

\bibitem[\protect\citeauthoryear{{Dalcanton} \& {Bernstein}}{{Dalcanton} \&
  {Bernstein}}{2002}]{2002AJ....124.1328D}
{Dalcanton}, J.~J.,  \& {Bernstein}, R.~A.\ 2002, AJ, 124, 1328

\bibitem[Dinescu et al.\ (1999)]{dinescu99} Dinescu, D., Girard, T.\ M.,
\& van~Altena, W.\ F.\ 1999, \aj, 117, 1792

\bibitem[Dinescu et al.\ (2003)]{dinescu03} Dinescu, D., Girard, T.\ M.,
van~Altena, W.\ F., \& L\'{o}pez, C.\ E.\ 2003, \aj, 125, 1373

\bibitem[Dinescu et al.\ (2000)]{dinescu00} Dinescu, D., 
Majewski, S.\ R., Girard, T.\ M., \& Cudworth, K.\ M.\ 2000, \aj, 120, 1892

\bibitem[Dinescu et al.\ (2001)]{dinescu01} Dinescu, D., 
Majewski, S.\ R., Girard, T.\ M., \& Cudworth, K.\ M.\ 2001, \aj, 122, 1916

\bibitem[\protect\citeauthoryear{{Edvardsson} et~al.}{{Edvardsson}
  et~al.}{1993}]{1993A&A...275..101E}
{Edvardsson}, B., {Andersen}, J., {Gustafsson}, B., {Lambert}, D.~L., {Nissen},
  P.~E.,  \& {Tomkin}, J.\ 1993, A\&A, 275, 101

\bibitem[\protect\citeauthoryear{{Feltzing}, {Bensby}, \& {Lundstr{\"
  o}m}}{{Feltzing} et~al.}{2003}]{2003A&A...397L...1F}
{Feltzing}, S., {Bensby}, T.,  \& {Lundstr{\" o}m}, I.\ 2003, A\&A, 397, L1

\bibitem[Freeman \& Bland-Hawthorn (2002)]{freeman02} Freeman, K.\ C., \&
Bland-Hawthorn, J.\ 2002, \araa, 40, 487

\bibitem[Fry et al.\ (1999)]{fry99} Fry, A.\ M., Morrison, H.\ L.,
Harding, P., \& Boroson, T.\ A.\ 1999, \aj, 118, 1209

\bibitem[Fuchs et al.\ (2001)]{fuchs01} Fuchs, B., Dettbarn, C., Jahreiss, H., \&
Wielen, R.\ 2001, in STARS2000: Dynamics of Star Clusters and the
Milky Way, ed.\ S.\ Deiters et al., ASP Conf.\ Ser.\, 228, p.\ 235

\bibitem[\protect\citeauthoryear{{Fuhrmann}}{{Fuhrmann}}{1998}]{1998A&A...338.%
.161F}
{Fuhrmann}, K.\ 1998, A\&A, 338, 161

\bibitem[Fuhrmann (2004)]{fuhrmann2004} Fuhrmann, K.\ 2004, {\em AN}, 325, 3

\bibitem[\protect\citeauthoryear{{Fulbright}}{{Fulbright}}{2000}]{2000AJ....12%
0.1841F}
{Fulbright}, J.~P.\ 2000, AJ, 120, 1841

\bibitem[\protect\citeauthoryear{{Gilmore}}{{Gilmore}}{1984}]{1984MNRAS.207..2%
23G}
{Gilmore}, G.\ 1984, MNRAS, 207, 223

\bibitem[\protect\citeauthoryear{{Gilmore} \& {Reid}}{{Gilmore} \&
  {Reid}}{1983}]{1983MNRAS.202.1025G}
{Gilmore}, G.,  \& {Reid}, I.\ N.\ 1983, MNRAS, 202, 1025

\bibitem[\protect\citeauthoryear{{Gilmore} \& {Wyse}}{{Gilmore} \&
  {Wyse}}{1985}]{1985AJ.....90.2015G}
{Gilmore}, G.,  \& {Wyse}, R.~F.~G.\ 1985, AJ, 90, 2015

\bibitem[\protect\citeauthoryear{{Gilmore}, {Wyse}, \& {Jones}}{{Gilmore}
  et~al.}{1995}]{1995AJ....109.1095G}
{Gilmore}, G., {Wyse}, R.~F.~G.,  \& {Jones}, J.~B.\ 1995, AJ, 109, 1095

\bibitem[\protect\citeauthoryear{{Gratton}}{{Gratton}}{1989}]{1989A&A...208..1%
71G}
{Gratton}, R.~G.\ 1989, A\&A, 208, 171

\bibitem[\protect\citeauthoryear{{Gratton} et~al.}{{Gratton}
  et~al.}{2003}]{2003A&A...404..187G}
{Gratton}, R.~G., {Carretta}, E., {Claudi}, R., {Lucatello}, S.,  \&
  {Barbieri}, M.\ 2003, A\&A, 404, 187

\bibitem[\protect\citeauthoryear{{Gratton} et~al.}{{Gratton}
  et~al.}{2000}]{2000A&A...358..671G}
{Gratton}, R.~G., {Carretta}, E., {Matteucci}, F.,  \& {Sneden}, C.\ 2000, A\&A,
  358, 671

\bibitem[Grebel et al.\ (2003)]{grebel03} Grebel, E.\ K., Gallagher,
J.\ S., III, \& Harbeck, D.\ 2003, \aj, 125, 1926

\bibitem[Grevesse \& Sauval (1998)]{grevesse1998} Grevesse, N., \&
Sauval, A.\ J.\ 1998, Space Sci.\ Rev., 85, 161

\bibitem[\protect\citeauthoryear{{Hesser} et~al.}{{Hesser}
  et~al.}{1987}]{1987PASP...99..739H}
{Hesser}, J.~E., {Harris}, W.~E., {Vandenberg}, D.~A., {Allwright}, J.~W.~B.,
  {Shott}, P.,  \& {Stetson}, P.~B.\ 1987, PASP, 99, 739

\bibitem[\protect\citeauthoryear{{Holweger} \& {Mueller}}{{Holweger} \&
  {Mueller}}{1974}]{1974SoPh...39...19H}
{Holweger}, H.,  \& {Mueller}, E.~A.\ 1974, \solphys, 39, 19

\bibitem[\protect\citeauthoryear{{James} et~al.}{{James}
  et~al.}{2004}]{2004A&A...414.1071J}
{James}, G., Fran\c{c}ois, P., Bonifacio, P., Bragaglia, A., Carretta, E.,
Centuri\'{o}n, M., Clementini, G., Desidera, S., Gratton, R.\ G.,
Grundahl, F., Lucatello, S., Molaro, P., Pasquini, L., Sneden, C.,
\& Spite, F.\ 2004, A\&A, 414, 1071

\bibitem[Jenkins (1992)]{jenkins92} Jenkins, A.\ 1992, \mnras, 257, 620

\bibitem[Kaluzny et al.\ (1998)]{kaluzny98} Kaluzny, J., Wysocka, A.,
Stanek,K.\ Z., \& Krzemi\'{n}ski, W.\ 1998, Acta Astron.\, 48, 439

\bibitem[\protect\citeauthoryear{{Kim} et~al.}{{Kim}
  et~al.}{2002}]{2002ApJS..143..499K}
{Kim}, Y., {Demarque}, P., {Yi}, S.~K.,  \& {Alexander}, D.~R.\ 2002, ApJS, 143,
  499

\bibitem[Kisleman (1993)]{kiselman1993} Kiselman, D.\ 1993, \aap, 275, 269

\bibitem[Kisleman \& Nordlund (1995)]{kiselman1995} Kiselman, D., \&
Nordlund, A.\ 1995, \aap, 302, 578

\bibitem[Koch \& Edvardsson (2002)]{koch02} Koch, A., \& Edvardsson,
B.\ 2002, \aap, 381, 500

\bibitem[Korn et al.\ (2003)]{korn2003} Korn, A.\ J., Shi, J., \&
Gehren, T.\ 2003, \aap, 407, 691

\bibitem[\protect\citeauthoryear{{Krauss} \& {Chaboyer}}{{Krauss} \&
  {Chaboyer}}{2003}]{2003Sci...299...65K}
{Krauss}, L.~M.,  \& {Chaboyer}, B.\ 2003, Science, 299, 65

\bibitem[\protect\citeauthoryear{{Kurucz}}{{Kurucz}}{1993}]{1993KurCD..13....%
.K}
{Kurucz}, R.\ 1993, ATLAS9 Stellar Atmosphere Programs and 2 km/s grid.~Kurucz
  CD-ROM No.~13.~ Cambridge, Mass.: Smithsonian Astrophysical Observatory

\bibitem[\protect\citeauthoryear{{Kurucz} \& {Bell}}{{Kurucz} \&
  {Bell}}{1995}]{1995KurCD..23.....K}
{Kurucz}, R.,  \& {Bell}, B.\ 1995, Atomic Line Data (R.L.~Kurucz and B.~Bell)
  Kurucz CD-ROM No.~23.~Cambridge, Mass.: Smithsonian Astrophysical
  Observatory

\bibitem[\protect\citeauthoryear{{Lamb} et~al.}{{Lamb}
  et~al.}{1977}]{1977ApJ...217..213L}
{Lamb}, S.~A., {Howard}, W.~M., {Truran}, J.~W.,  \& {Iben}, I.\ 1977, ApJ, 217,
  213

\bibitem[\protect\citeauthoryear{{Latham} et~al.}{{Latham}
  et~al.}{2002}]{2002AJ....124.1144L}
{Latham}, D.~W., {Stefanik}, R.~P., {Torres}, G., {Davis}, R.~J., {Mazeh}, T.,
  {Carney}, B.~W., {Laird}, J.~B.,  \& {Morse}, J.~A.\ 2002, AJ, 124, 1144

\bibitem[\protect\citeauthoryear{{Majewski}}{{Majewski}}{1993}]{1993ARA&A..31.%
.575M}
{Majewski}, S.~R.\ 1993, ARA\&A, 31, 575

\bibitem[\protect\citeauthoryear{{Martin} \& {Morrison}}{{Martin} \&
  {Morrison}}{1998}]{1998AJ....116.1724M}
{Martin}, J.~C.,  \& {Morrison}, H.~L.\ 1998, AJ, 116, 1724

\bibitem[\protect\citeauthoryear{{Mashonkina} \& {Gehren}}{{Mashonkina} \&
  {Gehren}}{2000}]{2000A&A...364..249M}
{Mashonkina}, L.,  \& {Gehren}, T.\ 2000, A\&A, 364, 249

\bibitem[\protect\citeauthoryear{{Mashonkina} \& {Gehren}}{{Mashonkina} \&
  {Gehren}}{2001}]{2001A&A...376..232M}
{Mashonkina}, L.,  \& {Gehren}, T.\ 2001, A\&A, 376, 232

\bibitem[\protect\citeauthoryear{{Mashonkina} et~al.}{{Mashonkina}
  et~al.}{2003}]{2003A&A...397..275M}
{Mashonkina}, L., {Gehren}, T., {Travaglio}, C.,  \& {Borkova}, T.\ 2003, A\&A,
  397, 275

\bibitem[\protect\citeauthoryear{{Matteucci} \& {Recchi}}{{Matteucci} \&
  {Recchi}}{2001}]{2001ApJ...558..351M}
{Matteucci}, F.,  \& {Recchi}, S.\ 2001, ApJ, 558, 351

\bibitem[McWilliam (1998)]{mcwilliam1998} McWilliam, A.\ 1998, \aj, 115, 1640

\bibitem[\protect\citeauthoryear{{McWilliam} \& {Rich}}{{McWilliam} \&
  {Rich}}{1994}]{1994ApJS...91..749M}
{McWilliam}, A.,  \& {Rich}, R.~M.\ 1994, ApJS, 91, 749

\bibitem[Mishenina et al.\ (2004)]{mishenina04} Mishenina, T.\ V., Soubiran, C.,
Kovtyukh, V.\ V., \& Korotin, S.\ A.\ 2004, \aap, 418, 551

\bibitem[\protect\citeauthoryear{{Morrison}, {Boroson}, \&
  {Harding}}{{Morrison} et~al.}{1994}]{1994AJ....108.1191M}
{Morrison}, H.~L., {Boroson}, T.~A.,  \& {Harding}, P.\ 1994, AJ, 108, 1191

\bibitem[\protect\citeauthoryear{{Morrison} et~al.}{{Morrison}
  et~al.}{1997}]{1997AJ....113.2061M}
{Morrison}, H.~L., {Miller}, E.~D., {Harding}, P., {Stinebring}, D.~R.,  \&
  {Boroson}, T.~A.\ 1997, AJ, 113, 2061

\bibitem[Mould (2005)]{mould2005} Mould, J.\ 2005, \aj, 129, 698

\bibitem[Nauomov (1999)]{nauomov99} Naumov, S.\ 1999, unpublished Ph.\ D.\ thesis,
University of North Carolina

\bibitem[Newberg et al.\ 2002]{newberg02} Newberg, H.\ J., 
Yanny, B., Rockosi, C.,
Grebel, E.\ K., Rix, H.-H., Brinkmann, J., Csabai, I., Hennessy, G.,
Hindsley, R.\ B., Ibata, R., Ivez\`{i}\'{c}, Z., Lamb, D., Nash, E.\ T.,
Odenkirchen, M., Rave, H.\ A., Schneider, D.\ P., Smyth, J.\ A., Stolte, A.,
\& York, D.\ G.\ 2002, \apj, 569, 245

\bibitem[\protect\citeauthoryear{{Nissen} et~al.}{{Nissen}
  et~al.}{2000}]{2000A&A...353..722N}
{Nissen}, P.~E., {Chen}, Y.~Q., {Schuster}, W.~J.,  \& {Zhao}, G.\ 2000, A\&A,
  353, 722

\bibitem[\protect\citeauthoryear{{Nissen} \& {Schuster}}{{Nissen} \&
  {Schuster}}{1997}]{1997A&A...326..751N}
{Nissen}, P.~E.,  \& {Schuster}, W.~J.\ 1997, A\&A, 326, 751

\bibitem[\protect\citeauthoryear{{Norris}, {Bessell}, \& {Pickles}}{{Norris}
  et~al.}{1985}]{1985ApJS...58..463N}
{Norris}, J., {Bessell}, M.~S.,  \& {Pickles}, A.~J.\ 1985, ApJS, 58, 463

\bibitem[\protect\citeauthoryear{{Olsen}}{{Olsen}}{1993}]{1993A&AS..102...89O}
{Olsen}, E.~H.\ 1993, A\&AS, 102, 89

\bibitem[Pardi et al.\ (1995)]{pardi95} Pardi, M.\ C., Ferrini, F., \&
Matteucci, F.\ 1995, \apj, 444, 207

\bibitem[\protect\citeauthoryear{{Perryman} et~al.}{{Perryman}
  et~al.}{1997}]{1997A&A...323L..49P}
{Perryman}, M.~A.~C., et~al.\ 1997, A\&A, 323, L49

\bibitem[\protect\citeauthoryear{{Pomp{\' e}ia}, {Barbuy}, \&
  {Grenon}}{{Pomp{\' e}ia} et~al.}{2002}]{2002ApJ...566..845P}
{Pomp{\' e}ia}, L., {Barbuy}, B.,  \& {Grenon}, M.\ 2002, ApJ, 566, 845

\bibitem[\protect\citeauthoryear{{Pomp{\' e}ia}, {Barbuy}, \&
  {Grenon}}{{Pomp{\' e}ia} et~al.}{2003}]{2003ApJ...592.1173P}
{Pomp{\' e}ia}, L., {Barbuy}, B.,  \& {Grenon}, M.\ 2003, ApJ, 592, 1173

\bibitem[\protect\citeauthoryear{{Prochaska} et~al.}{{Prochaska}
  et~al.}{2000}]{2000AJ....120.2513P}
{Prochaska}, J.~X., {Naumov}, S.~O., {Carney}, B.~W., {McWilliam}, A.,  \&
  {Wolfe}, A.~M.\ 2000, AJ, 120, 2513

\bibitem[\protect\citeauthoryear{{Quinn}, {Hernquist}, \& {Fullagar}}{{Quinn}
  et~al.}{1993}]{1993ApJ...403...74Q}
{Quinn}, P.~J., {Hernquist}, L.,  \& {Fullagar}, D.~P.\ 1993, ApJ, 403, 74

\bibitem[\protect\citeauthoryear{{Raiteri}, {Gallino}, \& {Busso}}{{Raiteri}
  et~al.}{1992}]{1992ApJ...387..263R}
{Raiteri}, C.~M., {Gallino}, R.,  \& {Busso}, M.\ 1992, ApJ, 387, 263

\bibitem[\protect\citeauthoryear{{Ram{\'{i}}rez} \& {Cohen}}{{Ram{\'{i}}rez} \&
  {Cohen}}{2002}]{2002AJ....123.3277R}
{Ram{\'{i}}rez}, S.~V.,  \& {Cohen}, J.~G.\ 2002, AJ, 123, 3277

\bibitem[\protect\citeauthoryear{{Reddy} et~al.}{{Reddy}
  et~al.}{2003}]{2003MNRAS.340..304R}
{Reddy}, B.~E., {Tomkin}, J., {Lambert}, D.~L.,  \& {Allende~Prieto}, C.\ 2003,
  MNRAS, 340, 304

\bibitem[\protect\citeauthoryear{{Robin} et~al.}{{Robin}
  et~al.}{1996}]{1996A&A...305..125R}
{Robin}, A.~C., {Haywood}, M., {Cr\'{e}z\'{e}}, M., {Ojha}, D.~K.,  \& {Bienaym\'{e}}, O.\
  1996, A\&A, 305, 125

\bibitem[Rose et al.\ (2004)]{rose04} Rose, J.\ A., Arimoto, N.,
Caldwell, N., Schiavon, R.\ P., Vazdekis, A., \& Yamada, Y.\ 2004,
\aj, in press

\bibitem[\protect\citeauthoryear{{Sandage}}{{Sandage}}{1990}]{1990JRASC..84...%
70S}
{Sandage}, A. 1990, J.\ R.\ Astron.\ Soc.\ Can., 84, 70

\bibitem[Schlegel et al.\ (1998)]{schlegel98} Schlegel, D.\ J.,
Finkbeiner, D.\ P., Davis, M.\ 1998, \apj, 500, 525

\bibitem[Schiavon et al.\ 2004]{schiavon04} Schiavon, R.\ P.,
Caldwell, N., \& Rose, J.\ A.\ 2004, \aj, 127, 1513

\bibitem[\protect\citeauthoryear{{Schuster} \& {Nissen}}{{Schuster} \&
  {Nissen}}{1989}]{1989A&A...221...65S}
{Schuster}, W.~J.,  \& {Nissen}, P.~E.\ 1989, A\&A, 221, 65

\bibitem[Sellwood et al.\ (1998)]{sellwood98} Sellwood, J.\ A., Nelson, R.\ W.,
\& Tremaine, S.\ 1998, \apj, 506, 590

\bibitem[\protect\citeauthoryear{{Shetrone} et~al.}{{Shetrone}
  et~al.}{2003}]{2003AJ....125..684S}
{Shetrone}, M., {Venn}, K.~A., {Tolstoy}, E., {Primas}, F., {Hill}, V.,  \&
  {Kaufer}, A.\ 2003, AJ, 125, 684

\bibitem[\protect\citeauthoryear{{Shetrone}, {C{\^ o}t{\' e}}, \&
  {Sargent}}{{Shetrone} et~al.}{2001}]{2001ApJ...548..592S}
{Shetrone}, M.~D., {C{\^ o}t{\' e}}, P.,  \& {Sargent}, W.~L.~W.\ 2001, ApJ,
  548, 592

\bibitem[\protect\citeauthoryear{{Sneden}}{{Sneden}}{1973}]{1973PhDT.........2%
N}
{Sneden}, C.\ 1973, Ph.D.~Thesis, University of Texas

\bibitem[Soubiran \& Girard (2005)]{soubiran2005} Soubiran, C., \&
Girard, P.\ 2005, astro-ph/0503498

\bibitem[\protect\citeauthoryear{{Statler}}{{Statler}}{1988}]{1988ApJ...331...%
71S}
{Statler}, T.~S.\ 1988, ApJ, 331, 71

\bibitem[Th\'{e}venin \& Idiart (1999)]{thevenin1999} Th\'{e}venin, F., \&
Idiart, T.\ P.\ 1999, \apj, 521, 753

\bibitem[\protect\citeauthoryear{{Timmes}, {Woosley}, \& {Weaver}}{{Timmes}
  et~al.}{1995}]{1995ApJS...98..617T}
{Timmes}, F.~X., {Woosley}, S.~E.,  \& {Weaver}, T.~A.\ 1995, ApJS, 98, 617

\bibitem[\protect\citeauthoryear{{Tody}}{{Tody}}{1986}]{1986SPIE..627..733T}
{Tody}, D.\ 1986, in Instrumentation in astronomy VI; Proceedings of the
  Meeting, Tucson, AZ, Mar. 4-8, 1986. Part 2 (A87-36376 15-35). Bellingham,
  WA, Society of Photo-Optical Instrumentation Engineers, 1986, p. 733

\bibitem[Uns\"{o}ld (1955)]{unsold1955} Uns\"{o}ld, A., Physik der
Sternatmosph\"{a}ren (Berlin: Springer-Verlag)

\bibitem[Venn et al.\ (2004)]{venn04} Venn, K.\ A., Irwin, M., Shetrone, M.\ D.,
Tout, C.\ A., Hill, V., \& Tolstoy, E.\ 2004, \aj, 128, 1177

\bibitem[Walker et al.\ (1996)]{walker96} Walker, I.\ R., Mihos, J.\ C., 
\& Hernquist, L.\ 1996, \apj, 460, 121

\bibitem[\protect\citeauthoryear{{Wasserburg} \& {Qian}}{{Wasserburg} \&
  {Qian}}{2000}]{2000ApJ...529L..21W}
{Wasserburg}, G.~J.,  \& {Qian}, Y.-Z.\ 2000, ApJ, 529, L21

\bibitem[\protect\citeauthoryear{{Woosley} \& {Weaver}}{{Woosley} \&
  {Weaver}}{1995}]{1995ApJS..101..181W}
{Woosley}, S.~E.,  \& {Weaver}, T.~A.\ 1995, ApJS, 101, 181

\bibitem[Worthey (2004)]{worthey04} Worthey, G.\ 2004, \aj, 128, 2826

\bibitem[\protect\citeauthoryear{{Wyse} \& {Gilmore}}{{Wyse} \&
  {Gilmore}}{1988}]{1988AJ.....95.1404W}
{Wyse}, R.~F.~G.,  \& {Gilmore}, G.\ 1988, AJ, 95, 1404

\bibitem[Yoachim \& Dalcanton (2005)]{yoachim2005} Yoachim, P., \&
Dalcanton, J.\ J.\ 2005, \apj, 624, 701

\bibitem[\protect\citeauthoryear{{Yong} et~al.}{{Yong}
  et~al.}{2003}]{2003A&A...402..985Y}
{Yong}, D., {Grundahl}, F., {Lambert}, D.~L., {Nissen}, P.~E.,  \& {Shetrone},
  M.~D.\ 2003, A\&A, 402, 985

\bibitem[\protect\citeauthoryear{{Yoshii}, {Ishida}, \& {Stobie}}{{Yoshii}
  et~al.}{1987}]{1987AJ.....93..323Y}
{Yoshii}, Y., {Ishida}, K.,  \& {Stobie}, R.~S.\ 1987, AJ, 93, 323

\bibitem[\protect\citeauthoryear{{Zhao} \& {Magain}}{{Zhao} \&
  {Magain}}{1990}]{1990A&A...238..242Z}
{Zhao}, G.,  \& {Magain}, P.\ 1990, A\&A, 238, 242

\bibitem[Zibetti et al.\ (2004)]{zibetti04} Zibetti, S., White, S.\ D.\ M.,
\& Brinkmann, J.\ 2004, \mnras, 347, 556

\end{thebibliography}
\end{document}